\documentclass[%
prc,
superscriptaddress,
reprint,
showpacs,
showkeys,
nofootinbib,
nobibnotes,
 amsmath,amssymb,
 aps,
]{revtex4-1}

\usepackage{graphicx}% Include figure files
\usepackage{dcolumn}% Align table columns on decimal point
\usepackage{bm}% bold math
\usepackage{hyperref}% add hypertext capabilities
\usepackage{longtable} %tables of more than one page
\usepackage{array}

\newcolumntype{C}[1]{>{\centering\let\newline\\\arraybackslash\hspace{0pt}}m{#1}} %C columns auto break line 
\LTcapwidth=\textwidth

\begin{document}
\preprint{APS/123-QED}

\title{$\beta^-$-decay of $^{65}$Mn to $^{65}$Fe}

\author {B.~Olaizola}
\email{olaizola@nuclear.fis.ucm.es}
\affiliation{Grupo de F\'isica Nuclear, Facultad de F\'isicas, Universidad Complutense - CEI Moncloa, E-28040 Madrid, Spain}

\author{L.M.~Fraile}
\affiliation{Grupo de F\'isica Nuclear, Facultad de F\'isicas, Universidad Complutense - CEI Moncloa, E-28040 Madrid, Spain}

\author{H.~Mach}
\altaffiliation[Permanent address: ]{Medical and Scientific Time Imaging Consulting, MASTICON - Fru\"angsgatan 56 E, Nyk\"oping, 61130 Sweden}
\affiliation{Grupo de F\'isica Nuclear, Facultad de F\'isicas, Universidad Complutense - CEI Moncloa, E-28040 Madrid, Spain}

\author{A.~Aprahamian}
\affiliation{Department of Physics, University of Notre Dame, Notre Dame, Indiana 46556, USA}

\author{J.A.~Briz}
\affiliation{Instituto de Estructura de la Materia, CSIC, E-28006 Madrid, Spain}

\author{J.~Cal-Gonz\'alez}
\affiliation{Grupo de F\'isica Nuclear, Facultad de F\'isicas, Universidad Complutense - CEI Moncloa, E-28040 Madrid, Spain}

\author{D.~Ghi\c{t}a}
\affiliation{Horia Hulubei National Institute of Physics and Nuclear Engineering , Bucharest, Romania}

\author{U.~K\"oster}
\affiliation{Institut Laue-Langevin, B.P. 156, 38042 Grenoble Cedex 9, France}

\author{W.~Kurcewicz}
\affiliation{Institute of Experimental Physics, Warsaw University, Ho\.za 69, PL 00-681 Warsaw, Poland}

\author{S.R.~Lesher}
\affiliation{Department of Physics, University of Wisconsin - La Crosse, La Crosse, WI 54601, USA}
\affiliation{Department of Physics, University of Notre Dame, Notre Dame, Indiana 46556, USA}
%\altaffiliation[also at ]{Department of Physics, University of Notre Dame, Notre Dame, Indiana 46556, USA}

\author{D.~Pauwels}
\affiliation{K.U. Leuven, IKS, Celestijnenlaan 200 D, 3001 Leuven, Belgium }

\author{E.~Picado}
\affiliation{Grupo de F\'isica Nuclear, Facultad de F\'isicas, Universidad Complutense - CEI Moncloa, E-28040 Madrid, Spain}
\affiliation{Secci\'on de Radiaciones, Departamento de F\'isica, Universidad Nacional, Heredia, Costa Rica}

\author{A.~Poves}
\affiliation{Departamento de F\'isica Te\'orica e IFT-UAM/CSIC, Universidad Aut\'onoma de Madrid, E-28049 Madrid, Spain}

\author{D.~Radulov}
\affiliation{K.U. Leuven, IKS, Celestijnenlaan 200 D, 3001 Leuven, Belgium }

\author{G.S.~Simpson}
\affiliation{LPSC, Universit\'e Joseph Fourier Grenoble 1, CNRS/IN2P3, Institut National Polytechnique de Grenoble, F-38026 Grenoble Cedex, France}

\author{ J.M.~Ud\'ias}
\affiliation{Grupo de F\'isica Nuclear, Facultad de F\'isicas, Universidad Complutense - CEI Moncloa, E-28040 Madrid, Spain}

\date{\today}

\begin{abstract}
The low energy structure of $^{65}$Fe has been studied by means of $\gamma$- and fast-timing spectroscopy. A level scheme of $^{65}$Fe populated following the $\beta ^-$decay of $^{65}$Mn was established for the first time. It includes 41 levels and 85 transitions. The excitation energy of the $\beta$-decaying isomer in $^{65}$Fe has been precisely determined at 393.7(2) keV. The $\beta$ delayed neutron emission branch was measured as $P_n$ = 7.9(12) $\%$, which cannot be reconciled with the previously reported value of 21.0(5) $\%$. Four $\gamma$-rays and four excited states in $^{64}$Fe were identified as being populated following the $\beta$-n decay. Four lifetimes and five lifetime limits in the subnanosecond range have been measured using the Advanced Time-Delayed Method. The level scheme is compared with shell-model calculations. Tentative spin and parity assignments are proposed based on the observed transition rates, the calculations and the systematics of the region. 
\end{abstract}


\maketitle

\section{Introduction \label{sec:Introduction}}

In the neutron-rich nuclei protons and neutrons occupy different orbitals than in stable nuclei. This may lead to a modification of the single particle energies and the appearance of strong quadrupole correlations, which in turn may neutralize the spherical mean-field shell gaps. The quadrupole correlations make deformed intruder configurations energetically favourable. Therefore some of the shell closures established along the valley of stability break down in nuclei with high isospin values, resulting in new shell structures. 

\begin{figure}%[h!]
\includegraphics[width=\columnwidth]{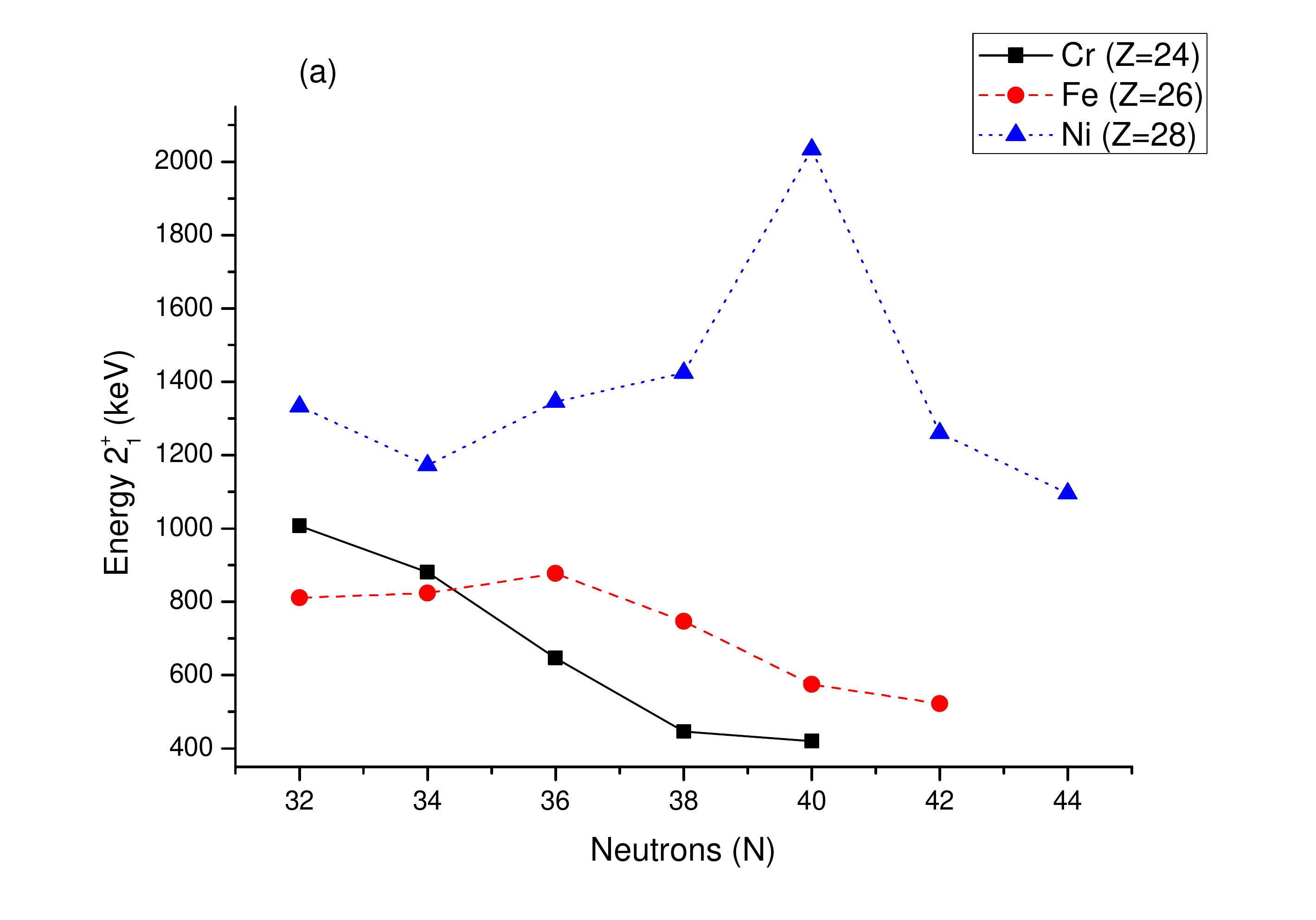}
\includegraphics[width=\columnwidth]{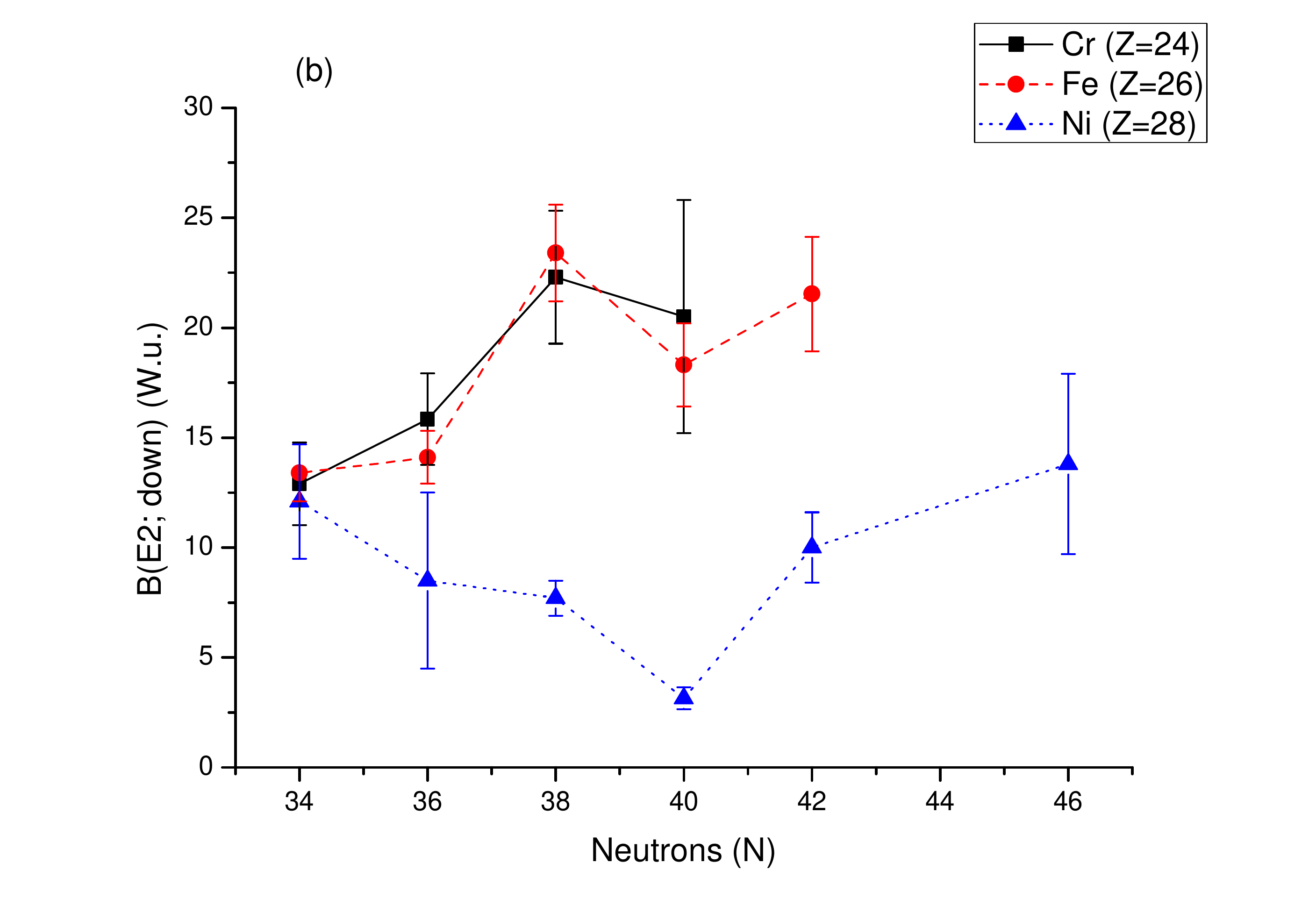}
\caption{\label{fig:systematics} \textit{(Color online)} (a) Systematics of the $E(2^+_1)$ energies. (b) $B(E2)$ values in the $^{68}$Ni region. Data was taken from \cite{PRI12}, except $^{64}$Cr, $^{66,68}$Fe values from \cite{MAC13} and $^{58,60,62}$Cr values from \cite{BAU12}.}
\end{figure}

The interplay of spherical and collective configurations is observed in the neutron-rich nuclei below $Z=28$ and $N=40$. Figure \ref{fig:systematics} shows the systematics of the excitation $E(2^+_1)$ energies and $B(E2)$ values in the heavy even-even Ni, Fe and Cr nuclei. It shows a contrast between the systematics of Ni and of both Fe and Cr. At $N=40$ Ni shows the highest excitation energy of $E(2^+_1)$ and the lowest $B(E2)$ value for the subregion, while these parameters take opposite characteristics for Fe and Cr. 

Beyond $N=34$, the $g_{9/2}$ neutron orbital starts playing an important role. However, shell-model calculations using only the $pfg$ neutron valence space fail to correctly describe the collectivity at $N=40$ \cite{kan08,LUN07}, and specifically the low energy of the 573 keV $2^+_1$ state in $^{66}$Fe \cite{HAN99}. As pointed out in \cite{CAU02}, a proper description of the strong quadrupole collectivity in this region requires also an inclusion of the neutron $1d_{5/2}$ orbital. Recently Lenzi \textit{et al.}~\cite{LEN10} have developed shell-model calculations in a large valence space that encompasses the $pf$ shell for protons and the $1p_{3/2}$, $1p_{1/2}$ $0f_{5/2}$, $0g_{9/2}$ and $1d_{5/2}$ orbitals for neutrons, by using a new effective interaction and with the monopole part empirically tuned to reproduce the experimental single-particle energies. With this approach a very good agreement with the available experimental data was obtained, not only for excitation energies but also for transition rates. 

The $N=40$ nucleus $^{68}$Ni ($Z=28$) has a large $E(2^+_1)$ energy above 2 MeV \cite{BRO95} and a small value of $B(E2;0^+_1\rightarrow2^+_1)$ = 265 $e^2fm^4$ (3.2 W.u) \cite{SOR02}. However, mass measurements have showed that the shell gap at $N=40$ is weak for $^{68}$Ni \cite{GUE07}, implying that the small $B(E2)$ value does not indicate a sub-shell gap. As protons are removed from the $f_{7/2}$ orbital, the energies of the $2^+_1$ states drop sharply and the $B(E2)$ values increase, as seen in Fig.~\ref{fig:systematics}. The lowest-lying $2^+_1$ level experimentally reported so far in the region is the 420-keV state in $^{64}$Cr \cite{GAD10}. It is in line with theoretical calculations \cite{LEN10}, which also predict a large value for the $B(E2;2^+\rightarrow0^+)$ in $^{64}$Cr. Recently this value was measured at NSCL by the Coulex method to be 21(5) W.u.~\cite{MAC13}.

Compared to $^{66}$Fe, the $2^+_1$ energy is lower in $^{68}$Fe \cite{ADR08}, at 517~keV. This behaviour suggests a rapid development of collectivity in the heavier neutron rich Fe isotopes where the $\nu g_{9/2}$ orbit plays an important role (see for example Ref.~\cite{LUN07}). This is consistent with a decreasing excitation energy of the $9/2^+$ isomers in the odd Fe isotopes. Precise experimental information on the Fe isotopes is therefore crucial to better understand of the nuclear structure in this region and to map the onset of collectivity. Of particular interest are transition rates, which provide stringent tests of theoretical models and probe the collective and single particle nature of states.

Our present study includes the $\gamma$ and fast timing spectroscopy of the levels in $^{65}$Fe populated following the $\beta$ decay of $^{65}$Mn. The Advanced Time Delayed $\beta \gamma \gamma$(t) method \cite{MAC89,MOS89} has been employed to measure excited level lifetimes in the subnanosecond range. Prior to this study, there was scarce information on the $\beta$ decay of $^{65}$Mn and on the levels in $^{65}$Fe. The half-life of the $^{65}$Mn ground state has been measured as 92(1)~ms \cite{BRO10}. A $P_n=21.0(5)\:\%$ for the $\beta$-n branch of $^{65}$Mn was reported in \cite{HAN00}, who concluded that all the delayed neutron intensity directly feeds the ground state of $^{64}$Fe.

A 420(13)-ns isomer was identified at 396.8 keV, which feeds the 363.3(5)-keV level by a 33.5(5)-keV transition  \cite{GRZ98,DAU10,DAU11}. The 33.5-keV transition was assigned as E1 \cite{DAU06} although no evidence was provided. A second isomer in $^{65}$Fe was identified using Penning trap mass spectrometry. The excitation energy of this $\beta^-$decaying state was measured at 402(10)~keV \cite{BLO08,FER10}. Precise $\gamma$-spectroscopic studies established that the two $\beta$-decaying  states in $^{65}$Fe populate two sets of mutually independent level schemes in $^{65}$Co \cite{PAU09}. The half-lives of the $\beta$-decaying states were measured as $T_{1/2}$ = 1.12(15)~s for the 402-keV $(9/2^+)$ isomer and $T_{1/2}$ = 0.81(5) s for the $(1/2^-)$ ground state \cite{PAU09}. Studies using multinucleon transfer reactions \cite{LUN07} revealed two transitions in $^{65}$Fe which most likely form a high-spin yrast cascade. These transitions are not expected to be observed in the $\beta$ decay of $^{65}$Mn. 

A tentative partial level scheme of $^{65}$Fe following the $\beta$ decay of $^{65}$Mn was proposed by Gaudefroy \textit{et al.} \cite{GAU05PhD}, where ten transitions in the $\beta$ decay of $^{65}$Mn were identified and five of them were placed in a tentative level scheme. Since no $\gamma \gamma$ coincidences were observed two new levels at excitation energies of 455 and 1089 keV were proposed based on the energy matching. 

\section{Experimental details \label{sec:Technical Details}}

The activity of $^{65}$Mn was produced at the ISOLDE facility at CERN by the bombardment of a UC$_x$/graphite target with 1.4-GeV protons. The reaction products diffused out of the target matrix, were ionized by the selective resonant ionization laser-ion source RILIS \cite{FED12} and accelerated to 30 kV. The $A = 65$ ions were then mass separated by the General Purpose Separator (GPS) \cite{KUG00} and implanted into a thin aluminium foil in the center of the experimental setup. The proton beam was pulsed, with proton packets separated in time by multiples of 1.2 s. During the proton impact the ion beam was deflected for 6 ms by an electrostatic gate, preventing contamination with other masses. After a predefined period of time of 400~ms the beam was deflected again in order to block the collection of long-lived activities released from the target and the collected sample was allowed to decay out. In our experiment the ions were deposited on the collection foil creating a saturated source that included short- and long-lived decay products from the decay of $^{65}$Mn and its daughters. No old activity was removed from the source. The strongly produced 15 min $^{65}$Ga was also present in the data as a contaminant that was surface ionized in the hot tungsten \cite{KOS02} tube where the laser beams interact with the effusing manganese atoms. Manganese and other isobars (Fe, Co, Ni, Cu, Zn) have higher ionization potentials and show negligible surface ionization.

The measurement station included five detectors positioned in a close geometry around the beam deposition point. The fast timing $\beta$ detector was a 3-mm thick NE111A plastic scintillator placed directly behind the radioactive source. The $\gamma$-ray detectors included two fast-response LaBr$_3$(Ce) scintillators in the shape of truncated cones (38.1~mm in height, 38.1~mm diameter at the bottom and 25.4~mm diameter at the entrance window), which were coupled to the Photonis XP20D0 photomultipliers, as well as two HPGe detectors with relative efficiencies of 60$\%$. The experimental set-up and data collection was optimized for the application of the Advanced Time-Delayed $\beta\gamma\gamma$(t) method described in \cite{MAC89,MOS89,MAC91}, so only a few details are given below. 

The data were collected using a digital data acquisition system which consisted of four Digital Gamma Finder (DGF) Pixie-4 modules Revision C \cite{XIA}. Ten parameters were collected in an independent ungated mode without external triggers: energies from five individual detectors, four time differences between $\beta$ and each $\gamma$ detector signals, and the time of arrival of the proton pulse on the target. A PIXIE time stamp was added to each collected parameter. Coincident events between detectors were sorted off-line. For precise timing information we had set four analog time-delayed $\beta\gamma$(t) coincidence systems each started by a signal from the $\beta$ detector and stopped by a fast signal from one of the $\gamma$ detectors. The time range was 50 ns for coincidences between the scintillator detectors and 2~$\mu$s for the $\beta$-HPGe coincidences. The analog outputs from the Time-to-Amplitude Conversion (TAC) units were fed to the Pixie-4 modules. 

The energy and efficiency calibrations of the HPGe detectors were made using the sources of $^{152}$Eu, $^{24}$Na, $^{88}$Rb and $^{140}$Ba. During the whole measurement the energy shifts remained within the $\pm$0.1 keV limit. At the energies below 120 keV relative efficiency calibrations were provided by the decay of $^{63}$Co to $^{63}$Ni observed in the same experiment.

\subsection*{Time response calibrations of the scintillator detectors}

The lifetime measurements in subnanosecond range were performed using the $\beta\gamma\gamma$(t) method \cite{MAC89,MOS89,MAC91}. The time-delayed $\beta\gamma$(t) coincidences were started by a signal from the $\beta$ detector and stopped by one of the LaBr$_3$(Ce) $\gamma$ detectors. An additional coincidence with the HPGe detector was used to select the desired $\gamma$ decay branch. A comparison of the $\gamma\gamma$ coincidences using HPGe-HPGe and HPGe-LaBr$_3$(Ce) detectors revealed the exact composition of the LaBr$_3$(Ce) coincident spectra which were characterized by worse energy resolution. The time response calibration of the fast timing detectors were performed using the $\beta$-decay sources of $^{140}$Ba/$^{140}$La, $^{88}$Rb and $^{24}$Na. We have corrected for the $\beta$ walk-curve non-linearity using our standard procedures. The residual differences from a flat constant response were on the average less than 5 ps.

\begin{figure}
\includegraphics[width=\columnwidth]{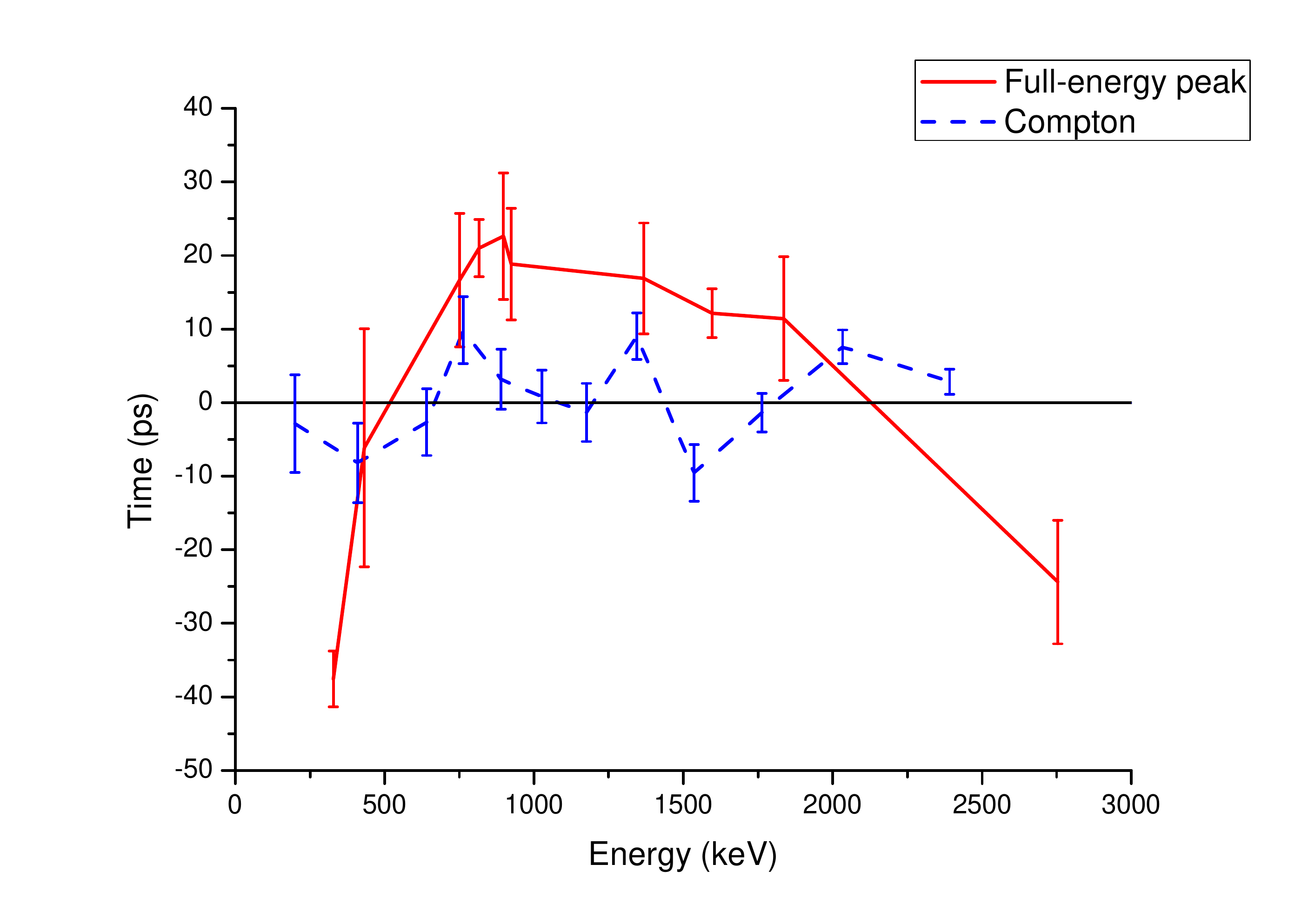}
\caption{\label{fig:time-response} \textit{(Color online)} Residual Difference curves showing relative time response for LaBr$_3$(Ce)-1 detector as a function of energy for the FEP (solid line) and Compton events (dashed line).}
\end{figure}

The time responses of the LaBr$_3$(Ce) $\gamma$ detectors were calibrated to within 10 picosecond precision separately for the Compton and the Full-Energy-Peak (FEP) $\gamma$ events detected in the crystals. The time response of a Compton events is different from FEP of the same energy \cite{MAC91}. The procedure started with the construction of an ``Approximate Prompt Curve" (APC) using the Compton events for transitions following the decay of the $^{24}$Na source, and then using this curve to extract ``Residual Differences" (RD) between the FEP events and the APC curve using different sources. Our procedure is very similar to the one described in detail in Ref.~\cite{MAC91}. The Residual Differences for the Compton and FEP events are shown in Fig. \ref{fig:time-response}, which is equivalent to Fig.~4 in Ref.~\cite{MAC91}. 

\section{Results}

\begin{figure}
\centering
\begin{tabular}{c}
\includegraphics[width=\columnwidth]{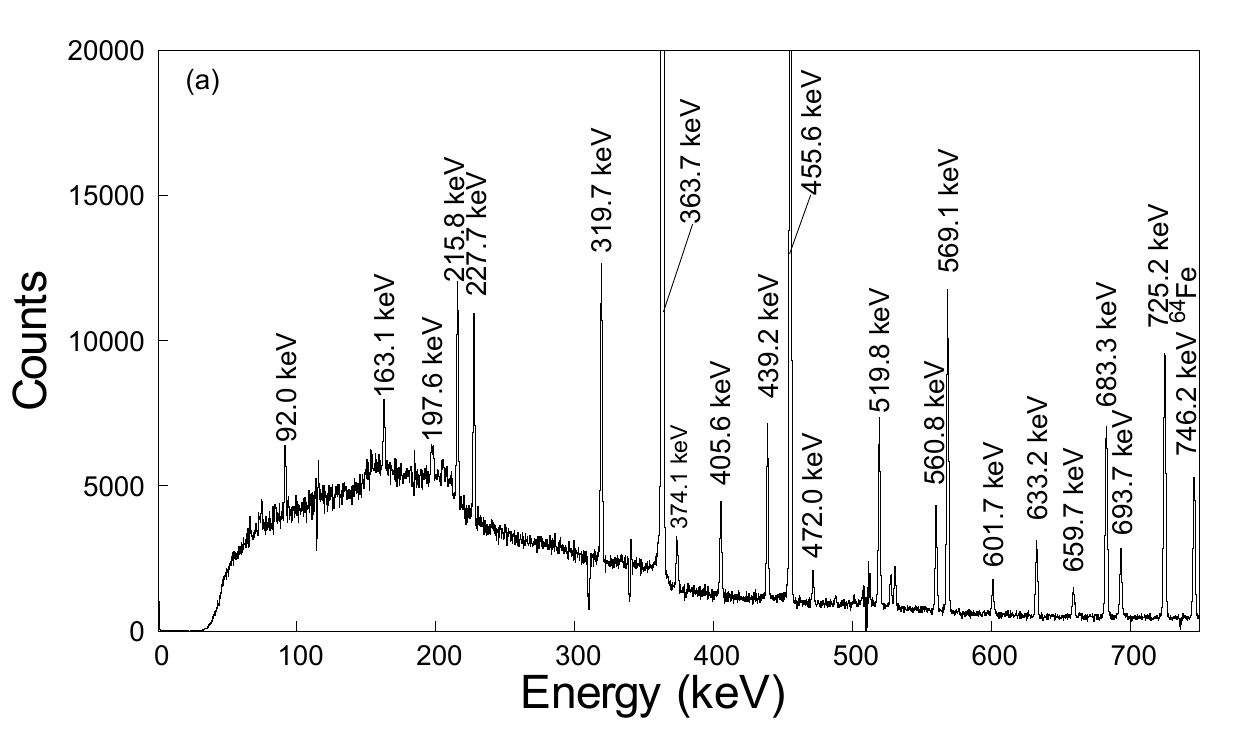}\\
\includegraphics[width=\columnwidth]{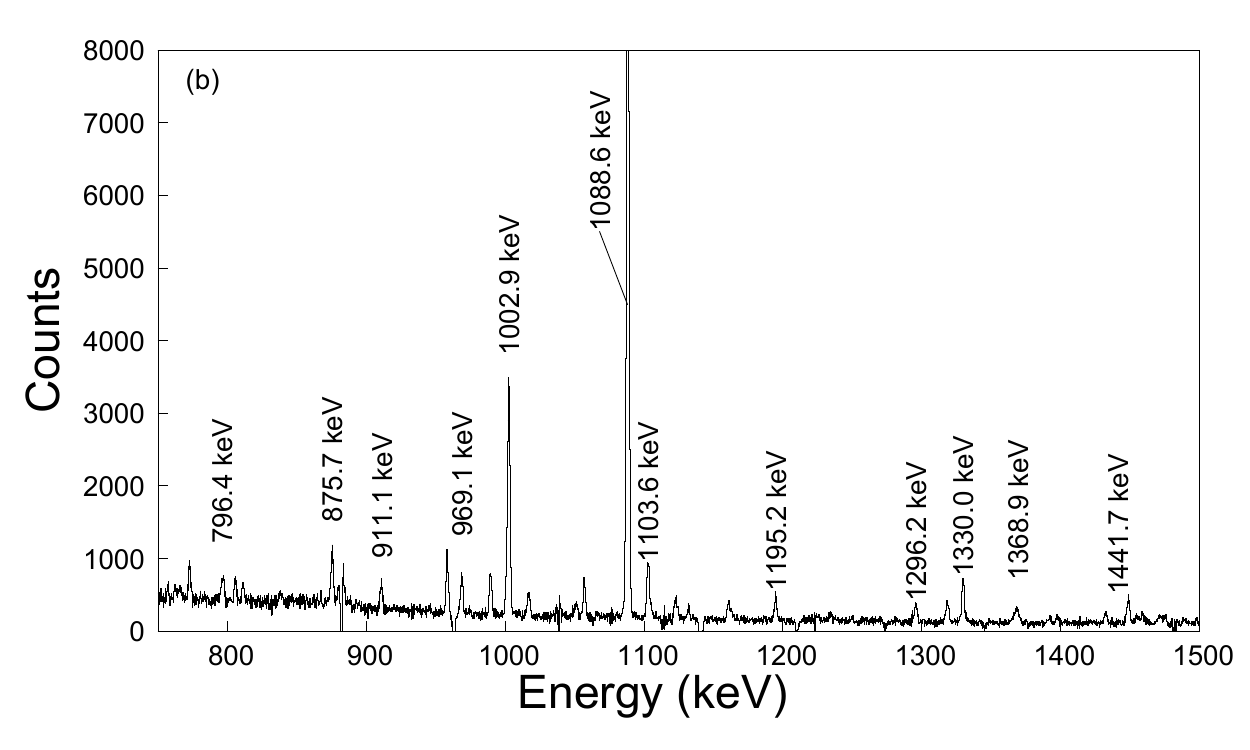}\\
\end{tabular}
\caption{\label{fig:HPGe-spectra} Lower (a) and higher (b) energy spectrum recorded in a HPGe detector sorted from the two-parameter data set involving HPGe events and the time elapsed between the event and the last proton pulse. A time gate was set on the proton time spectrum in order to enhance the $^{65}$Mn activity. Longer lived activities were subtracted as discussed in the text. }
\end{figure}

\begin{figure}
\includegraphics[width=\columnwidth]{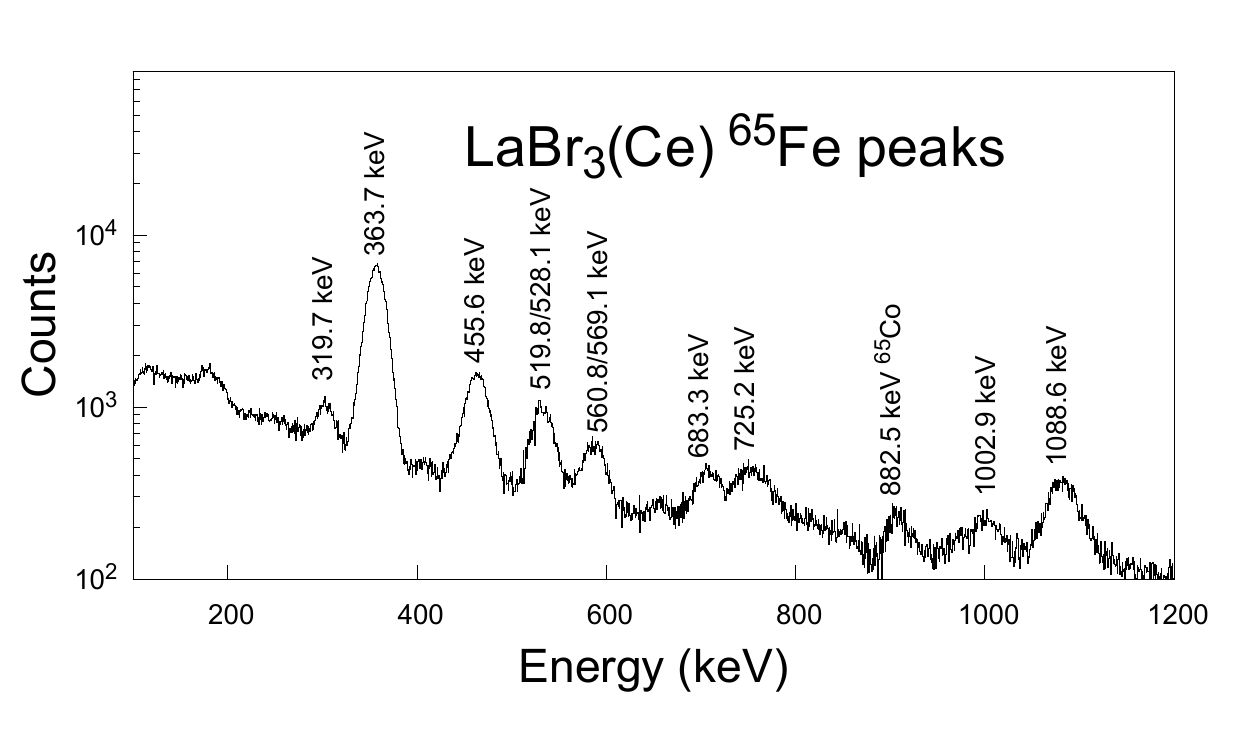}
\caption{\label{fig:LaBr-energy-spectrum} LaBr$_3$(Ce) energy spectrum obtained in an equivalent way to the HPGe energy spectrum. The most intense transition energies in $^{65}$Fe are labelled. }
\end{figure}

The identification of $\gamma$-rays belonging to the $\beta$ decay of $^{65}$Mn was made in three different ways. First, a two-parameter data set, which included the energies of events recorded in the HPGe detector and the time elapsed between the event and the last proton pulse, was analyzed. The short-lived activity of $^{65}$Mn was enhanced in the off-line sorting by selecting a time window from 10 to 450 ms after the proton pulse. Longer-lived components, including lines from $^{65}$Ga, have been subtracted using an equivalent time window above 800 ms, which contained virtually no $^{65}$Mn activity. An almost pure $^{65}$Mn HPGe energy spectrum is shown in Fig.~\ref{fig:HPGe-spectra}. Some of the longer lived decay products were oversubstracted and appear in the spectrum as small negative peaks. Figure \ref{fig:LaBr-energy-spectrum} shows an equivalent energy spectrum observed in a LaBr$_3$(Ce) detector, which was characterized by 3.3$\%$ energy resolution at 900 keV.

Then, using the same data set, gates were set on the full energy peaks in the HPGe spectrum and projected on the proton time spectra. The $\gamma$ rays were identified as belonging to the $^{65}$Mn decay when their time spectrum was consistent with the decay half-life of 92 ms \cite{BRO10}. Finally the identification of a line from the decay of $^{65}$Mn was made based on firm $\gamma\gamma$ coincidences with the Mn lines already identified in the first two steps. 

\subsection{$^{65}$Mn half-life}

\begin{figure}
\includegraphics[width=8.6cm,keepaspectratio=true]{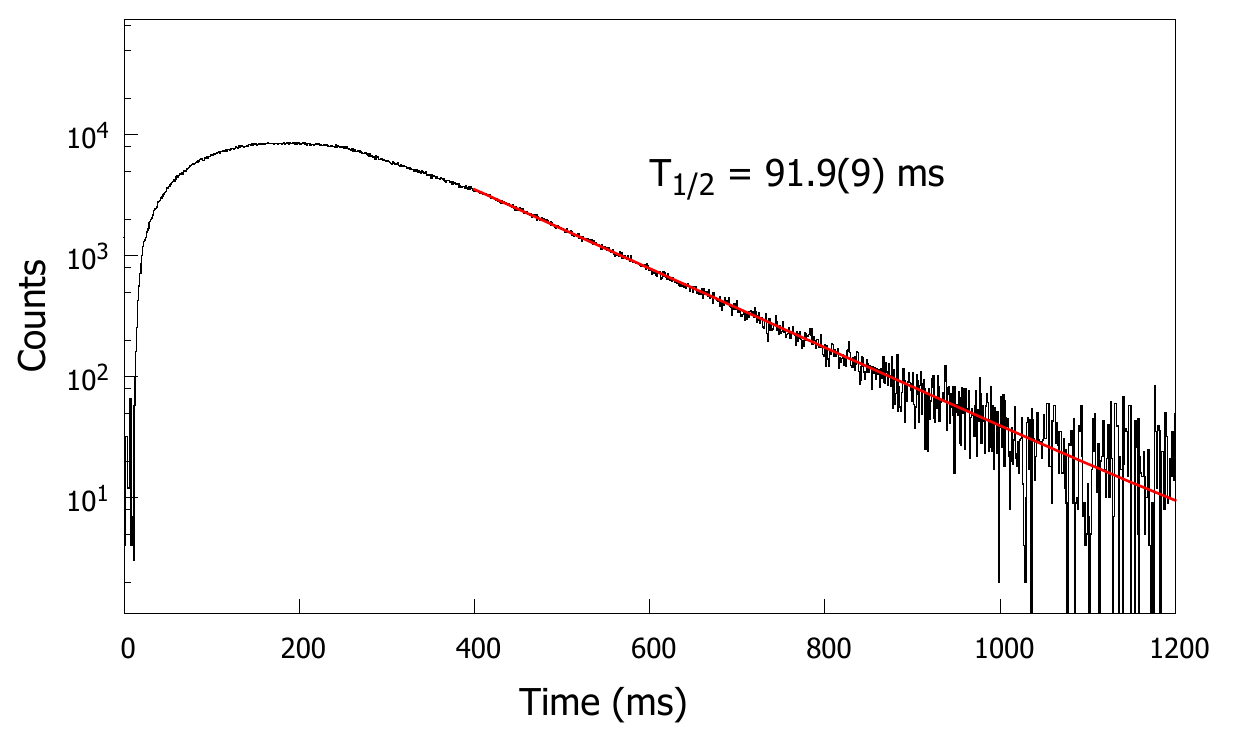}
\caption{\label{fig:Beta_half-life_singles} \textit{(Color online)} The time spectrum for the time elapsed from the last proton pulse for the HPGe events gated by the 363.7-keV transition. The fit between 400 and 1200 ms gives the half-life of $^{65}$Mn; see the text for details.}
\end{figure}

The $^{65}$Mn half-life was obtained by fitting a proton time spectrum sorted out from the aforementioned two-parameter data set. It was gated by the 363.7-keV $\gamma$ ray and projected onto the time elapsed from the last proton pulse shown in Fig.~\ref{fig:Beta_half-life_singles}. A portion of the spectrum, from 400 to 1200 ms, was fitted to an exponential decay plus a constant background. Note, that at 400 ms after the proton pulse the beam gate was closed and radioactive sample was left free to decay. The fitted slope gives $T_{1/2}$ = 91.9(9) ms in very good agreement with the adopted half-life of 92(1) ms \cite{BRO10}. A similar analysis was performed on the five transitions at 455.6, 569.1, 683.3 (a doublet), 725.2 and 1002.9 keV. Their weighted average gives a comparable value $T_{1/2} = 92.0(13)\;\text{ms}$. 

\subsection{$\gamma\gamma$ coincidences}

\begin{figure}
\includegraphics[width=\columnwidth]{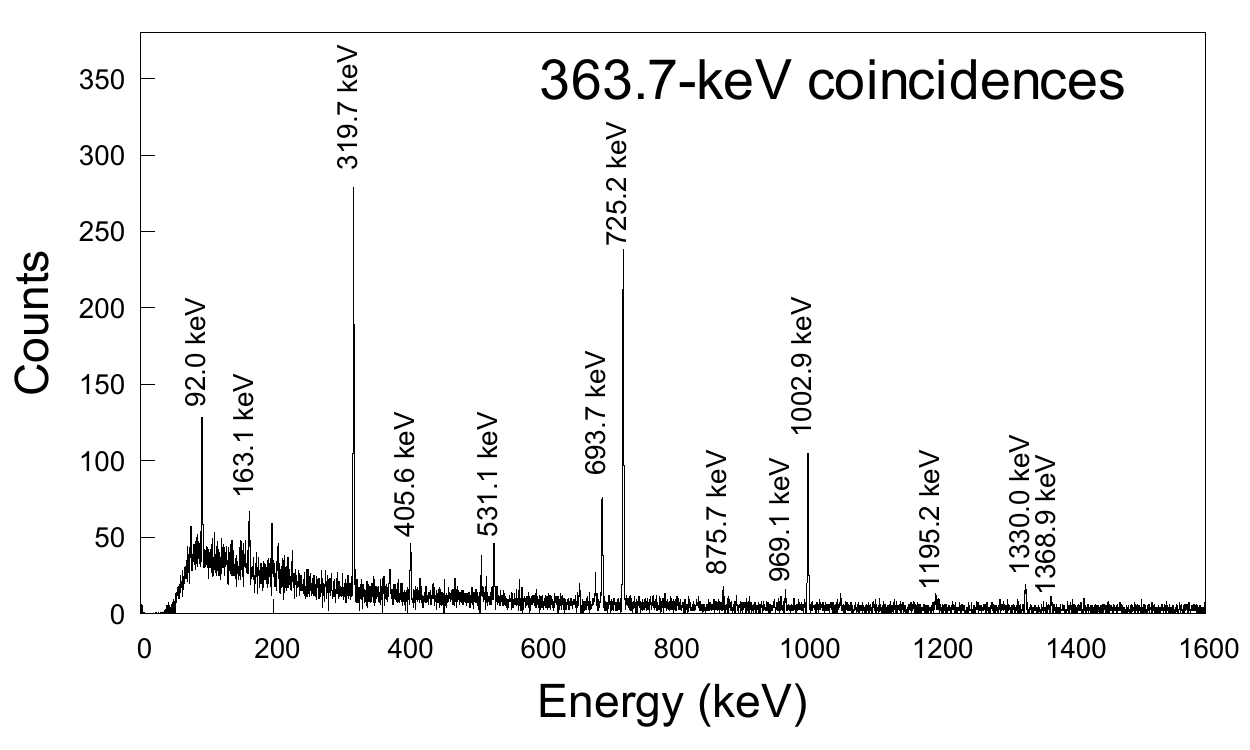}
\caption{\label{fig:363-gate-coincidences} A $\gamma \gamma$ coincident spectrum gated on the intense 363.7-keV line in a HPGe detector and projected onto the second HPGe; see the text for details. }
\end{figure}

The level scheme was constructed using the HPGe-HPGe $\gamma\gamma$ coincidences. In order to enhance the $^{65}$Mn activity a gate was also set on the proton time spectrum between 10 to 450 ms. Figure \ref{fig:363-gate-coincidences} shows $\gamma$ rays in coincidence with the 363.7-keV line.

\begin{figure*}
\includegraphics[height=\textheight]{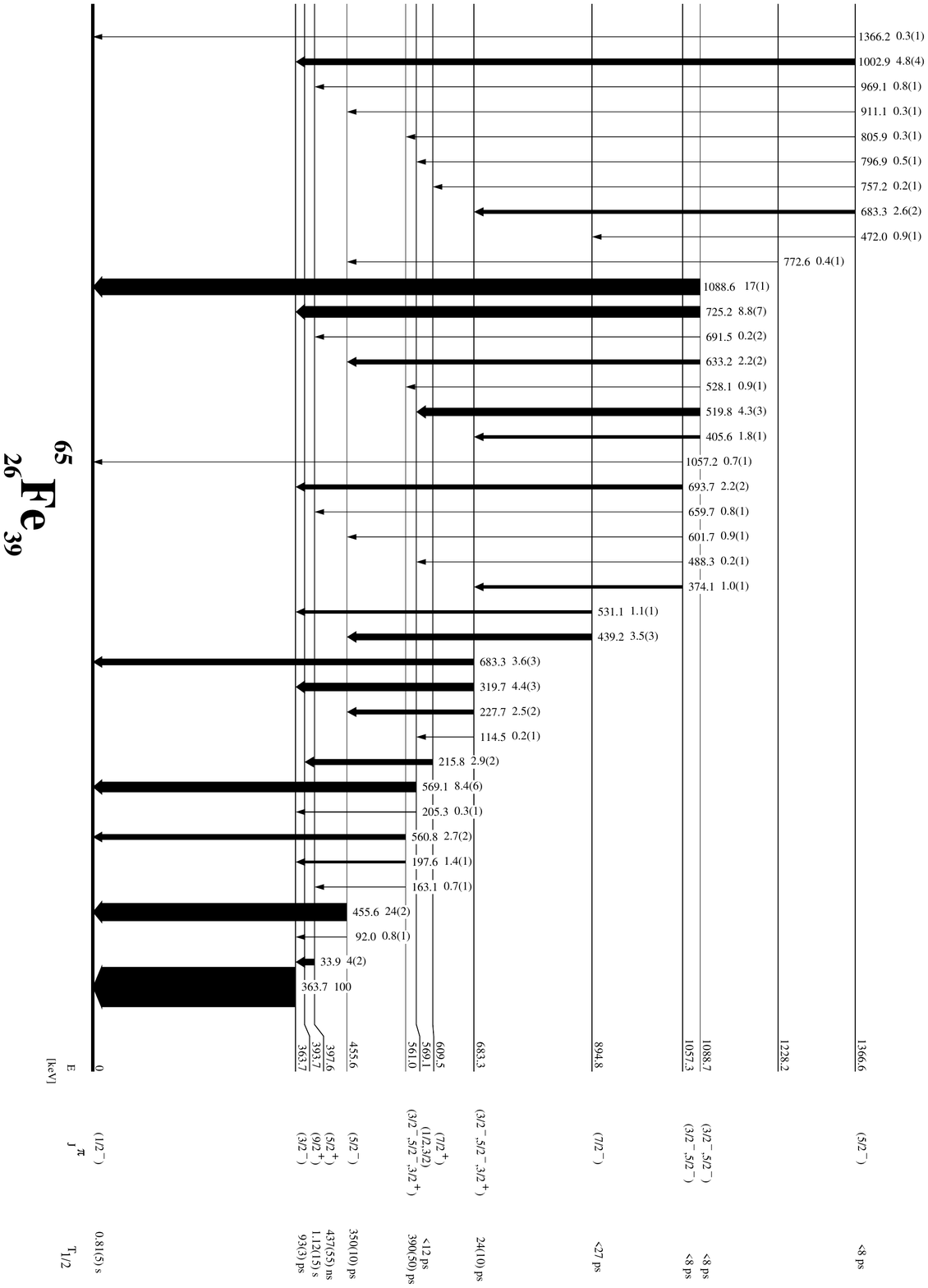}
\caption{\label{fig:Fe65_low-energies} Low energy $^{65}$Fe level scheme populated following the $\beta$-decay of $^{65}$Mn. All results are from this work except for the half-lives of the $\beta$-decaying states.}
\end{figure*}

\begin{figure*}
\includegraphics[height=\textheight]{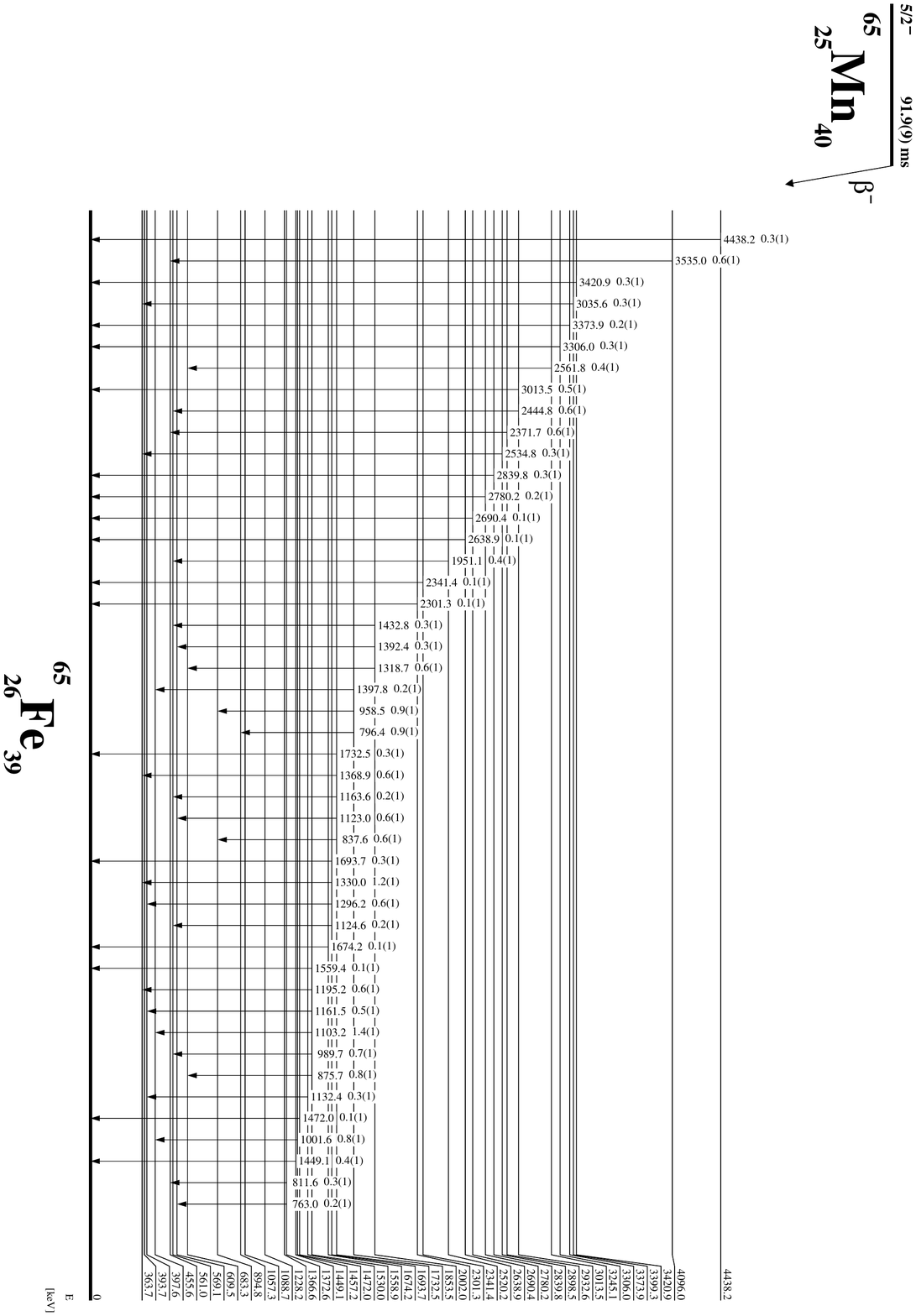}
\caption{\label{fig:Fe65_high-energies} High energy level scheme of $^{65}$Fe populated in the $\beta$-decay of $^{65}$Mn.}
\end{figure*}

The level scheme for the $\beta$ decay of $^{65}$Mn to $^{65}$Fe is summarized in Figs.~\ref{fig:Fe65_low-energies} and \ref{fig:Fe65_high-energies} and Tables~\ref{tab:List_gammas} and \ref{tab:List_levels}. It includes 85 transitions and 41 excited states, while a level scheme in $^{64}$Fe populated in the $\beta$-n branch is presented in Table \ref{tab:b-n_gammas}. The $\gamma$-ray intensities were obtained from the $\gamma$-ray singles spectra, except for the $\gamma$ rays that were part of unresolved multiplets or were too weak to clearly show up in the singles spectra. In those cases their energies and intensities were obtained from the $\gamma\gamma$ coincidences. There are a few high energy transitions for which no coincidences were observed. They are assigned to $^{65}$Fe, although there is a minor possibility that they belong $^{64}$Fe. Without a detailed $^{64}$Fe level scheme, this possibility cannot be completely excluded.

Out of 10 transitions for the decay of $^{65}$Mn to $^{65}$Fe identified in \cite{GAU05PhD} we confirm 7 lines at the energy of 92, 363, 455, 684, 724, 1004, and 1089 keV. It is not certain whether the 214-keV line listed in Ref.~\cite{GAU05PhD} is our 215.8-keV transition, but we definitely did not identify transitions of the energies 1550 and 1843 keV as belonging to the decay of $^{65}$Mn. Based on our $\gamma\gamma$ coincidences we confirm the very tentative and partial level scheme proposed in \cite{GAU05PhD} based on the energy matching.

\begingroup
\squeezetable
\begin{longtable*}{ccccC{10cm}}
\caption{\label{tab:List_gammas} A list of $\gamma$-ray energies, relative intensities, placement and the strongest $\gamma\gamma$ coincident transitions from the $\beta$ decay of $^{65}$Mn. }\\

\hline\hline
E$_\gamma$ (keV) & E$^{level}_{initial}$ (keV) & E$^{level}_{final}$ (keV) & I$^{rel}_\gamma$ & Strongest $\gamma\gamma$ coincidences (keV) \\
\hline
\endfirsthead
\multicolumn{5}{c}%
{\tablename\ \thetable\ -- \textit{Continued from previous page}} \\
\hline
E$_\gamma$ (keV) & E$^{level}_{initial}$ (keV) & E$^{level}_{final}$ (keV) & I$^{rel}_\gamma$ & Strongest $\gamma\gamma$ coincidences (keV) \\
\hline
\endhead
\hline \multicolumn{5}{c}{\textit{Continued on next page}} \\
\endfoot
\hline
\multicolumn{5}{l}{{\footnotesize $^a$~Total intensity obtained from the time-delayed component of the 363.7-keV transition, see the text for details.}}\\
\multicolumn{5}{l}{{\footnotesize $^b$~Intensity obtained from $\gamma \gamma$ coincidence spectra.}}\\
\multicolumn{5}{l}{{\footnotesize $^c$~$\gamma$ rays assigned to the $^{65}$Mn decay but not placed in the level scheme.}}\\
\multicolumn{5}{l}{{\footnotesize $^d$~Doublet in $^{65}$Fe, the intensity is obtained from coincidences with 374.1- and 405.6-keV transitions.}}\\
\multicolumn{5}{l}{{\footnotesize $^e$~Only observed in the delayed coincidences of the 363.7-keV transition.}}\\
\endlastfoot
\hline
33.9(2)	&	397.6(2)	&	363.7(1)	&	4(1)$^a$	&	-	\\
92.0(1)	&	455.6(1)	&	363.7(1)	&	0.8(1)	&	228.1, 363.7, 439.5, 683.8, 1103.6\\
114.5(3)	&	683.3(1)	&	569.1(1)	&	0.2(1)$^b$	&	205.7, 363.6, 569.3 \\
163.1(1)	&	561.0(2)	&	455.6(1)	&	0.7(1)	&	363.7, 456.0, 528.5	\\
197.6(3)	&	561.0(2)	&	363.7(1)	&	1.4(1)$^b$	&	363.7, 528.2, 2371.7	\\
205.3 (2)	&	569.1(1)	&	363.7(1)	&	0.3(1)	&	363.7, 455.0, 520.4, 633.3, 725.9, 1318.3	\\
215.8(1)	&	609.5(3)	&	393.7(2)	&	2.9(2)	&	757.5, 763.4, 1123.0, 1392.5	\\
227.7(1)	&	683.3(1)	&	455.6(1)	&	2.5(2)	&	92.8, 363.7, 374.0, 405.9, 455.8	\\
319.7(1)	&	683.3(1)	&	363.7(1)	&	4.4(3)	&	363.9, 374.6, 405.9, 683.8, 724.6, 875.8, 1319.0	\\	
363.7(1)	&	363.7(1)	&	g.s.	&	100	&	92.4, 163.5, 197.7, 205.9, 227.9, 319.9, 374.4, 405.8, 439.3, 531.5, 659.7, 683.7, 693.9, 725.4, 875.4, 969.1, 1002.9,1195.2, 1318.3, 1330.1, 1368.7, 2372.1 \\	
374.1(1)	&	1057.3(1)	&	683.3(1)	&	1.0(1)	&	227.8, 319.6, 363.9, 455.9, 683.5, 796.7, 2341.5	\\	
405.6(1)	&	1088.7(1)	&	683.3(1)	&	1.8(1)	&	227.9, 320.0, 363.9, 455.9, 683.6	\\	
439.2(1)	&	894.8(1)	&	455.6(1)	&	3.5(3)	&	92.2, 363.5, 455.9, 472.4, 958.7	\\	
455.6(1)	&	455.6(1)	&	g.s.	&	24.4(1.8)	&	227.9, 374.1, 405.8, 439.4, 472.3, 601.9, 633.5, 683.7, 875.8, 911.0, 958.6, 1001.6, 1103.2	\\	
472.0(1)	&	1366.6(2)	&	894.8(1)	&	0.9(1)	&	92.9, 363.7, 439.4, 455.9	\\	
488.3(2)	&	1057.3(1)	&	569.1(1)	&	0.2(1)	&	569.2	\\	
501.3(5)$^c$	&	-	&	-	&	0.2(1)	&	-	\\	
519.8(1)	&	1088.7(1)	&	569.1(1)	&	4.3(3)	&	205.9, 569.4\\	
528.1(1)	&	1088.7(1)	&	561.0(2)	&	0.9(1)	&	163.6, 197.6, 363.7, 561.1	\\	
531.1(1)	&	894.8(1)	&	363.7(1)	&	1.1(1)	&	363.8, 472.0, 837.7, 958.5	\\	
560.8(1)	&	561.0(2)	&	g.s.	&	2.7(2)	&	528.4, 806.2, 811.6, 1163.2, 1559.4, 2444.8, 3535.1	\\	
569.1(1)	&	569.1(1)	&	g.s.	&	8.4(6)	&	114.4, 488.2, 520.0, 796.8, 989.7 1123.9, 1163.7, 1731.9, 1950.7, 2371.2	\\	
601.7(1)	&	1057.3(1)	&	455.6(1)	&	0.9(1)	&	92.5, 363.8, 455.9	\\	
633.2(1)	&	1088.7(1)	&	455.6(1)	&	2.2(2)	&	92.3, 363.9, 455.9	\\	
659.7(1)	&	1057.3(1)	&	397.6(1)	&	0.8(1)	&	363.9	\\	
683.3(1)$^d$	&	683.3(1)	&	g.s.	&	3.6(3)	&	92.4, 228.1, 319.9, 363.7, 374.5, 405.8, 455.9, 683.5, 875.9	\\	
683.3(1)$^d$	&	1366.6(2)	&	683.3(1)	&	2.6(2)	&	See above	\\	
691.5(5)$^e$	&	1088.7(1)	&	397.6(1)	&	0.2(2)	&	363.5	\\	
693.7(1)	&	1057.3(1)	&	363.7(1)	&	2.2(2)	&	363.7	\\	
725.2(1)	&	1088.7(1)	&	363.7(1)	&	8.8(7)	&	363.7	\\	
757.2(2)	&	1366.6(2)	&	609.5(3)	&	0.2(1)	&	215.9	\\	
763.0(3)	&	1372.6(3)	&	609.5(3)	&	0.2(1)	&	215.8	\\	
772.6(2)	&	1228.2(2)	&	455.6(1)	&	0.4(1)	&	92.3, 363.5, 455.9 \\	
796.4(4)	&	1853.5(4)	&	569.1(1)	&	0.9(2)	&	569.4	\\	
796.9(1)	&	1366.6(2)	&	1057.3(1)	&	0.5(1)$^b$	&	693.9	\\	
805.9(2)	&	1366.6(2)	&	561.0(2)	&	0.3(1)	&	163.5, 197.8, 363.7, 560.1	\\	
811.6(1)	&	1372.6(3)	&	561.0(2)	&	0.3(1)	&	197.5, 561.9	\\	
837.6(5)	&	1732.5(4)	&	894.8(1)	&	0.6(1)$^b$	&	439.9, 455.2	\\	
875.7(1)	&	1558.9(5)	&	683.3(1)	&	0.8(1)	&	115.6, 227.9, 319.9, 363.9, 455.9, 683.4	\\	
911.1(2)	&	1366.6(2)	&	455.6(1)	&	0.3(1)	&	92.8, 364.3, 455.9	\\	
958.5(2)	&	1853.5(4)	&	894.8(1)	&	0.9(1)	&	364.4, 439.4, 455.9, 531.5	\\	
969.1(1)	&	1366.6(2)	&	397.6(1)	&	0.8(1)	&	363.8	\\	
989.7(1)	&	1558.9(5)	&	569.1(1)	&	0.7(1)	&	205.4, 363.4, 569.3	\\	
1001.6(5)	&	1457.2(5)	&	455.6(1)	&	0.8(1)$^b$	&	92.0, 363.9, 455.8	\\	
1002.9(1)	&	1366.6(2)	&	363.7(1)	&	4.8(4)	&	364.0	\\	
1051.5(5)$^c$	&	-	&	-	&	0.2(1)	&	-	\\	
1057.2(1)	&	1057.3(1)	&	g.s.	&	0.7(1)	&	-	\\	
1088.6(1)	&	1088.7(1)	&	g.s.	&	16.9(1.3)	&	-	\\	
1103.2(1)	&	1558.9(5)	&	455.6(1)	&	1.4(1)	&	92.5, 363.4, 455.8	\\	
1123.0(2)	&	1732.5(4)	&	609.5(3)	&	0.6(1)	&	216.3	\\	
1124.6(5)	&	1693.7(1)	&	569.1(1)	&	0.2(1)$^b$	&	569.0	\\	
1132.4(5)	&	1530.0(5)	&	397.6(2)	&	0.2(1)	&	363.5	\\	
1161.5(3)	&	1558.9(5)	&	397.6(1)	&	0.5(1)	&	364.0	\\	
1163.6(3)	&	1732.5(4)	&	569.1(1)	&	0.2(1)$^b$	&	569.0	\\	
1195.2(2)	&	1558.9(5)	&	363.7(1)	&	0.6(1)	&	364.0	\\	
1296.2(2)	&	1693.7(1)	&	397.6(1)	&	0.6(1)	&	364.0	\\	
1318.7(2)	&	2002.0(3)	&	683.3(1)	&	0.6(1)	&	319.9, 363.7, 683.8	\\	
1330.0(1)	&	1693.7(1)	&	363.7(1)	&	1.2(1)	&	364.1 	\\	
1366.2(4)	&	1366.6(2)	&	g.s.	&	0.3(1)	&	-	\\	
1368.9(3)	&	1732.5(4)	&	363.7(1)	&	0.6(1)	&	363.8	\\	
1392.4(4)	&	2002.0(3)	&	609.5(3)	&	0.2(1)	&	215.9	\\	
1397.8(3)	&	1853.5(4)	&	455.6(1)	&	0.2(1)	&	92.1, 364.0, 455.0	\\	
1432.8(3)	&	2002.0(3)	&	569.1(1)	&	0.3(1)	&	205.5, 363.7, 569.6	\\	
1449.1(4)	&	1449.1(4)	&	g.s.	&	0.4(1)	&	-	\\	
1472.0(6)	&	1472.0(6)	&	g.s.	&	0.1(1)	&	-	\\	
1559.4(4)	&	1558.9(5)	&	g.s.	&	0.1(1)	&	-	\\	
1674.2(7)	&	1674.2(7)	&	g.s.	&	0.1(1)	&	-	\\	
1693.7(4)	&	1693.7(1)	&	g.s.	&	0.3(1)	&	-	\\	
1732.5(4)	&	1732.5(4)	&	g.s.	&	0.3(1)	&	-	\\	
1951.1(4)	&	2520.2(4)	&	569.1(1)	&	0.4(1)	&	205.6, 363.6, 569.4	\\	
2301.3(8)	&	2301.3(8)	&	g.s.	&	0.1(1)	&	-	\\	
2341.4(7)	&	2341.4(7)	&	g.s.	&	0.1(1)	&	-	\\	
2371.7(4)	&	2932.6(4)	&	561.0(2)	&	0.6(1)	&	197.6, 363.8, 561.1	\\	
2444.8(7)	&	3013.5(7)	&	569.1(1)	&	0.3(1)	&	205.5, 363.6, 569.0	\\	
2534.8(8)	&	2898.5(8)	&	363.7(1)	&	0.3(1)	&	363.3	\\	
2561.8(7)	&	3245.1(7)	&	683.3(1)	&	0.4(1)	&	320, 364, 683	\\	
2638.9(8)	&	2638.9(8)	&	g.s.	&	0.1(1)	&	-	\\	
2690.4(8)	&	2690.4(8)	&	g.s.	&	0.2(1)	&	-	\\	
2780.2(8)	&	2780.2(8)	&	g.s.	&	0.2(1)	&	-	\\	
2839.8(8)	&	2839.8(8)	&	g.s.	&	0.3(1)	&	-	\\	
3013.1(6)	&	3013.5(7)	&	g.s.	&	0.5(1)	&	-	\\	
3035.6(5)	&	3399.3(5)	&	363.7(1)	&	0.3(1)	&	363.3, 569.3	\\	
3306.0(9)	&	3306.0(9)	&	g.s.	&	0.3(1)	& -	\\	
3373.9(8)	&	3373.9(8)	&	g.s.	&	0.2(1)	&	-	\\	
3420.9(9)	&	3420.9(9)	&	g.s.	&	0.3(1)	&	-	\\	
3535.0(4)	&	4096.0(4)	&	561.0(2)	&	0.6(1)	&	561	\\	
4438.2(9)	&	4438.2(9)	&	g.s.	&	0.3(1)	&	-	\\

\hline\hline

\end{longtable*}
\endgroup

\begingroup
\squeezetable
\begin{table}
\caption{\label{tab:List_levels}A list of level energies, level half-lives, $\beta$ feeding, log$ft$, and spin/parity assignments for the levels populated in the $\beta$ decay of $^{65}$Mn to $^{65}$Fe. The log$ft$ values were calculated using T$_{1/2}$=91.9(9)~ms and Q$_\beta$=10.254(6)~MeV \cite{NAI12}.}
\begin{ruledtabular}
\begin{tabular}{ccccc}
E$_{level}$ (keV) & $\beta$ feeding & log$\left(ft\right)$ & T$_{1/2}$ &J$_{\pi}$ \\
\colrule
0	&	$<8.8$	&	$>6.0$	&	0.81(5) s	&	$\left(1/2^-\right)$	\\
363.7(1)	&	42(5)	&	4.63(7)	&	93(3) ps	&	$\left(3/2^-\right)$	\\
393.7(2)	&		&		&	1.12(15) s	&	$\left(9/2^+\right)$	\\
397.6(2)	&	0.1(6)	&	$>6.4$	&	437(55) ns	&	$\left(5/2^+\right)$	\\
455.6(1)	&	8.1(1.1)	&	5.33(5)	&	350(10) ps	&	$\left(5/2^-\right)$	\\
561.0(2)	&	1.5(2)	&	6.06(8)	& 390(50) ps &	$\left(3/2^+,3/2^-,5/2^-\right)$	\\
569.1(1)	&	0.9(4)	&	6.31(5)	&	$<12$ ps	&	$\left(1/2,3/2\right)$	\\
609.5(3)	&	1.0(1)	&	6.2(6)	&		&	$\left(7/2^+\right)$	\\
683.3(1)	&	2.1(3)	&	5.86(8)	&	24(12) ps	&	$\left(3/2^+,3/2^-,5/2^-\right)$\\
894.8(1)	&	1.4(2)	&	5.99(8)	&	$<27$ ps	&	$\left(7/2^-\right)$	\\
1057.3(1)	&	2.9(2)	&	5.64(6)	&	$<8$ ps\footnotemark[1]	&	$\left(3/2^-,5/2^-\right)$	\\
1088.7(1)	&	20.7(9)	&	4.78(5)	&	$<8$ ps\footnotemark[1]	&	$\left(3/2^-,5/2^-\right)$	\\
1228.2(2)	&	0.2(1)	&	6.76(23)	&		&		\\
1366.6(2)	&	6.3(3)	&	5.23(5)	&	$<8$ ps\footnotemark[1]	&	$\left(5/2^-\right)$	\\
1372.6(3)	&	0.3(1)	&	6.55(15)	&		&		\\
1449.1(4)	&	0.3(1)	&	6.31(10)	&		&		\\
1457.2(5)	&	0.5(1)	&	5.59(5)	&		&		\\
1472.0(6)	&	0.1(1)	&	7.0(5)	&		&		\\
1530.0(5)	&	0.2(1)	&	6.74(22)	&		&		\\
1558.9(5)	&	2.4(1)	&	5.61(5)	&		&		\\
1674.2(7)	&	0.1(1)	&	7.0(5)	&		&		\\
1693.7(1)	&	1.3(1)	&	5.84(6)	&		&		\\
1732.5(4)	&	1.4(1)	&	5.8(6)	&		&		\\
1853.5(4)	&	1.2(1)	&	5.84(6)	&		&		\\
2002.0(3)	&	0.7(1)	&	6.04(8)	&		&		\\
2301.3(8)	&	0.1(1)	&	6.8(5)	&		&		\\
2341.4(7)	&	0.1(1)	&	6.8(5)	&		&		\\
2520.2(4)	&	0.2(1)	&	6.45(23)	&		&		\\
2638.9(8)	&	0.1(1)	&	6.7(5)	&		&		\\
2690.4(8)	&	0.1(1)	&	6.7(5)	&		&		\\
2780.2(8)	&	0.1(1)	&	6.7(5)	&		&		\\
2839.8(8)	&	0.2(1)	&	6.4(2)	&		&		\\
2898.5(8)	&	0.2(1)	&	6.4(2)	&		&		\\
2932.6(4)	&	0.4(1)	&	6.04(12)	&		&		\\
3013.5(7)	&	0.5(1)	&	5.92(10)	&		&		\\
3245.1(7)	&	0.3(1)	&	6.07(15)	&		&		\\
3306.0(9)	&	0.2(1)	&	6.23(23)	&		&		\\
3373.9(8)	&	0.1(1)	&	6.5(5)	&		&		\\
3399.3(5)	&	0.2(1)	&	6.2(23)	&		&		\\
3420.9(9)	&	0.2(1)	&	6.2(23)	&		&		\\
4096.0(4)	&	0.3(1)	&	5.81(15)	&		&		\\
4438.2(9)	&	0.2(1)	&	5.87(23)	&		&		\\

\end{tabular}
\end{ruledtabular}
\footnotetext[1]{Indicates time calibration level, see the text for details.}
\end{table}
\endgroup

\subsection{The exact energy of the $\beta$-decaying isomer}

\begin{figure}
\includegraphics[width=\columnwidth]{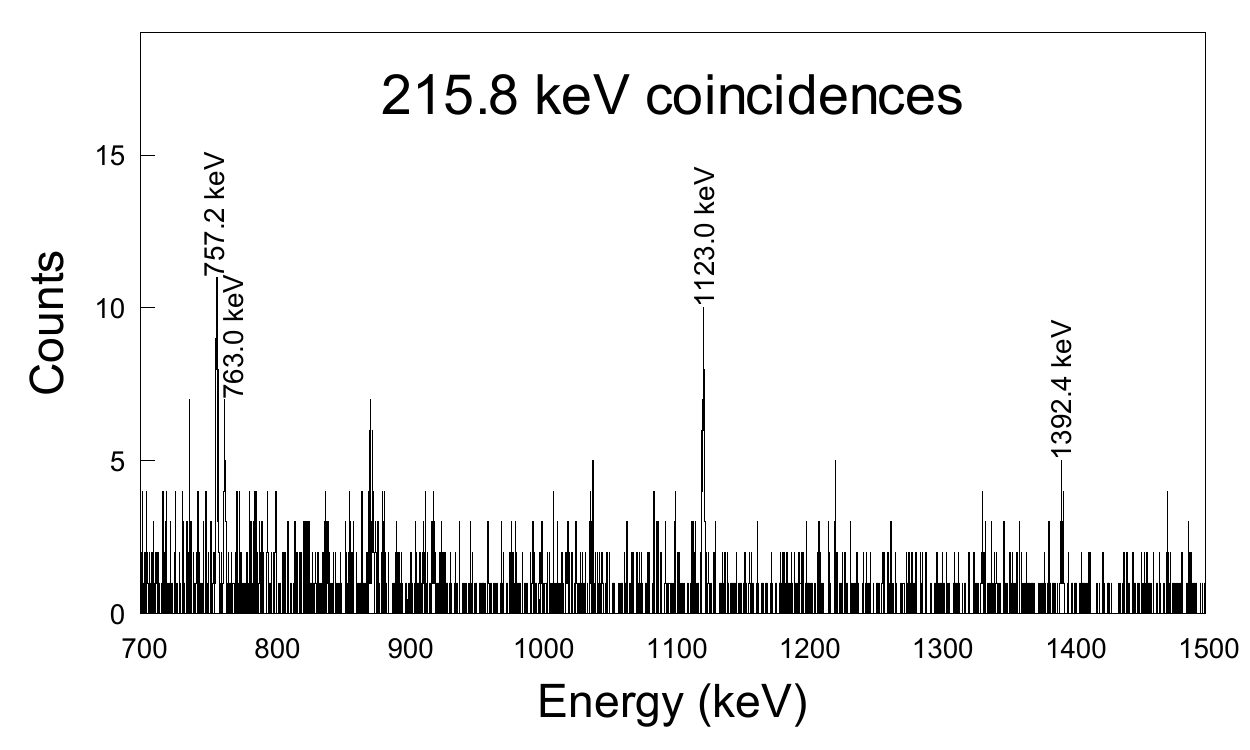}
\caption{\label{fig:216-gate-coincidences} $\gamma \gamma$ coincident spectrum gated by the 215.8-keV transition, the only $\gamma$ ray directly feeding the $\beta$-decaying isomer at 393.7 keV.} 
\end{figure}

The 215.8-keV transition is in firm mutual coincidences with the 757.2-, 763.0-, 1123.0-, and 1392.4-keV $\gamma$ rays (see Fig. \ref{fig:216-gate-coincidences}). If the 215.8-keV line feeds a level at 393.7(2) keV, then one defines a new level at 609.5 keV de-excited by the 215.8-keV transition, and then all four coincident lines de-excite levels already assigned to the level scheme based on other $\gamma\gamma$ coincidences. The 393.7-keV level is identified as the isomeric level at 402(10) keV reported in \cite{BLO08}. Thus the energy of the $\beta$-decaying isomer is precisely determined at 393.7(2) keV. We found no $\gamma$ rays de-exciting the 393.7-keV level. In particular for the 393.7-keV transition an upper limit of intensity was established at 0.15 relative to 100 for the 363.7-keV transition.

\subsection{The 420-ns isomer}

\begin{figure}
\includegraphics[width=\columnwidth]{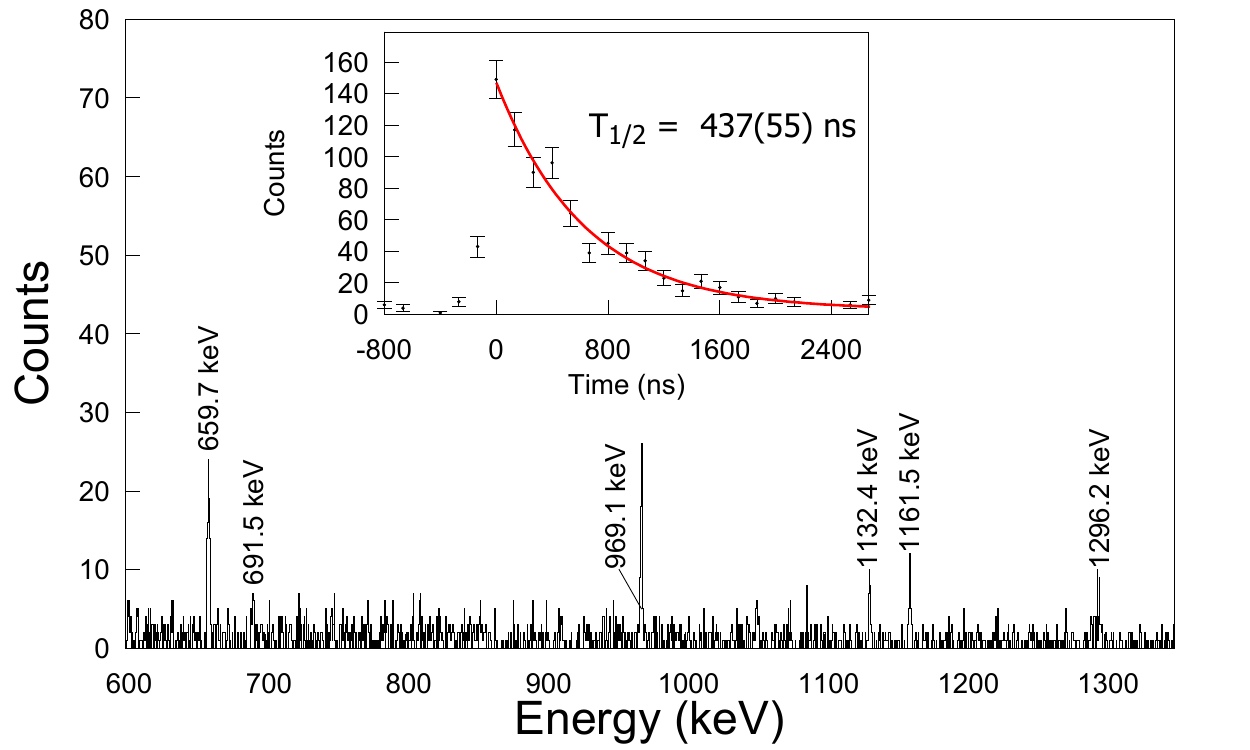}
\caption{\label{fig:33-keV_delayed-fit} \textit{(Color online)} Time-delayed coincidence HPGe $\gamma$ spectrum gated by the 363.7-keV transition in $^{65}$Fe. These $\gamma$ rays populate the 397.6-keV isomer and are in coincidence with 363.7-keV transition through the 33.9-keV line. The insert shows the time difference between the $\gamma$-rays feeding the isomer and the coincident 363.7-keV transition; see text for details.}
\end{figure}

Using the time-delayed $\gamma\gamma$(t) coincidences between two HPGe detectors we have confirmed the excitation energy and the half-life of the 396.8-keV 420(13)-ns isomer reported in \cite{DAU10}. Figure \ref{fig:33-keV_delayed-fit} shows the coincident HPGe spectrum gated by the 363.7-keV transition with two additional gates. One gate was set on the proton time between 10 to 450 ms to enhance the $^{65}$Mn decay, while the second gate was set on the $\sim$400 ns slope part (outside of the semi-prompt region) of the $\gamma\gamma$(t) time-delayed spectrum where the 363.7-keV transition was selected in the HPGe-STOP detector. The transitions shown in the spectrum are those which are feeding the 420-ns isomer from above and have started the time measurement. These are the transitions at 163.1, 659.7, 969.1, 1132.4, 1161.5, and 1296.2 keV, which feed the 397.6-keV level and are coincident to the 363.7-keV line via the unobserved 33.9-keV transition. All these $\gamma$ rays feeding the isomer de-excite established levels in $^{65}$Fe. The energy of the de-exciting transition is 33.9(2) keV based on the energies of the levels involved. Unfortunately, it is not directly observed by us as the energy threshold on each of our HPGe detectors was above this energy. By resorting the data and constructing a time-delayed spectrum started by the feeding $\gamma$ rays and stopped by the 363.7-keV line, one obtains a spectrum shown in the insert to Fig.~\ref{fig:33-keV_delayed-fit}. The slope in the time spectrum gives a half-life of 437(55)~ns in agreement with the known half-life of the isomeric state. 

In order to get the relative total intensity of the 33.9-keV $\gamma$-ray, we have sorted time-delayed $\beta\gamma$(t) coincidences using $\beta$ and HPGe detectors with two gates: one set on the 363.7-keV transition in the HPGe and the other on the proton time spectrum between 10 to 450 ms in order to enhance the $^{65}$Mn activity. The $\beta\gamma$(t) time spectrum shows a large prompt peak and a sloping delayed component with a half-life of about $\sim$400~ns. The area of the delayed component corresponds to the intensity of the 33.9-keV transition while the total area of the time spectrum represents the total intensity of the 363.7-keV line. The total intensity of the 33.9-keV transition was estimated to be 4(1) in relative units. This intensity includes conversion electron contribution and is independent of the multipolarity of the transition.

The intensity sum of the observed $\gamma$ rays feeding the 397.6-keV isomer is 3.9(2) in relative units, which compared to the total intensity of the 33.9-keV transition of 4(1) implies that this isomer is weakly, if at all, directly populated in $\beta$ decay of $^{65}$Mn. No indication of any other $\gamma$ ray de-exciting this isomer was found. In particular no $\gamma$ ray of energy 397.6 keV was identified and an upper limit of its intensity is 0.15 in relative units.

\subsection{Absolute $\beta$ and $\gamma$ intensities}

 In the following discussion we will use the term ``relative intensity'' to define an intensity normalized to 100 for the 363.7~keV line and ``absolute intensity'' to define the intensity per 100 $\beta$ decays of $^{65}$Mn expressed in $\%$.

The $A=65$ measurement was run for 18 hours in saturation mode and as a consequence an intensity balance was established in the $A=65$ decay chain and in the $\beta$-n branch along the $A=64$ chain. Measurement of the absolute decay branches for the decay of $^{65}$Mn was then possible using absolutely calibrated $\gamma$ transitions from other members of the decay chain. We assume here that only one $\beta$-decaying state of $^{65}$Mn is present in this measurement and that the ion source does not surface ionize other atoms in the $\beta$-decaying $A=65$ chain (Fe, Co, Ni and Cu) owing to their larger ionization potentials. In this analysis we used unrestricted $\gamma$-ray spectra observed in the HPGe detectors. 

In order to determine the ground-state $\beta$ feeding and the $\beta$-delayed neutron emission probability, we divide the total intensity of the $\beta$ decay of $^{65}$Mn into the following four decay branches: $\beta$-delayed neutron emission branch, $\beta$-n; direct $\beta$ feeding to the ground state of $^{65}$Fe, $\beta ^{g.s.}$; $\beta$ feeding to the low-spin levels in $^{65}$Fe, which then directly or indirectly feed the low-spin ground state in $^{65}$Fe, $\beta ^{levels}_{LS}$; and the portion of the $\beta$ feeding that ends up in the second $\beta$-decaying state in $^{65}$Fe at 393.7 keV, $\beta ^{levels}_{HS}$. 

Two of these branches, $\beta ^{g.s.}$ and  $\beta ^{levels}_{LS}$, feed the $J^\pi$=(1/2$^-$) g.s.~in $^{65}$Fe, which is then followed by $\beta$-decay to $^{65}$Co. This decay has no direct $\beta$ feeding to the ground state of $^{65}$Co and all of the $\beta$ intensity is then carried out by three $\gamma$ rays of the energy of 882.5, 1222.7, and 1996.6 keV \cite{PAU09}. In our measurement these transitions have relative intensities summed to $I$ = 163.6(61). On the other hand, the intensity of the $\beta ^{levels}_{LS}$ feeding is simply the sum of intensities of $\gamma$ rays directly feeding the ground state in $^{65}$Fe and is equal to 162.1(135) in relative units. The difference between these intensity values gives the direct g.s.~feeding in $^{65}$Fe of 1.5(148) in relative units. 

The $\beta$-n branch feeds the states in $^{64}$Fe. The ground state of $^{64}$Fe $\beta$ decays to $^{64}$Co, which then decays to $^{64}$Ni. In $^{64}$Ni the strongest $\gamma$ transition is the 1345.8-keV line, which represents $7.54\;\%$ of the total decay intensity \cite{PAU12}. We observe a transition at 1345.1 keV with a relative intensity of 1.3(1). This transition includes an impurity component from the decay of $^{65}$Ga to $^{65}$Ni with the energy of 1343.9 keV and the intensity of 0.18(2) (determined using other known lines in the decay of $^{65}$Ga and the branching ratios listed in \cite{BRO10}). The resulting relative intensity of the 1345.8-keV transition in $^{64}$Ni is 1.1(1), which gives a relative intensity of 14.6(21) for the $\beta$-n branch.

A direct $\beta$ feeding of the isomeric state at 393.7-keV must be very weak due to the spin/parity difference between the isomer, with the expected $J^\pi$ = $(9/2^+)$, and the ground state of $^{65}$Mn with $J^\pi$ = $(5/2^-)$. As for the indirect feeding, we have identified only one $\gamma$ ray, which directly feeds this state. The relative intensity of the 215.8-keV transition is only 2.9(2). 

The 393.7-keV isomer $\beta$ decays to levels in $^{65}$Co, which are different than those observed in the ground state decay of $^{65}$Fe \cite{PAU09}. In the following analysis we use the absolute intensities re-evaluated in the Nuclear Data Sheets for $A=65$ \cite{BRO10}. We observe $^{65}$Co transitions of energies 1412.7, 1642.2, 2443.7 and 2558.0 keV. Their summed relative intensity is 4.5(3). The total intensity for these $\gamma$-rays is 62.0(76) $\%$ of the isomer decay, giving an intensity of 7.3(10) in relative units for the population of the isomer in $^{65}$Fe. Independently, we looked into the intensities from the decay of $^{65}$Ni to $^{65}$Cu. Using the intensities of the 366.2-, and 1481.5-keV transitions (the latter corrected for a $4\;\%$ impurity contribution from the 1479.0-keV line in $^{65}$Cu), we obtain the total feeding intensity to the $\beta$ decaying levels in $^{65}$Fe as 172(10) relative units, which must be compared to the individual contributions of 163.6(61) and 7.3(10) for the ground state and isomer, respectively, giving together 170.9(62) in excellent agreement with the $^{65}$Cu data.

The total $\beta^-$ decay intensity of $^{65}$Mn is the sum of 14.6(21), 163.6(61) (sum of two components) and 7.3(10), giving a total relative intensity of 185.5(65). This intensity is equal to 100$\%$ in absolute units giving a renormalization factor of 0.539(19), which converts relative intensity units into absolute intensities. The $\beta$-n branch is 7.9(12) $\%$ and the direct ground state feeding is $\le8.8\;\%$. We note the excellent agreement with the preliminary value of $<10\;\%$ for the ground state $\beta$ feeding reported in \cite{GAU05PhD}.

The absolute $\beta$ feeding to each level was calculated as the difference of $\gamma$ transition absolute intensities feeding and depopulating the level. Two of the levels, at 363.7 and 1088.7~keV, receive very strong $\beta$ feeding which together amounts to more than 60$\%$. A significant $\beta$ feeding goes also to the levels at 455.6, 683.3, 1057.3, 1366.6 and 1558.9 keV. A summary of the properties of levels in $^{65}$Fe is given in Table \ref{tab:List_levels}.

\subsection{$\beta$-n branch directly feeding the ground sate of $^{64}$Fe}

The total $\beta$-delayed neutron emission branch in the decay of $^{65}$Mn is 7.9(12) $\%$ and a significant portion of the $\beta$-n feeding goes to the excited states in $^{64}$Fe. We observe the 746.4(1)-keV transition in coincidence with the 1017.4-, 1105.8- and 1370.7-keV lines, all previously assigned to the level scheme of $^{64}$Fe \cite{HOT06,HAN99,HAN00}

A fit to the slope of the time-delayed proton spectrum gated by the 746.4-keV line gives a half-life of 99(9)~ms which firmly assigns this line to the $\beta$-n decay of $^{65}$Mn. Table \ref{tab:b-n_gammas} provides a summary on the information of these transitions. The absolute intensity of the 746.4~keV transition is 2.4(1)$\%$. There must be other ground state transitions as well in this $\beta$-n decay that remain unobserved by us. This leads to the conclusion that the direct $\beta$-n feeding to the g.s.~of $^{64}$Fe is less than $5.5\;\%$. 

Our $P_n$ value of $7.9(12)\;\%$ contradicts the value of $21.0(5)\;\%$ reported in \cite{HAN00} as our result is about three times lower. Our result on the direct g.s.~feeding is even more different, since in ref.~\cite{HAN00} the entire $\beta$-n intensity of $21\;\%$ was assigned to directly feed the ground state of $^{64}$Fe, while in our work a significant portion goes to the excited states. Thus for the direct ground-state feeding our measured intensity is almost four times lower.

\begingroup
\squeezetable
\begin{table}
\caption{\label{tab:b-n_gammas} $\gamma$-rays assigned to $^{64}$Fe from the $\beta$-n decay of $^{65}$Mn. The intensities are normalized to the 363.7-keV intensity of 100.}
\begin{ruledtabular}
\begin{tabular}{cccccc}
E$_\gamma$ (keV) & E$^{level}_{initial}$ (keV) & E$^{level}_{final}$ (keV) & I$^{rel}_\gamma$ & $\gamma$-$\gamma$ coincidences (keV) \\
\colrule
746.4(1)	&	746.4(1)	&	g. s.	&	4.4(2)	&	1017.4, 1105.8, 1370.7  \\
1017.4(3)	&	1763.8(3)	&	746.4(1)	&	0.4(1)	&	746.4 \\
1105.8(5)	&	1852.2(5)	&	746.4(1)	&	0.5(1)	&	746.4 \\
1370.7(5)	&	2117.1(5)	&	746.4(1)	&	0.2(1)	&	746.4 \\
\end{tabular}

\end{ruledtabular}
\end{table}
\endgroup

\section{Fast Timing measurements}

The shape deconvolution technique was used for the half-lives longer than 80 ps when time spectra have shown a slope on the delayed side, while the centroid shift analysis was applied for shorter lifetimes. A summary of lifetime measurements is given in Table \ref{tab:transition-rates}. The fast timing analysis involved $\beta$-LaBr$_3$(Ce)(t) and $\beta$-HPGe-LaBr$_3$(Ce)(t) coincidences. The contribution from the two HPGe detectors were summed together, while those from the LaBr$_3$(Ce) detectors were analyzed separately since their time responses were different.

Our preliminary analysis of the level lifetimes in $^{65}$Fe indicated that only three levels, those at 363.7, 455.6 and 561.0 keV, have sufficiently long lifetimes to be measurable by the deconvolution method. These lifetimes were deduced from double $\beta$-LaBr$_3$(Ce)(t) coincidences. The first two levels are de-excited by strong $\gamma$-rays, which are well separated in the LaBr$_3$(Ce) spectrum, see Fig. \ref{fig:LaBr-energy-spectrum}. 

\begin{figure}
\includegraphics[width=\columnwidth]{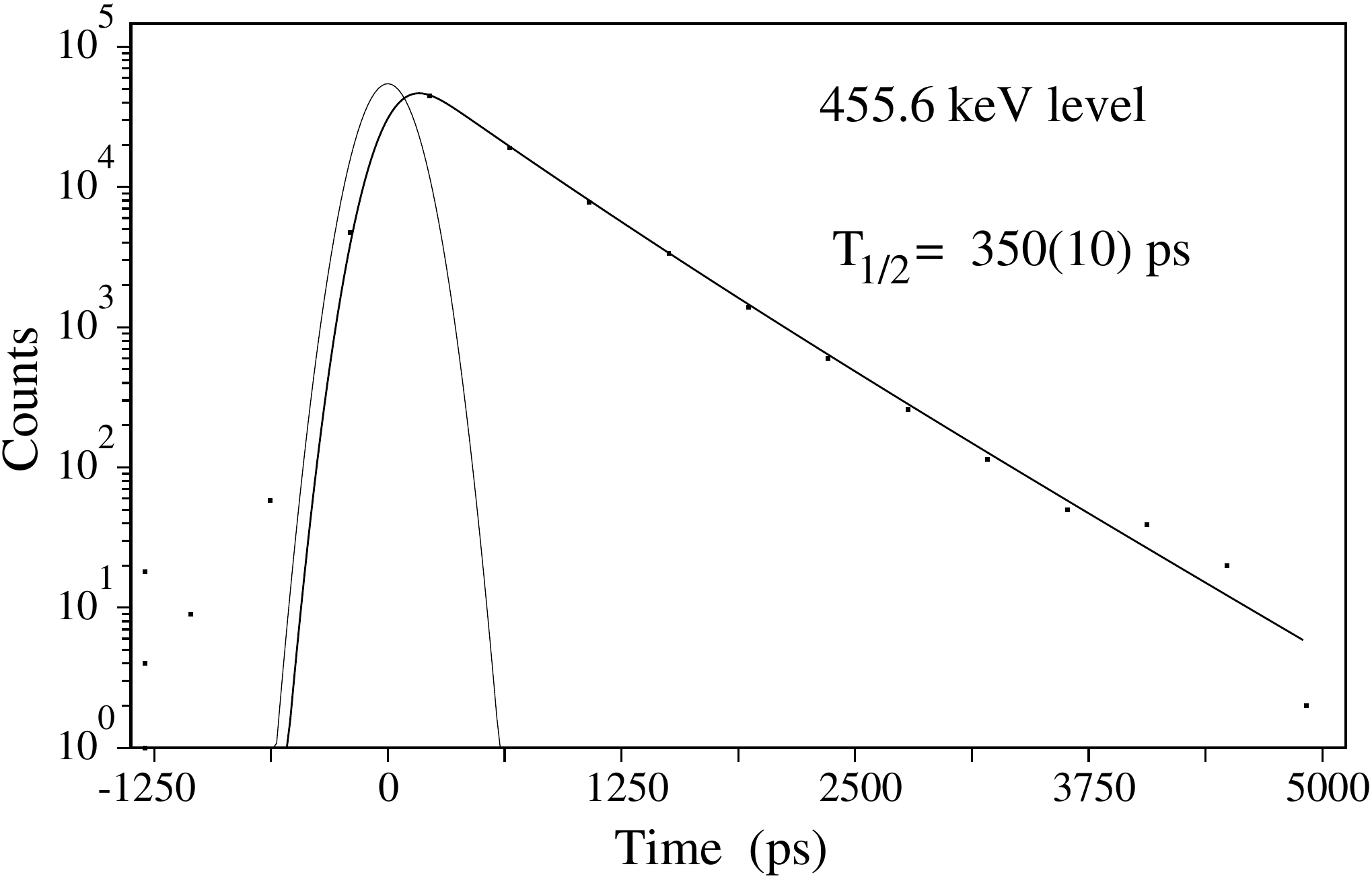}
\caption{\label{fig:455-keV_level_half-life} Time-delayed $\beta\gamma$(t) spectrum gated by the 455.6-keV transition. Its slope is due to the half-life of the 455.6-keV level. A fit to the exponential decay yields $T_{1/2}$ =  350(10) ps. }
\end{figure}

{\bf 455.6-keV level:} 
A narrow gate was set on the 455.6-keV peak in the LaBr$_3$(Ce) spectrum to ensure that the contribution from the 439.2-keV transition is minimized. In any case, the 894.8-keV level has a very short half-life (see results below) so that a small presence of the 439.3-keV $\gamma$ ray in the energy gate does not influence the present slope fitting. The $\beta\gamma$ time-delayed spectrum gated by the 455.6-keV transition is shown in Fig.~\ref{fig:455-keV_level_half-life}. Fitting the slope with an exponential decay yielded the half-life of 350(10) ps. This result is the weighted average of the two lifetimes obtained independently using each of the LaBr$_3$(Ce) detectors.

\begin{figure}
\includegraphics[width=\columnwidth]{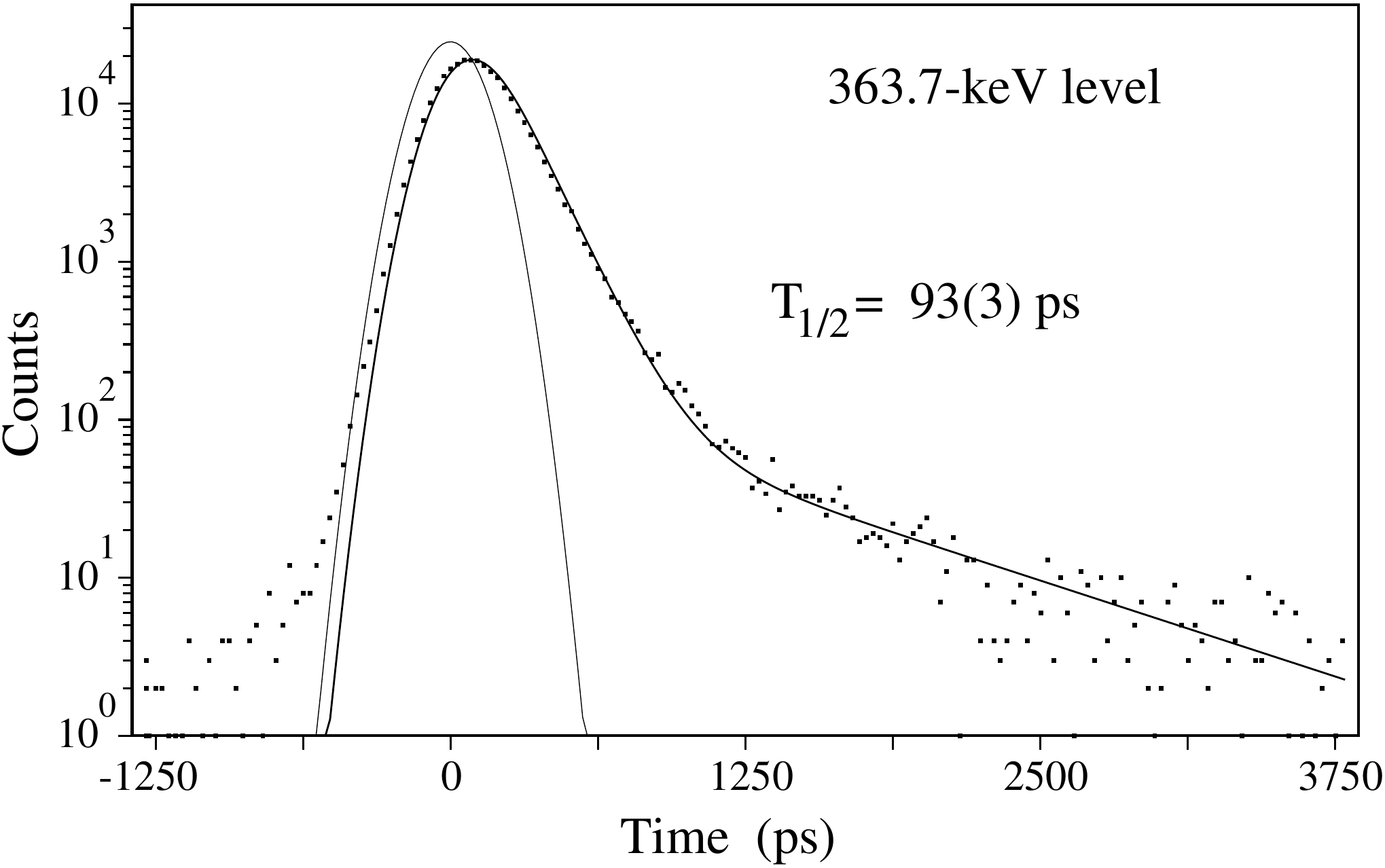}
\caption{\label{fig:363-keV_level_half-life} 
Time-delayed $\beta\gamma$(t) spectrum gated by the 363.7-keV transition. Its shorter slope is due to the 93(3) ps half-life of the 363.7-keV level; see text for details. }
\end{figure}

{\bf 363.7-keV level:}
The time-delayed spectrum gated by the 363.7-keV transition shown in Fig.~\ref{fig:363-keV_level_half-life}, has two lifetime components. The longer one, with a half-life consistent with 350 ps is due to the level half-life of the 455.6-keV state that feeds the 363.7-keV level via the 92.0-keV line. The time distribution was fit to a prompt Gaussian plus two exponential decays. The shorter component gives 93(3) ps for the half-life of the 363.7-keV level. One should note that on a much longer scale, not relevant to this analysis, this time spectrum has also a 420 ns component due to the 397.6-keV isomer feeding the 363.7-keV level via the 33.9-keV transition. Over a few nanoseconds range, this long component provides merely a flat background contribution.

{\bf 561.0-keV level:}
In the LaBr$_3$(Ce) energy spectrum of the $\beta\gamma$ data set, the 560.8- and 569.1-keV transitions can not resolved. Thus, the time spectrum gated on that double peak represents a combination of the half-lives of both levels at 561.0 and 569.1 keV. The time spectrum shows a short and a long component, the latter with a well defined slope in its time-delayed part. The 569.1-keV $\gamma$ ray is 3 times more intense than the 560.8-keV transition, but its half-life is short enough to appear as a semi-prompt time distribution (as determined in the following section). Consequently the observed long half-life was assigned to the 561.0-keV level. A fit of a two-component exponential decay gives a half-life of the 561.0-keV level as 390(30) ps.

The half-lives of the 569.1-, 683.3-, and 894.8-keV levels were obtained by the centroid shift method using the triple coincident fast timing $\beta\gamma\gamma$ data.

{\bf 569.1-keV level:}
For the 569.1-keV level we have used the two $\gamma$-ray cascade of the 519.8-569.1 keV transitions. The centroid of the time-delayed $\beta\gamma$(t) spectrum gated by the 519.8-keV $\gamma$ ray in LaBr$_3$(Ce) detector with an additional HPGe gate at 569.1 keV gives a time reference spectrum. By reversing the gates and taking the 569.1-keV line in LaBr$_3$(Ce) and the 519.8-keV transition in HPGe, one obtains a time spectrum, whose centroid is shifted from the reference point by the mean-life of the 569.1-keV level. In all centroid-shift analyses we make standard corrections for the Compton continuum under the full-energy peaks in the LaBr$_3$(Ce) spectrum and the energy dependent time response for the full-energy peaks (the RD-corrections) \cite{MAC91}. For the first LaBr$_3$(Ce) detector the centroid shift gives the value of T$_{1/2}$ = 0.0(106) ps, and for the second the value of 8.7(104) ps, giving the average value of $T_{1/2}$ = 4.3(74) ps or a limit of $\le$12 ps.

{\bf 683.3-keV level:}
For the 683.3- and 894.8-keV levels we do not have easily defined reference points for the centroid shift analysis. Therefore, we define an approximate prompt position at 363.7 keV. First we construct three $\beta\gamma$(t) time spectra gated by the 693.7-, 725.2-, and 1002.9-keV transitions in the HPGe detector, respectively, and by the 363.7-keV line in the LaBr$_3$(Ce). These time spectra are shifted from the prompt time position by $\Delta\tau$ = $\tau_0$ + $\tau_{top level}$ + $\tau_{363}$, where $\tau_0$ is a constant shift from the prompt due to the electronics, $\tau_{top level}$ is the mean life of the top level, which in this case is either the 1057.7-, 1088.7-, or 1366.6-keV level, respectively, while $\tau_{363}$ is the meanlife of the 363.7-keV state, which has a known value of 134.2(43) ps ($T_{1/2}$ = 93(3) ps) measured by us. The three time spectra for the first LaBr$_3$(Ce) detector differ by less than 3 ps from their average position, while for for the second detector they differ by less than 1.6 ps. It means the lifetimes of the top levels are the same within a few ps. Since these states do not have lifetimes manifested by slopes, and it is very unlikely that all of them have the same lifetime of tens of ps, therefore the equality of their centroids really means their lifetimes are short with $T_{1/2}$ below a few ps. Consequently, as a correction we arbitrarily assume for these levels a half-life of $T_{1/2}$ = 0(8) ps, which effectively increases the uncertainty in the lifetime determinations due to the our inability to determine the lifetimes of the ``upper levels" assumed here to be below a few ps.

Using the triple coincidence fast timing $\beta\gamma\gamma$ data, a gate was set on the HPGe detector on the 363.7-keV transition and a $\gamma$-ray peak was selected in the LaBr$_3$(Ce) detector at 319.7 keV. The centroid of the selected fast timing $\beta\gamma\gamma$ spectrum is equal to $\tau_0$ + $\tau_{683}$. The difference from the approximate prompt position at 363.7 keV in the half-life units is 31(12) ps for the first LaBr$_3$ detector and 18(12) ps for the second one. The weighted average of these results with the subtracted correction for the reference levels of 0(8) ps, gives a half-life for the 683.3-keV level equal to 24(12) ps.

{\bf 894.8-keV level:}
In a similar way we have measured the lifetime of the 894.8-keV state. Using the fast timing $\beta\gamma\gamma$ data, a gate was set in the HPGe detector on the 455.6-keV transition and a $\gamma$-ray peak was selected in the LaBr$_3$(Ce) detector at 439.2 keV. The centroid of the selected fast timing $\beta\gamma$ spectrum is equal to $\tau_0$ + $\tau_{894}$. The differences from the approximate prompt position at 363.7 keV in the half-life units are 8(18) and 16(16) ps for the first and second detectors, respectively. The averaged result, minus the correction of 0(8) ps, gives the half-life of the 894.8-keV level equal to 12(15) ps or $T_{1/2}$ $\le$27 ps.

\section{Discussion}

As the starting point for the spin/parity assignments to levels in $^{65}$Fe we take the assignments adopted by the evaluators in the Nuclear Data Sheets for $A=65$ \cite{BRO10} for the following $\beta$-decaying states: $J^{\pi}$ = $(5/2)^-$ for the ground state of $^{65}$Mn and $J^{\pi}$ = $(1/2)^-$ and $(9/2)^+$ for the ground state and the 393.7-keV states in $^{65}$Fe, respectively. We do not adopt other assignments proposed in \cite{BRO10} as our new results correct them.

\begingroup
\squeezetable
\begin{table}[h]
\caption{\label{tab:transition-rates} Summary of the half lives and transition rates measured in $^{65}$Fe.}
\begin{ruledtabular}
\begin{tabular}{ccccc}

Level (keV)	&	T$_{1/2}$ (ps)	&	E$_\gamma$ (keV)	&	Multipolarity	&	B(XL) W.u.	\\	\hline
363.7	&	93(3)	&	363.7	&	M1	&	$4.9(2)\cdot10^{-3}$	\\	\hline
397.6	&	437(55) ns	&	33.3	&	E1	&	$9.2(71)\cdot10^{2}$	\\	\hline
455.6	&	350(10)	&	92.0	&	M1	&	$2.6(4)\cdot10^{-3}$	\\	
	&		&	455.6	&	E2	&	5.1(6)	\\	\hline
561.0	&	390(30)	&	163.1	&	M1	&	$6.4(8)\cdot10^{-4}$	\\	
	&		&		&	E1	&	$3.9(4)\cdot10^{-5}$	\\	
	&		&	197.6	&	M1	&	$2.2(3)\cdot10^{-3}$	\\	
	&		&		&	E1	&	$7.7(1)\cdot10^{-7}$	\\	
	&		&	561.0	&	M1	&	$1.8(2)\cdot10^{-4}$	\\	
	&		&		&	E2	&	$11(1)$	\\	\hline
569.1	&	$< 12$	&	205.3	&	M1	&	$>7.9\cdot10^{-3}$	\\	
	&		&		&	E1	&	$>1.5\cdot10^{-4}$	\\	
	&		&	569.1	&	M1	&	$>9.6\cdot10^{-3}$	\\	
	&		&		&	E1	&	$>1.8\cdot10^{-4}$	\\	\hline
683.3	&	24(12)	&	227.7	&	M1	&	$1.8(9)\cdot10^{-2}$	\\	
	&		&		&	E1	&	$3.5(18)\cdot10^{-4}$	\\	
	&		&	319.7	&	M1	&	$1.2(6)\cdot10^{-2}$	\\	
	&		&		&	E1	&	$2.2(11)\cdot10^{-4}$	\\	
	&		&	683.3	&	M1	&	$9(5)\cdot10^{-4}$	\\	
	&		&		&	E1	&	$1.8(3)\cdot10^{-5}$	\\	\hline
894.8	&	$< 27$	&	439.2	&	M1	&	$>7.1\cdot10^{-3}$	\\	
	&		&		&	E1	&	$>1.4\cdot10^{-4}$	\\	
	&		&	531.1	&	M1	&	$>1.3\cdot10^{-3}$	\\	
	&		&		&	E1	&	$>2.5\cdot10^{-5}$	\\	
	&		&		&	E2	&	$>7$	\\
\end{tabular}
\end{ruledtabular}
\end{table}
\endgroup

{\bf Ground state, $\bm{J^{\pi}=(1/2^-)}$:}
The tentative spin parity assignments of $(5/2)^-$ and $(1/2)^-$ to the ground states of $^{65}$Mn and $^{65}$Fe would imply a very weak direct $\beta$ feeding between these states as the feeding would go via a second forbidden $\beta$ transition. Our limit on a direct $\beta$ feeding is $\le$8.8 $\%$, supporting these assignments.

{\bf 363.7-keV level, $\bm{J^{\pi}=(3/2^-)}$:}
This state receives a very high direct $\beta$ feeding with the intensity of 42(5)$\%$ and log$ft$ = 4.6, which implies an allowed $\beta$ transition between the $(5/2)^-$ g.s.~of $^{65}$Mn and the 363.7-keV state, the same parity for the states, and the spin/parity assignments of $(3/2)^-$, $(5/2)^-$ and $(7/2)^-$.

The first excited state has a half-life of 93(3) ps and it decays to the ground state by the 363.7-keV line. If this $\gamma$ ray would have an $E2$ multipolarity its $B(E2)$ value would be 62(2) W.u., thus about three times higher than the collective $B(E2;2^+_1 \to 0^+_1)$ values in the neighboring even-even Fe isotopes. Consequently the 363.7-keV transition is not an $E2$. This transition could be either E1 or M1 in character, with the $B(E1)$ value of 9.3$\times$10$^{-5}$ W.u.~or $B(M1)$ = 4.9$\times$10$^{-3}$ W.u. Due to the parity considerations, this transition could be only $M1$ in character, limiting the spin/parity for the level to $(3/2)^-$. The lifetime of the level does not exclude a small $E2$ component in the transition. If we assumed a small $E2$ component in the 363.7-keV transition with $B(E2)$ of 5 W.u.~then the branching ratio for $E2$ would be 8$\%$ while for $M1$ it would be 92$\%$.

{\bf 393.7-keV level, $\bm{J^{\pi}=(9/2^+)}$:}
The second excited state is a $\beta$ decaying isomer with $T_{1/2}$ = 1.12(15) s. We could precisely determine its excitation energy and we could also investigate possible $\gamma$-decay branches to the lower-lying states; the ground state and the 363.7-keV state. The isomer has a weak direct and indirect $\beta$ feeding, which amounts to only 3.9(5) $\%$, out of which 1.6$\%$ goes via a single 215.8-keV $\gamma$-ray feeding the state. There are likely a few more unobserved $\gamma$ rays feeding the isomer. Direct $\beta$ feeding is probably much smaller owing to the spin/parity differences between $J^{\pi}$ = $(9/2^+)$ for the isomer and $J^{\pi}$ = $(5/2^-)$ for the ground state of $^{65}$Mn.

We do not observe any $\gamma$ rays de-exciting the isomer. The upper limit on intensity for the 393.7-keV transition is 0.15 in relative units, while the relative intensity in the $\beta$-decay channel is 7.3(1.0). Thus using the $\gamma$ branch of $\le$ 0.021 one obtains for a $M4$ transition the $B(M4)\le3900$ W.u., which is consistent with no observation of any $\gamma$ transition, since it means that the true $\gamma$ branching is much lower than the measured limit. The other possible parity changing multipolarities are $E1$, $M2$ or $E3$. Yet, they have exceptionally low and unrealistic rates except perhaps for the $E3$ multipolarity.

This isomer may also decay to the 363.7-keV level by a 30.0-keV transition, which would be $E3$ in character. But even with the $B(E3)$ value of $\sim$10 W.u.~the total intensity of this transition would be negligible.

{\bf 397.6-keV level, $\bm{J^{\pi}=(5/2^+)}$:}
The 397.6-keV state is also an isomer with the half-life of 420(13) ns. It decays to the $(3/2^-)$ state by the 33.9-keV line, for which an $E1$ multipolarity was proposed albeit no evidence was provided for this assignment \cite{DAU06}. The observed $\gamma$ intensity feeding the isomer is 3.9(2) in relative units, while the total intensity of the 33.9-keV transition is 4(1). This intensity balance implies that the isomer is weakly, if at all, directly populated in the $\beta$-decay of $^{65}$Mn. We do not observe any direct $\gamma$ transition to the ground state at the energy 397.6 keV, for which an upper limit of intensity was established to be 0.15 in relative units. Another possible branch of decay is a 3.9-keV transition to the $(9/2^+)$ state at 393.7 keV.

The 33.9-keV transition must be a dipole either $E1$ or $M1$ in character making the spin of the 397.6-keV state either $1/2$, $3/2$ or $5/2$. The $B(E1)$ and $B(M1)$ values would be 1.2$\times$10$^{-5}$ W.u.~and 7.9$\times$10$^{-4}$ W.u., respectively. The transition cannot be $E2$ since then its $B(E2)$ value would be 64 W.u., a transition rate too collective for $^{65}$Fe. Moreover, its electron conversion coefficient of 29.5 would make this transition virtually undetectable in low-intensity experiments and yet the transition was previously observed. 

If the unobserved 397.6-keV transition would be a dipole $E1$ or $M1$ transition then the $B(E1)$ and $B(M1)$ rates would be $\le$ 6$\times$10$^{-10}$ W.u.~and $\le$ 3$\times$10$^{-8}$ W.u., respectively. These rates are too low to be realistic, therefore we reject these options. Consequently, this transition must be either $E2$ or $M2$ in character. The corresponding $B(E2)$ and $B(M2)$ rates are $\le$ 3$\times$10$^{-4}$ W.u.~and $\le$ 1.7$\times$10$^{-2}$ W.u., respectively. The $B(E2)$ rate is exceptionally low, while the $B(M2)$ rate is just about the average value one could expect for this nucleus \cite{END79,END81}. Consequently, the most likely spin/parity assignment for the 397.6-keV state is $(5/2^+)$. The difference in the parity between this state and the ground state of $^{65}$Mn explains very weak, if any, direct $\beta$ feeding to this state.

The proposed spin/parity assignments would imply that the unobserved 3.9-keV transition is $E2$ in character. By assuming that its $B(E2)$ value is below 20 W.u.~it would imply the total intensity for this transition to be less than 0.05 in relative units, thus negligible. On the other hand, if the 3.9-keV transition would be $M1$ or $E1$ in character, a significant portion de-exciting the 397.6-keV level would feed the state at 393.7-keV. That portion would be detectable via missing intensity in the 33.9-keV line. The balance of intensities feeding and de-exciting the 397.6-keV level excludes this possibility. Thus the 397.6-keV level is de-excited almost exclusively by the 33.9-keV transition, which supports the proposed spin/parity assignments to this level.

{\bf 455.6-keV level, $\bm{J^{\pi}=(5/2^-)}$:}
This level receives a strong $\beta$ feeding with log$ft$ = 5.3 from the $(5/2^-)$ ground state in $^{65}$Mn, which implies a spin/parity of $(3/2^-)$, $(5/2^-)$ or $(7/2^-)$ for the 455.6-keV state. On the other hand, this level has a relatively long half-life of 350(10) ps and is de-excited by $\gamma$-rays to the $(1/2^-)$ ground state and the $(3/2^-)$ 363.7-keV state. The 455.6-keV ground state transition can be either $M1$ with $B(M1)$ = 6.4(8)$\times$10$^{-4}$~W.u., or $E2$ with $B(E2)$ = 5.2(6) W.u., while the 92.0-keV transition can only be $M1$ with $B(M1)$ = 2.5(4)$\times$10$^{-3}$~W.u. The 92.0-keV $\gamma$ ray cannot be $E2$ since then its $B(E2)$ would have an unrealistic value of 500 W.u.

The transition rates allow for a choice of only two spin/parities: $(3/2^-)$ or $(5/2^-)$. We adopt the second alternative, since it allows for a $B(E2)$ value for the ground state transition consistent with a core coupled state.

This level can also decay by a 58.0-keV transition to the $(5/2^+)$ state at 397.6 keV. Assuming a typical $B(E1)$ value of about 1$\times$10$^{-5}$ W.u. for this transition its total relative intensity would be only about 0.2.

{\bf 561.0 keV, $\bm{J^{\pi}=(3/2^-, 5/2^-, 3/2^+)}$:}
This level has a small $\beta$-feeding with log$ft$ = 6.1 and populates levels of spin $(1/2^-)$ and $(3/2^-)$ and $(5/2^+)$, but not the $9/2^+$ isomer nor the $(5/2^-)$ state. One of the $\gamma$ transitions must be parity changing $E1$. If the 163.1-keV line is E1 then the possible spin/parity for the 561.0-keV level is either $(3/2^-)$, $(5/2^-)$ or $(7/2^-)$. Taking the measured T$_{1/2}$ we obtain a $B(E1)$ value of $4\times$10$^{-5}$ W.u. The 197.6-keV line must be M1, with a $B(M1)$ value of 2.2$\times$10$^{-3}$ W.u., as a $B(E2)$ rate of above 93 W.u.~which would be too high. The 560.8-keV transition could be then $M1$ ($B(M1)$ $\sim$1.8$\times$10$^{-4}$ W.u.) or $E2$ ($B(E2)$ $\sim$1.0 W.u.), but $M3$ is firmly excluded. This scenario allows only for the spin/parity of $(3/2^-)$ and $(5/2^-)$ for the 561.0-keV state.
 
If the 197.6-keV line is $E1$, then the 163.1-keV $\gamma$ ray could be $M1$ ($B(M1)$ $\sim$1.9$\times$10$^{-3}$ W.u.), however $E2$ is excluded. While the 560.8-keV transition could be $E1$ ($B(E1)$ $\sim$3$\times$10$^{-6}$ W.u.), $M2$ is excluded. This scenario allows only for a spin/parity of $(3/2^+)$.

{\bf 569.1 keV, $\bm{J=(1/2, 3/2)}$:}
This level receives a very low $\beta$-feeding with log$ft$ = 6.3, which may indicate a forbidden transition from the parent $^{65}$Mn. It de-excites to the ground state and the first excited state. With the half-life limit of $\le$ 12 ps, the 569.1-keV ground state transition can be either $E1$ with $B(E1)$ $\ge$1.8$\times$10$^{-4}$~W.u.~or $M1$ with $B(M1)$ $\ge$9.6$\times$10$^{-3}$~W.u. Both $M2$ and $E2$ multipolarities are excluded for this transition. A similar situation occurs for the 205.3-keV line. Thus the spin for the 569.1-keV level can be either $(1/2)$ or $(3/2)$ with either parity, although a negative parity would be favoured.

Due to the very short lifetime limit and energy factors, even a moderately fast 171.5-keV transition to the $(5/2^+)$ 397.6-keV level would be undetectable in our experiment, regardless of whether it would be $M1$ or $E1$ in character.

{\bf 609.5 keV, $\bm{J^{\pi}=(7/2^+)}$:}
This level is the only one that is known to de-excite to the $\beta$-isomer at 393.7 keV. It has very weak $\beta$ feeding if any, with an intensity of $\le$1.0$\%$. It is mainly $\gamma$ fed from higher lying levels. The only reasonable spin/parity assignment would be $(7/2^+)$. A lower spin or negative parity would make transitions possible to other levels beside the $(9/2^+)$ isomer (the $(5/2^-)$ at 455.l keV, for example). A higher spin would cause the levels, whose transitions feed the 609.5-keV level, to directly $\gamma$ feed the $\beta$-decaying isomer. We found no evidence for that.

We do not observe the 211.9-keV transition to the $(5/2^+)$ $\gamma$-ray isomer. An upper limit of intensity for this transition is 0.2 in relative units.

{\bf 683.3 keV, $\bm{J^{\pi}=(3/2^-, 5/2^-, 3/2^+)}$:}
The 683.3-keV level is weakly $\beta$ fed. It de-excites to four levels with three of them having spin/parity $(1/2^-)$, $(3/2^-)$ and $(5/2^-)$, respectively. All of the de-exciting $\gamma$ rays could be only of either $E1$ or $M1$ type, with the exception of the ground state transition which can be also $E2$ in character. This makes the possible spin/parity assignments for this level as $(3/2^-)$, $(5/2^-)$ and $(3/2^+)$.

{\bf 894.8 keV, $\bm{J^{\pi}=(7/2^-)}$:}
This level seems to get some $\beta$ feeding, which would imply negative parity. Moreover it feeds two excited states of spin/parity $(3/2^-)$ and $(5/2^-)$, but it does not feed the ground state, which is energetically favoured. The transition rates imply that each of the de-exciting transitions could be either $E1$ or $M1$ in character, but the 531.1-keV transition could be also of the $E2$ type with $B(E2)$ $\ge$8 W.u. The lack of any ground state feeding almost definitely excludes the $(1/2^-)$ and $(3/2^-)$ cases, and makes even the $(5/2^-)$ assignment unlikely. As a result the most likely spin/parity assignment for this level is $(7/2^-)$.

{\bf 1057.2 keV, $\bm{J^{\pi}=(3/2^-, 5/2^-)}$:}
This level has a $\beta$ feeding of $2.9\;\%$ and log$ft$ =5.6, which favours negative parity for this state. It populates six states mainly of negative parity and also the $(5/2^+)$ state. As discussed before, our evidence indicates that the half-life of this state is very short in the range of a few ps, thus it was used as an internal ``semi-prompt" reference. We have placed a limit of $\le$8 ps on the half-life of this level. A check on the lifetime limit is provided by the $B(E1)$ value which we expect to be in the range of about 5$\times$10$^{-5}$ W.u.

Indeed the half-life limit gives the 659.7-keV E1 transition a $B(E1)$ value of $\ge$2.5$\times$10$^{-5}$ W.u.~and firmly excludes this transition to be $M2$. The $E1$ nature of this $\gamma$-ray feeding the $(5/2^+)$ state limits the possible spin/parity assignments to the 1057.2-keV level to  $(3/2^-)$, $(5/2^-)$ and $(7/2^-)$. However, the direct feeding to the $(1/2^-)$ ground state definitely excludes the $(7/2^-)$ case. The 1057.2-keV transition could be either $M1$ ($B(M1)$ $\ge$2.8$\times$10$^{-4}$ W.u.) or $E2$ ($B(E2)$ $\ge$0.42 W.u.), but not $M3$.

The most intense transition is the 693.7-keV one feeding the $(3/2^-)$ state. This transition can be either $M1$ with $B(M1)$ $\ge$3.1$\times$10$^{-3}$ W.u.~or $E2$ with $B(E2)$ $\ge$11 W.u., or a mixture of both. On the other hand the fastest transition is the 374.1-keV line, which could be either $E1$ ($B(E1)$ $\ge$1.7$\times$10$^{-4}$ W.u.) or $M1$ ($B(M1)$ $\ge$1.6$\times$10$^{-2}$ W.u.) but for sure not $E2$ in character. The transition rates for $\gamma$ rays de-exciting this state are thus consistent with the proposed spin/parity assignment.

{\bf 1088.7 keV, $\bm{J^{\pi}=(3/2^-, 5/2^-)}$:}
This level has the second highest $\beta$ feeding. This rules out a positive parity for this state and limits the spin/parity to $(3/2^-)$, $(5/2^-)$ and $(7/2^-)$ as the $\beta$ transition is clearly allowed. In similarity to the 1057.3-keV state this level was also used as ``semi-prompt" reference and a limit of $\le$8 ps was set on its half-life. With this limit the 1088.6-keV transition feeding the $(1/2^-)$ ground state can be either $M1$ or $E2$, with the corresponding $B(M1)$ $\ge$1.0$\times$10$^{-3}$ W.u.~and $E2$ with $B(E2)$ $\ge$1.4 W.u., but definitely not $M3$. This limits the spin/parity assignment for this level as $3/2^-$ or $5/2^-$. All transition rates for other $\gamma$ rays de-exciting this level are consistent with this assignment. The strong $\beta$-feeding to the $(3/2^-)$ levels in $^{65}$Fe makes a $(3/2^-)$ the most reasonable choice for the 1088.7-keV state, but $(5/2^-)$ cannot be completely excluded.

{\bf 1366.6 keV, $\bm{J^{\pi}=(5/2^-)}$:}
This level has a significant $\beta$ feeding of $6.3\;\%$ with a small log$ft$ = 5.2, which implies an allowed transition making the possible spin/parity assignments of $(3/2^-)$, $(5/2^-)$ and $(7/2^-)$. This state was also used as a ``semi-prompt" time reference and a limit of $\le$8 ps was set on its half-life. The $\gamma$ rays feeding the $(5/2^+)$ and $(7/2^+)$ states are $E1$ in character with the $B(E1)$ values of $\ge$4.3$\times$10$^{-6}$ W.u.~and $\ge$2.3$\times$10$^{-6}$ W.u., respectively, while the $M2$ character is definitely excluded. This allows only the $J^{\pi}$ assignments of $(5/2^-)$ and $(7/2^-)$. On the other hand, the ground state 1366.2-keV transition can be either $M1$ ($B(M1)$ $\ge$3.0$\times$10$^{-5}$ W.u.) or $E2$ ($B(E2)$ $\ge$2.7$\times$10$^{-2}$ W.u.), but definitely not $M3$ in character. This excludes the $J^{\pi}$ assignment of $(7/2^-)$, leaving $(5/2^-)$ as the only alternative. The absence of a transtion linking this state to the $(9/2^+)$ one further supports this assignment.

\section{Calculations}

Shell-model calculations were performed using the Lenzi-Nowacki-Poves-Sieja (LNPS) effective interaction \cite{LEN10}. This interaction works in a large shell model space, employing $^{48}$Ca as a core and including $pf$ orbitals for protons and $pfgd$ ($1p_{3/2}$, $1p_{1/2}$, $0f_{5/2}$, $0g_{9/2}$ and $1d_{5/2}$) orbitals for neutrons. For further details, refer to \cite{LEN10}. 

\begin{figure}
\includegraphics[width=\columnwidth]{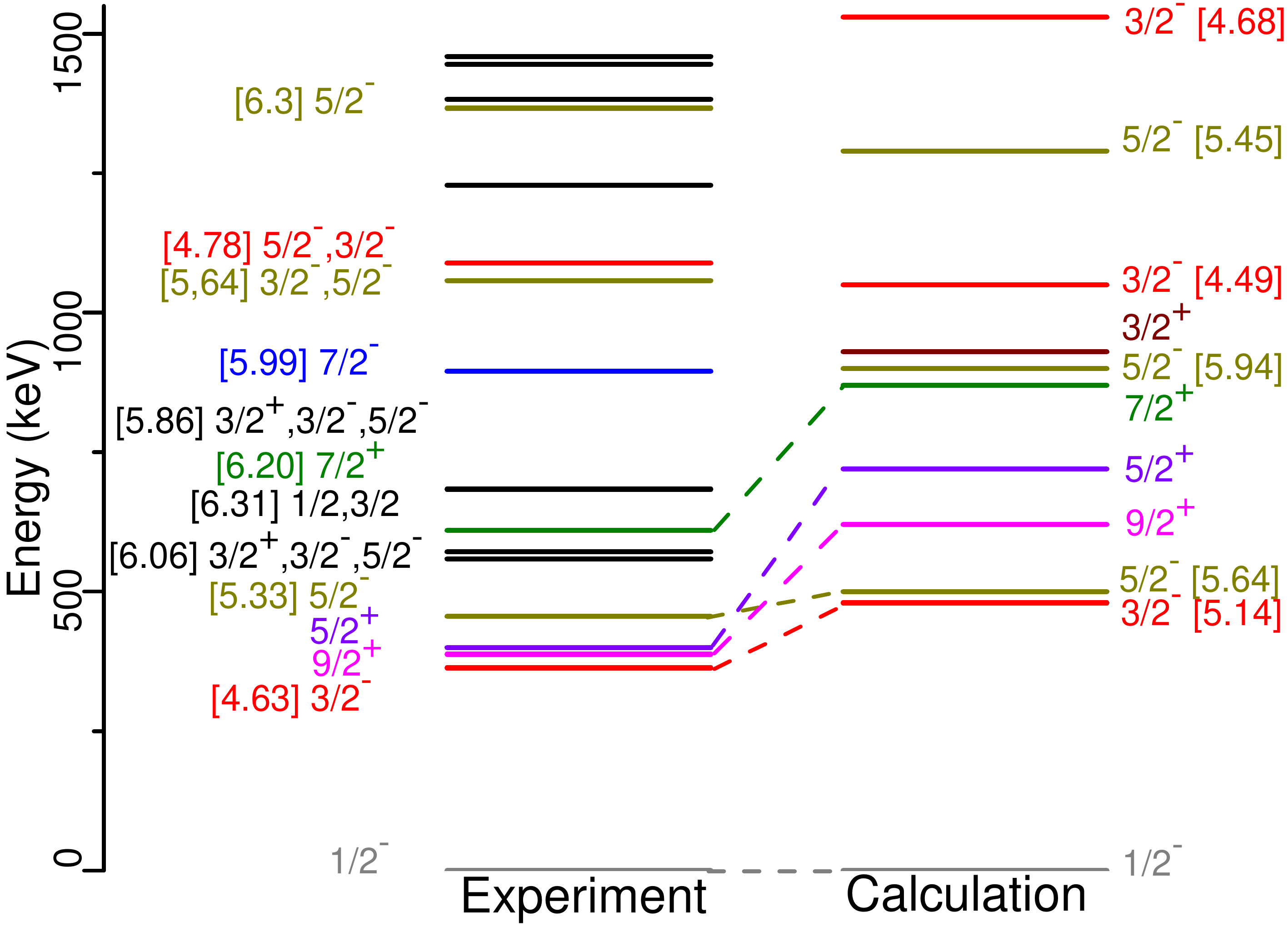}
\caption{\label{fig:Exp-the} \textit{(Color online)} A comparison of the experimental and calculated levels (see text for details). Log$ft$ (in brackets) and spin/parities were calculated for the lower energy levels.}
\end{figure}

For the beta-decay we calculate the $^{65}$Mn $5/2^-$ ground state in this valence space using the LNPS interaction. The Gamow-Teller strength function is obtained by the Lanczos strength function method and from the individual B(GT)'s the log$ft$ values are calculated. The decay pattern obtained in the calculations is plotted together with the experimental results in Fig. \ref{fig:Exp-the} for energies below 1.6 MeV. Very good agreement is obtained for the $3/2^-$ and $5/2^-$ states when compared to the experimental results, easily identifying the first $3/2^-$ and $5/2^-$ with the 363.7- and 455.6-keV levels seen in the experiment. The log$ft$ values help identifying the assigned spins and parities. The calculated $7/2^-$ states are above 1.5 MeV with log$ft>6.5$. States with spin-parity $1/2^-$ were not calculated as there would not be allowed $\beta$ transitions to them. Sizeable Gamow-Teller strength above 4 MeV is reproduced in the calculations for the beta-decay to the negative parity states in $^{65}$Fe, corresponding to the tail of the giant resonance. For the lowest lying negative $1/2^-$ (g.s.), $3/2^-$ and $5/2^-$ states similar configurations are obtained, supporting their interpretation as the coupling of the single particle $p_{1/2}$ neutron orbit to the 0$^+$ and 2$^+$ states of the core.

The positive parity states have been also calculated for spins $3/2$, $5/2$, $7/2$ and $9/2$. They are in reasonable agreement with the experimentally measured states that have been assigned positive parity, although they are predicted at slightly higher energies. This fact arises from the dimension limitations of the calculations, which slows down the convergence of the energy values. The calculations show a level with spin parity $3/2^+$ at 930 keV, which would be above the $\gamma$-decaying isomer (at 720 keV in the calculations). This level has not been identified in the experiment, nor its expected $3/2^+\rightarrow5/2^+$ and $3/2^+\rightarrow5/2^+$ transitions observed.

 The lowest level structure in $^{65}$Fe seems to stem from the coupling of the $0^+$ ground state and the collective $2^+$ state in $^{64}$Fe to a single neutron in the $pf$ shell. In this way a $1/2^-$, $3/2^-$, $5/2^-$ sequence arises from the ground state to the first excited states. A $7/2^-$ state, which might arise from the coupling $|4^+\otimes1/2^-\rangle=7/2^-,\;9/2^-$, has been experimentally identified at 894.8 keV, but is predicted by the calculations at higher energies. No $9/2^-$ state was observed experimentally.

\section{Conclusion}

We report on the first detailed study of the $\beta^-$ decay of $^{65}$Mn to $^{65}$Fe. Both $\gamma$ and fast timing spectroscopy were used. The half-life of the ground state in $^{65}$Mn has been measured as $T_{1/2}$ = 91.9(9) ms confirming the previous value. The $^{65}$Fe level scheme now includes 87 $\gamma$ rays and 41 excited states. The $\beta$-delayed neutron branch has been measured as $P_n = 7.9(12)\;\%$ about three times smaller than the value previously measured. A substantial $\beta$-n feeding going to four excited states in $^{64}$Fe has been observed. 

We have made the first precise energy determination of the $\beta$-decaying isomer in $^{65}$Fe at 393.7 keV. The half-life of $T_{1/2}$ = 437(55)~ns measured for the 397.6-keV $\gamma$ isomer is in good agreement with the previously reported values.

By employing the ATD method 4 new level half-lives and a few lifetime limits in the picosecond range have been determined for the excited states in $^{65}$Fe. The measured transition rates allow spin/parity assignments to the low energy levels to be made. 

The experimental excitation energies of positive and negative parity states below the excitation energy of 1.6 MeV and log$ft$ values have been compared to the shell model calculations obtained with the LNPS effective interaction. The calculation reproduces quite well the observed level structure of negative parity states. With the experimental information on the decay and level half lives we have tentatively assigned spin-parity to levels in $^{65}$Fe below 1.5 MeV. Positive parity states with spins $9/2^+$, $5/2^+$ and $7/2^+$ are proposed at low excitation energy. These assignments are also supported by our shell-model calculations, and show the importance of the positive parity orbitals across the $N=40$ subshell and $N=50$ shell gaps. 

\begin{acknowledgments}
This work was supported by the Spanish MINECO (FPA2010-17142 and FPA2011-29854), CPAN (CSD-2007-00042@ Ingenio 2010), RA-NET NuPNET (PRI-PIMNUP-2011-1338 and PRI-PIMMNUP-2011-1361) and Comunidad de Madrid, Spain (HEPHACOS S2009/ESP-1473). Support by Grupo de F\'isica Nuclear (GFN-UCM) and by the European Union Seventh Framework through ENSAR (contract no. 262010) is also acknowledged. Fast timing electronics were provided by the Fast Timing Collaboration and MASTICON. B.~Olaizola acknowledges funding by the CPAN (CSD-2007-00042@ Ingenio 2010) project. A.~Aprahamian and S.R.~Lesher acknowledge funding by the US-NSF under contract PHY-07-58100. H.~Mach would like to acknowledge the hospitality and support from the Grupo de F\'isica Nuclear during his stay as CPAN Senior Expert at the University Complutense of Madrid. 

\end{acknowledgments}

\bibliography{65Mn_bibliography}% Produces the bibliography via BibTeX.

%merlin.mbs apsrev4-1.bst 2010-07-25 4.21a (PWD, AO, DPC) hacked
%Control: key (0)
%Control: author (8) initials jnrlst
%Control: editor formatted (1) identically to author
%Control: production of article title (-1) disabled
%Control: page (0) single
%Control: year (1) truncated
%Control: production of eprint (0) enabled
\def\url#1{}
\begin{thebibliography}{35}%
\makeatletter
\providecommand \@ifxundefined [1]{%
 \@ifx{#1\undefined}
}%
\providecommand \@ifnum [1]{%
 \ifnum #1\expandafter \@firstoftwo
 \else \expandafter \@secondoftwo
 \fi
}%
\providecommand \@ifx [1]{%
 \ifx #1\expandafter \@firstoftwo
 \else \expandafter \@secondoftwo
 \fi
}%
\providecommand \natexlab [1]{#1}%
\providecommand \enquote  [1]{``#1''}%
\providecommand \bibnamefont  [1]{#1}%
\providecommand \bibfnamefont [1]{#1}%
\providecommand \citenamefont [1]{#1}%
\providecommand \href@noop [0]{\@secondoftwo}%
\providecommand \href [0]{\begingroup \@sanitize@url \@href}%
\providecommand \@href[1]{\@@startlink{#1}\@@href}%
\providecommand \@@href[1]{\endgroup#1\@@endlink}%
\providecommand \@sanitize@url [0]{\catcode `\\12\catcode `\$12\catcode
  `\&12\catcode `\#12\catcode `\^12\catcode `\_12\catcode `\%12\relax}%
\providecommand \@@startlink[1]{}%
\providecommand \@@endlink[0]{}%
\providecommand \url  [0]{\begingroup\@sanitize@url \@url }%
\providecommand \@url [1]{\endgroup\@href {#1}{\urlprefix }}%
\providecommand \urlprefix  [0]{URL }%
\providecommand \Eprint [0]{\href }%
\providecommand \doibase [0]{http://dx.doi.org/}%
\providecommand \selectlanguage [0]{\@gobble}%
\providecommand \bibinfo  [0]{\@secondoftwo}%
\providecommand \bibfield  [0]{\@secondoftwo}%
\providecommand \translation [1]{[#1]}%
\providecommand \BibitemOpen [0]{}%
\providecommand \bibitemStop [0]{}%
\providecommand \bibitemNoStop [0]{.\EOS\space}%
\providecommand \EOS [0]{\spacefactor3000\relax}%
\providecommand \BibitemShut  [1]{\csname bibitem#1\endcsname}%
\let\auto@bib@innerbib\@empty
%</preamble>
\bibitem [{\citenamefont {Kaneko}\ \emph {et~al.}(2008)\citenamefont {Kaneko},
  \citenamefont {Sun}, \citenamefont {Hasegawa},\ and\ \citenamefont
  {Mizusaki}}]{kan08}%
  \BibitemOpen
  \bibfield  {author} {\bibinfo {author} {\bibfnamefont {K.}~\bibnamefont
  {Kaneko}}, \bibinfo {author} {\bibfnamefont {Y.}~\bibnamefont {Sun}},
  \bibinfo {author} {\bibfnamefont {M.}~\bibnamefont {Hasegawa}}, \ and\
  \bibinfo {author} {\bibfnamefont {T.}~\bibnamefont {Mizusaki}},\ }\href
  {\doibase 10.1103/PhysRevC.78.064312} {\bibfield  {journal} {\bibinfo
  {journal} {Phys. Rev. C}\ }\textbf {\bibinfo {volume} {78}},\ \bibinfo
  {pages} {064312} (\bibinfo {year} {2008})}\BibitemShut {NoStop}%
\bibitem [{\citenamefont {Lunardi}\ \emph {et~al.}(2007)\citenamefont
  {Lunardi}, \citenamefont {Lenzi}, \citenamefont {Vedova}, \citenamefont
  {Farnea}, \citenamefont {Gadea}, \citenamefont {M\ifmmode~\u{a}\else
  \u{a}\fi{}rginean}, \citenamefont {Bazzacco}, \citenamefont {Beghini},
  \citenamefont {Bizzeti}, \citenamefont {Bizzeti-Sona}, \citenamefont
  {Bucurescu}, \citenamefont {Corradi}, \citenamefont {Deacon}, \citenamefont
  {Angelis}, \citenamefont {Fioretto}, \citenamefont {Freeman}, \citenamefont
  {Ionescu-Bujor}, \citenamefont {Iordachescu}, \citenamefont {Mason},
  \citenamefont {Mengoni}, \citenamefont {Montagnoli}, \citenamefont {Napoli},
  \citenamefont {Nowacki}, \citenamefont {Orlandi}, \citenamefont {Pollarolo},
  \citenamefont {Recchia}, \citenamefont {Scarlassara}, \citenamefont {Smith},
  \citenamefont {Stefanini}, \citenamefont {Szilner}, \citenamefont {Ur},
  \citenamefont {Valiente-Dob\'on},\ and\ \citenamefont {Varley}}]{LUN07}%
  \BibitemOpen
  \bibfield  {author} {\bibinfo {author} {\bibfnamefont {S.}~\bibnamefont
  {Lunardi}}, \bibinfo {author} {\bibfnamefont {S.~M.}\ \bibnamefont {Lenzi}},
  \bibinfo {author} {\bibfnamefont {F.~D.}\ \bibnamefont {Vedova}}, \bibinfo
  {author} {\bibfnamefont {E.}~\bibnamefont {Farnea}}, \bibinfo {author}
  {\bibfnamefont {A.}~\bibnamefont {Gadea}}, \bibinfo {author} {\bibfnamefont
  {N.}~\bibnamefont {M\ifmmode~\u{a}\else \u{a}\fi{}rginean}}, \bibinfo
  {author} {\bibfnamefont {D.}~\bibnamefont {Bazzacco}}, \bibinfo {author}
  {\bibfnamefont {S.}~\bibnamefont {Beghini}}, \bibinfo {author} {\bibfnamefont
  {P.~G.}\ \bibnamefont {Bizzeti}}, \bibinfo {author} {\bibfnamefont {A.~M.}\
  \bibnamefont {Bizzeti-Sona}}, \bibinfo {author} {\bibfnamefont
  {D.}~\bibnamefont {Bucurescu}}, \bibinfo {author} {\bibfnamefont
  {L.}~\bibnamefont {Corradi}}, \bibinfo {author} {\bibfnamefont {A.~N.}\
  \bibnamefont {Deacon}}, \bibinfo {author} {\bibfnamefont {G.~d.}\
  \bibnamefont {Angelis}}, \bibinfo {author} {\bibfnamefont {E.}~\bibnamefont
  {Fioretto}}, \bibinfo {author} {\bibfnamefont {S.~J.}\ \bibnamefont
  {Freeman}}, \bibinfo {author} {\bibfnamefont {M.}~\bibnamefont
  {Ionescu-Bujor}}, \bibinfo {author} {\bibfnamefont {A.}~\bibnamefont
  {Iordachescu}}, \bibinfo {author} {\bibfnamefont {P.}~\bibnamefont {Mason}},
  \bibinfo {author} {\bibfnamefont {D.}~\bibnamefont {Mengoni}}, \bibinfo
  {author} {\bibfnamefont {G.}~\bibnamefont {Montagnoli}}, \bibinfo {author}
  {\bibfnamefont {D.~R.}\ \bibnamefont {Napoli}}, \bibinfo {author}
  {\bibfnamefont {F.}~\bibnamefont {Nowacki}}, \bibinfo {author} {\bibfnamefont
  {R.}~\bibnamefont {Orlandi}}, \bibinfo {author} {\bibfnamefont
  {G.}~\bibnamefont {Pollarolo}}, \bibinfo {author} {\bibfnamefont
  {F.}~\bibnamefont {Recchia}}, \bibinfo {author} {\bibfnamefont
  {F.}~\bibnamefont {Scarlassara}}, \bibinfo {author} {\bibfnamefont {J.~F.}\
  \bibnamefont {Smith}}, \bibinfo {author} {\bibfnamefont {A.~M.}\ \bibnamefont
  {Stefanini}}, \bibinfo {author} {\bibfnamefont {S.}~\bibnamefont {Szilner}},
  \bibinfo {author} {\bibfnamefont {C.~A.}\ \bibnamefont {Ur}}, \bibinfo
  {author} {\bibfnamefont {J.~J.}\ \bibnamefont {Valiente-Dob\'on}}, \ and\
  \bibinfo {author} {\bibfnamefont {B.~J.}\ \bibnamefont {Varley}},\ }\href
  {\doibase 10.1103/PhysRevC.76.034303} {\bibfield  {journal} {\bibinfo
  {journal} {Phys. Rev. C}\ }\textbf {\bibinfo {volume} {76}},\ \bibinfo
  {pages} {034303} (\bibinfo {year} {2007})}\BibitemShut {NoStop}%
\bibitem [{\citenamefont {Hannawald}\ \emph {et~al.}(1999)\citenamefont
  {Hannawald}, \citenamefont {Kautzsch}, \citenamefont {W\"ohr}, \citenamefont
  {Walters}, \citenamefont {Kratz}, \citenamefont {Fedoseyev}, \citenamefont
  {Mishin}, \citenamefont {B\"ohmer}, \citenamefont {Pfeiffer}, \citenamefont
  {Sebastian}, \citenamefont {Jading}, \citenamefont {K\"oster}, \citenamefont
  {Lettry}, \citenamefont {Ravn},\ and\ \citenamefont {the
  ISOLDE~Collaboration}}]{HAN99}%
  \BibitemOpen
  \bibfield  {author} {\bibinfo {author} {\bibfnamefont {M.}~\bibnamefont
  {Hannawald}}, \bibinfo {author} {\bibfnamefont {T.}~\bibnamefont {Kautzsch}},
  \bibinfo {author} {\bibfnamefont {A.}~\bibnamefont {W\"ohr}}, \bibinfo
  {author} {\bibfnamefont {W.~B.}\ \bibnamefont {Walters}}, \bibinfo {author}
  {\bibfnamefont {K.-L.}\ \bibnamefont {Kratz}}, \bibinfo {author}
  {\bibfnamefont {V.~N.}\ \bibnamefont {Fedoseyev}}, \bibinfo {author}
  {\bibfnamefont {V.~I.}\ \bibnamefont {Mishin}}, \bibinfo {author}
  {\bibfnamefont {W.}~\bibnamefont {B\"ohmer}}, \bibinfo {author}
  {\bibfnamefont {B.}~\bibnamefont {Pfeiffer}}, \bibinfo {author}
  {\bibfnamefont {V.}~\bibnamefont {Sebastian}}, \bibinfo {author}
  {\bibfnamefont {Y.}~\bibnamefont {Jading}}, \bibinfo {author} {\bibfnamefont
  {U.}~\bibnamefont {K\"oster}}, \bibinfo {author} {\bibfnamefont
  {J.}~\bibnamefont {Lettry}}, \bibinfo {author} {\bibfnamefont {H.~L.}\
  \bibnamefont {Ravn}}, \ and\ \bibinfo {author} {\bibnamefont {the
  ISOLDE~Collaboration}},\ }\href {\doibase 10.1103/PhysRevLett.82.1391}
  {\bibfield  {journal} {\bibinfo  {journal} {Phys. Rev. Lett.}\ }\textbf
  {\bibinfo {volume} {82}},\ \bibinfo {pages} {1391} (\bibinfo {year}
  {1999})}\BibitemShut {NoStop}%
\bibitem [{\citenamefont {Caurier}\ \emph {et~al.}(2002)\citenamefont
  {Caurier}, \citenamefont {Nowacki},\ and\ \citenamefont {Poves}}]{CAU02}%
  \BibitemOpen
  \bibfield  {author} {\bibinfo {author} {\bibfnamefont {E.}~\bibnamefont
  {Caurier}}, \bibinfo {author} {\bibfnamefont {F.}~\bibnamefont {Nowacki}}, \
  and\ \bibinfo {author} {\bibfnamefont {A.}~\bibnamefont {Poves}},\ }\href
  {\doibase 10.1140/epja/i2001-10243-7} {\bibfield  {journal} {\bibinfo
  {journal} {The European Physical Journal A - Hadrons and Nuclei}\ }\textbf
  {\bibinfo {volume} {15}},\ \bibinfo {pages} {145} (\bibinfo {year}
  {2002})}\BibitemShut {NoStop}%
\bibitem [{\citenamefont {Lenzi}\ \emph {et~al.}(2010)\citenamefont {Lenzi},
  \citenamefont {Nowacki}, \citenamefont {Poves},\ and\ \citenamefont
  {Sieja}}]{LEN10}%
  \BibitemOpen
  \bibfield  {author} {\bibinfo {author} {\bibfnamefont {S.~M.}\ \bibnamefont
  {Lenzi}}, \bibinfo {author} {\bibfnamefont {F.}~\bibnamefont {Nowacki}},
  \bibinfo {author} {\bibfnamefont {A.}~\bibnamefont {Poves}}, \ and\ \bibinfo
  {author} {\bibfnamefont {K.}~\bibnamefont {Sieja}},\ }\href {\doibase
  10.1103/PhysRevC.82.054301} {\bibfield  {journal} {\bibinfo  {journal} {Phys.
  Rev. C}\ }\textbf {\bibinfo {volume} {82}},\ \bibinfo {pages} {054301}
  (\bibinfo {year} {2010})}\BibitemShut {NoStop}%
\bibitem [{\citenamefont {Broda}\ \emph {et~al.}(1995)\citenamefont {Broda},
  \citenamefont {Fornal}, \citenamefont {Kr\'olas}, \citenamefont {Paw\l{}at},
  \citenamefont {Bazzacco}, \citenamefont {Lunardi}, \citenamefont
  {Rossi-Alvarez}, \citenamefont {Menegazzo}, \citenamefont {de~Angelis},
  \citenamefont {Bednarczyk}, \citenamefont {Rico}, \citenamefont {De~Acu\~na},
  \citenamefont {Daly}, \citenamefont {Mayer}, \citenamefont {Sferrazza},
  \citenamefont {Grawe}, \citenamefont {Maier},\ and\ \citenamefont
  {Schubart}}]{BRO95}%
  \BibitemOpen
  \bibfield  {author} {\bibinfo {author} {\bibfnamefont {R.}~\bibnamefont
  {Broda}}, \bibinfo {author} {\bibfnamefont {B.}~\bibnamefont {Fornal}},
  \bibinfo {author} {\bibfnamefont {W.}~\bibnamefont {Kr\'olas}}, \bibinfo
  {author} {\bibfnamefont {T.}~\bibnamefont {Paw\l{}at}}, \bibinfo {author}
  {\bibfnamefont {D.}~\bibnamefont {Bazzacco}}, \bibinfo {author}
  {\bibfnamefont {S.}~\bibnamefont {Lunardi}}, \bibinfo {author} {\bibfnamefont
  {C.}~\bibnamefont {Rossi-Alvarez}}, \bibinfo {author} {\bibfnamefont
  {R.}~\bibnamefont {Menegazzo}}, \bibinfo {author} {\bibfnamefont
  {G.}~\bibnamefont {de~Angelis}}, \bibinfo {author} {\bibfnamefont
  {P.}~\bibnamefont {Bednarczyk}}, \bibinfo {author} {\bibfnamefont
  {J.}~\bibnamefont {Rico}}, \bibinfo {author} {\bibfnamefont {D.}~\bibnamefont
  {De~Acu\~na}}, \bibinfo {author} {\bibfnamefont {P.~J.}\ \bibnamefont
  {Daly}}, \bibinfo {author} {\bibfnamefont {R.~H.}\ \bibnamefont {Mayer}},
  \bibinfo {author} {\bibfnamefont {M.}~\bibnamefont {Sferrazza}}, \bibinfo
  {author} {\bibfnamefont {H.}~\bibnamefont {Grawe}}, \bibinfo {author}
  {\bibfnamefont {K.~H.}\ \bibnamefont {Maier}}, \ and\ \bibinfo {author}
  {\bibfnamefont {R.}~\bibnamefont {Schubart}},\ }\href {\doibase
  10.1103/PhysRevLett.74.868} {\bibfield  {journal} {\bibinfo  {journal} {Phys.
  Rev. Lett.}\ }\textbf {\bibinfo {volume} {74}},\ \bibinfo {pages} {868}
  (\bibinfo {year} {1995})}\BibitemShut {NoStop}%
\bibitem [{\citenamefont {Sorlin}\ \emph {et~al.}(2002)\citenamefont {Sorlin},
  \citenamefont {Leenhardt}, \citenamefont {Donzaud}, \citenamefont {Duprat},
  \citenamefont {Azaiez}, \citenamefont {Nowacki}, \citenamefont {Grawe},
  \citenamefont {Dombr\'adi}, \citenamefont {Amorini}, \citenamefont {Astier},
  \citenamefont {Baiborodin}, \citenamefont {Belleguic}, \citenamefont
  {Borcea}, \citenamefont {Bourgeois}, \citenamefont {Cullen}, \citenamefont
  {Dlouhy}, \citenamefont {Dragulescu}, \citenamefont {G\'orska}, \citenamefont
  {Gr\'evy}, \citenamefont {Guillemaud-Mueller}, \citenamefont {Hagemann},
  \citenamefont {Herskind}, \citenamefont {Kiener}, \citenamefont {Lemmon},
  \citenamefont {Lewitowicz}, \citenamefont {Lukyanov}, \citenamefont {Mayet},
  \citenamefont {de~Oliveira~Santos}, \citenamefont {Pantalica}, \citenamefont
  {Penionzhkevich}, \citenamefont {Pougheon}, \citenamefont {Poves},
  \citenamefont {Redon}, \citenamefont {Saint-Laurent}, \citenamefont
  {Scarpaci}, \citenamefont {Sletten}, \citenamefont {Stanoiu}, \citenamefont
  {Tarasov},\ and\ \citenamefont {Theisen}}]{SOR02}%
  \BibitemOpen
  \bibfield  {author} {\bibinfo {author} {\bibfnamefont {O.}~\bibnamefont
  {Sorlin}}, \bibinfo {author} {\bibfnamefont {S.}~\bibnamefont {Leenhardt}},
  \bibinfo {author} {\bibfnamefont {C.}~\bibnamefont {Donzaud}}, \bibinfo
  {author} {\bibfnamefont {J.}~\bibnamefont {Duprat}}, \bibinfo {author}
  {\bibfnamefont {F.}~\bibnamefont {Azaiez}}, \bibinfo {author} {\bibfnamefont
  {F.}~\bibnamefont {Nowacki}}, \bibinfo {author} {\bibfnamefont
  {H.}~\bibnamefont {Grawe}}, \bibinfo {author} {\bibfnamefont
  {Z.}~\bibnamefont {Dombr\'adi}}, \bibinfo {author} {\bibfnamefont
  {F.}~\bibnamefont {Amorini}}, \bibinfo {author} {\bibfnamefont
  {A.}~\bibnamefont {Astier}}, \bibinfo {author} {\bibfnamefont
  {D.}~\bibnamefont {Baiborodin}}, \bibinfo {author} {\bibfnamefont
  {M.}~\bibnamefont {Belleguic}}, \bibinfo {author} {\bibfnamefont
  {C.}~\bibnamefont {Borcea}}, \bibinfo {author} {\bibfnamefont
  {C.}~\bibnamefont {Bourgeois}}, \bibinfo {author} {\bibfnamefont {D.~M.}\
  \bibnamefont {Cullen}}, \bibinfo {author} {\bibfnamefont {Z.}~\bibnamefont
  {Dlouhy}}, \bibinfo {author} {\bibfnamefont {E.}~\bibnamefont {Dragulescu}},
  \bibinfo {author} {\bibfnamefont {M.}~\bibnamefont {G\'orska}}, \bibinfo
  {author} {\bibfnamefont {S.}~\bibnamefont {Gr\'evy}}, \bibinfo {author}
  {\bibfnamefont {D.}~\bibnamefont {Guillemaud-Mueller}}, \bibinfo {author}
  {\bibfnamefont {G.}~\bibnamefont {Hagemann}}, \bibinfo {author}
  {\bibfnamefont {B.}~\bibnamefont {Herskind}}, \bibinfo {author}
  {\bibfnamefont {J.}~\bibnamefont {Kiener}}, \bibinfo {author} {\bibfnamefont
  {R.}~\bibnamefont {Lemmon}}, \bibinfo {author} {\bibfnamefont
  {M.}~\bibnamefont {Lewitowicz}}, \bibinfo {author} {\bibfnamefont {S.~M.}\
  \bibnamefont {Lukyanov}}, \bibinfo {author} {\bibfnamefont {P.}~\bibnamefont
  {Mayet}}, \bibinfo {author} {\bibfnamefont {F.}~\bibnamefont
  {de~Oliveira~Santos}}, \bibinfo {author} {\bibfnamefont {D.}~\bibnamefont
  {Pantalica}}, \bibinfo {author} {\bibfnamefont {Y.-E.}\ \bibnamefont
  {Penionzhkevich}}, \bibinfo {author} {\bibfnamefont {F.}~\bibnamefont
  {Pougheon}}, \bibinfo {author} {\bibfnamefont {A.}~\bibnamefont {Poves}},
  \bibinfo {author} {\bibfnamefont {N.}~\bibnamefont {Redon}}, \bibinfo
  {author} {\bibfnamefont {M.~G.}\ \bibnamefont {Saint-Laurent}}, \bibinfo
  {author} {\bibfnamefont {J.~A.}\ \bibnamefont {Scarpaci}}, \bibinfo {author}
  {\bibfnamefont {G.}~\bibnamefont {Sletten}}, \bibinfo {author} {\bibfnamefont
  {M.}~\bibnamefont {Stanoiu}}, \bibinfo {author} {\bibfnamefont
  {O.}~\bibnamefont {Tarasov}}, \ and\ \bibinfo {author} {\bibfnamefont
  {C.}~\bibnamefont {Theisen}},\ }\href {\doibase
  10.1103/PhysRevLett.88.092501} {\bibfield  {journal} {\bibinfo  {journal}
  {Phys. Rev. Lett.}\ }\textbf {\bibinfo {volume} {88}},\ \bibinfo {pages}
  {092501} (\bibinfo {year} {2002})}\BibitemShut {NoStop}%
\bibitem [{\citenamefont {Gu\'enaut}\ \emph {et~al.}(2007)\citenamefont
  {Gu\'enaut}, \citenamefont {Audi}, \citenamefont {Beck}, \citenamefont
  {Blaum}, \citenamefont {Bollen}, \citenamefont {Delahaye}, \citenamefont
  {Herfurth}, \citenamefont {Kellerbauer}, \citenamefont {Kluge}, \citenamefont
  {Libert}, \citenamefont {Lunney}, \citenamefont {Schwarz}, \citenamefont
  {Schweikhard},\ and\ \citenamefont {Yazidjian}}]{GUE07}%
  \BibitemOpen
  \bibfield  {author} {\bibinfo {author} {\bibfnamefont {C.}~\bibnamefont
  {Gu\'enaut}}, \bibinfo {author} {\bibfnamefont {G.}~\bibnamefont {Audi}},
  \bibinfo {author} {\bibfnamefont {D.}~\bibnamefont {Beck}}, \bibinfo {author}
  {\bibfnamefont {K.}~\bibnamefont {Blaum}}, \bibinfo {author} {\bibfnamefont
  {G.}~\bibnamefont {Bollen}}, \bibinfo {author} {\bibfnamefont
  {P.}~\bibnamefont {Delahaye}}, \bibinfo {author} {\bibfnamefont
  {F.}~\bibnamefont {Herfurth}}, \bibinfo {author} {\bibfnamefont
  {A.}~\bibnamefont {Kellerbauer}}, \bibinfo {author} {\bibfnamefont {H.-J.}\
  \bibnamefont {Kluge}}, \bibinfo {author} {\bibfnamefont {J.}~\bibnamefont
  {Libert}}, \bibinfo {author} {\bibfnamefont {D.}~\bibnamefont {Lunney}},
  \bibinfo {author} {\bibfnamefont {S.}~\bibnamefont {Schwarz}}, \bibinfo
  {author} {\bibfnamefont {L.}~\bibnamefont {Schweikhard}}, \ and\ \bibinfo
  {author} {\bibfnamefont {C.}~\bibnamefont {Yazidjian}},\ }\href {\doibase
  10.1103/PhysRevC.75.044303} {\bibfield  {journal} {\bibinfo  {journal} {Phys.
  Rev. C}\ }\textbf {\bibinfo {volume} {75}},\ \bibinfo {pages} {044303}
  (\bibinfo {year} {2007})}\BibitemShut {NoStop}%
\bibitem [{\citenamefont {Gade}\ \emph {et~al.}(2010)\citenamefont {Gade},
  \citenamefont {Janssens}, \citenamefont {Baugher}, \citenamefont {Bazin},
  \citenamefont {Brown}, \citenamefont {Carpenter}, \citenamefont {Chiara},
  \citenamefont {Deacon}, \citenamefont {Freeman}, \citenamefont {Grinyer},
  \citenamefont {Hoffman}, \citenamefont {Kay}, \citenamefont {Kondev},
  \citenamefont {Lauritsen}, \citenamefont {McDaniel}, \citenamefont
  {Meierbachtol}, \citenamefont {Ratkiewicz}, \citenamefont {Stroberg},
  \citenamefont {Walsh}, \citenamefont {Weisshaar}, \citenamefont {Winkler},\
  and\ \citenamefont {Zhu}}]{GAD10}%
  \BibitemOpen
  \bibfield  {author} {\bibinfo {author} {\bibfnamefont {A.}~\bibnamefont
  {Gade}}, \bibinfo {author} {\bibfnamefont {R.~V.~F.}\ \bibnamefont
  {Janssens}}, \bibinfo {author} {\bibfnamefont {T.}~\bibnamefont {Baugher}},
  \bibinfo {author} {\bibfnamefont {D.}~\bibnamefont {Bazin}}, \bibinfo
  {author} {\bibfnamefont {B.~A.}\ \bibnamefont {Brown}}, \bibinfo {author}
  {\bibfnamefont {M.~P.}\ \bibnamefont {Carpenter}}, \bibinfo {author}
  {\bibfnamefont {C.~J.}\ \bibnamefont {Chiara}}, \bibinfo {author}
  {\bibfnamefont {A.~N.}\ \bibnamefont {Deacon}}, \bibinfo {author}
  {\bibfnamefont {S.~J.}\ \bibnamefont {Freeman}}, \bibinfo {author}
  {\bibfnamefont {G.~F.}\ \bibnamefont {Grinyer}}, \bibinfo {author}
  {\bibfnamefont {C.~R.}\ \bibnamefont {Hoffman}}, \bibinfo {author}
  {\bibfnamefont {B.~P.}\ \bibnamefont {Kay}}, \bibinfo {author} {\bibfnamefont
  {F.~G.}\ \bibnamefont {Kondev}}, \bibinfo {author} {\bibfnamefont
  {T.}~\bibnamefont {Lauritsen}}, \bibinfo {author} {\bibfnamefont
  {S.}~\bibnamefont {McDaniel}}, \bibinfo {author} {\bibfnamefont
  {K.}~\bibnamefont {Meierbachtol}}, \bibinfo {author} {\bibfnamefont
  {A.}~\bibnamefont {Ratkiewicz}}, \bibinfo {author} {\bibfnamefont {S.~R.}\
  \bibnamefont {Stroberg}}, \bibinfo {author} {\bibfnamefont {K.~A.}\
  \bibnamefont {Walsh}}, \bibinfo {author} {\bibfnamefont {D.}~\bibnamefont
  {Weisshaar}}, \bibinfo {author} {\bibfnamefont {R.}~\bibnamefont {Winkler}},
  \ and\ \bibinfo {author} {\bibfnamefont {S.}~\bibnamefont {Zhu}},\ }\href
  {\doibase 10.1103/PhysRevC.81.051304} {\bibfield  {journal} {\bibinfo
  {journal} {Phys. Rev. C}\ }\textbf {\bibinfo {volume} {81}},\ \bibinfo
  {pages} {051304(R)} (\bibinfo {year} {2010})}\BibitemShut {NoStop}%
\bibitem [{\citenamefont {Macchiavelli}(2013)}]{MAC13}%
  \BibitemOpen
  \bibfield  {author} {\bibinfo {author} {\bibfnamefont {A.}~\bibnamefont
  {Macchiavelli}},\ }\href {\doibase 10.5506/APhysPolB.44.359} {\bibfield
  {journal} {\bibinfo  {journal} {Acta Physica Polonica B}\ }\textbf {\bibinfo
  {volume} {44}},\ \bibinfo {pages} {359} (\bibinfo {year} {2013})}\BibitemShut
  {NoStop}%
\bibitem [{\citenamefont {Pritychenko}\ \emph {et~al.}(2012)\citenamefont
  {Pritychenko}, \citenamefont {Choquette}, \citenamefont {Horoi},
  \citenamefont {Karamy},\ and\ \citenamefont {Singh}}]{PRI12}%
  \BibitemOpen
  \bibfield  {author} {\bibinfo {author} {\bibfnamefont {B.}~\bibnamefont
  {Pritychenko}}, \bibinfo {author} {\bibfnamefont {J.}~\bibnamefont
  {Choquette}}, \bibinfo {author} {\bibfnamefont {M.}~\bibnamefont {Horoi}},
  \bibinfo {author} {\bibfnamefont {B.}~\bibnamefont {Karamy}}, \ and\ \bibinfo
  {author} {\bibfnamefont {B.}~\bibnamefont {Singh}},\ }\href {\doibase
  10.1016/j.adt.2012.06.004} {\bibfield  {journal} {\bibinfo  {journal} {Atomic
  Data and Nuclear Data Tables}\ }\textbf {\bibinfo {volume} {98}},\ \bibinfo
  {pages} {798 } (\bibinfo {year} {2012})}\BibitemShut {NoStop}%
\bibitem [{\citenamefont {Baugher}\ \emph {et~al.}(2012)\citenamefont
  {Baugher}, \citenamefont {Gade}, \citenamefont {Janssens}, \citenamefont
  {Lenzi}, \citenamefont {Bazin}, \citenamefont {Brown}, \citenamefont
  {Carpenter}, \citenamefont {Deacon}, \citenamefont {Freeman}, \citenamefont
  {Glasmacher}, \citenamefont {Grinyer}, \citenamefont {Kondev}, \citenamefont
  {McDaniel}, \citenamefont {Poves}, \citenamefont {Ratkiewicz}, \citenamefont
  {McCutchan}, \citenamefont {Sharp}, \citenamefont {Stefanescu}, \citenamefont
  {Walsh}, \citenamefont {Weisshaar},\ and\ \citenamefont {Zhu}}]{BAU12}%
  \BibitemOpen
  \bibfield  {author} {\bibinfo {author} {\bibfnamefont {T.}~\bibnamefont
  {Baugher}}, \bibinfo {author} {\bibfnamefont {A.}~\bibnamefont {Gade}},
  \bibinfo {author} {\bibfnamefont {R.~V.~F.}\ \bibnamefont {Janssens}},
  \bibinfo {author} {\bibfnamefont {S.~M.}\ \bibnamefont {Lenzi}}, \bibinfo
  {author} {\bibfnamefont {D.}~\bibnamefont {Bazin}}, \bibinfo {author}
  {\bibfnamefont {B.~A.}\ \bibnamefont {Brown}}, \bibinfo {author}
  {\bibfnamefont {M.~P.}\ \bibnamefont {Carpenter}}, \bibinfo {author}
  {\bibfnamefont {A.~N.}\ \bibnamefont {Deacon}}, \bibinfo {author}
  {\bibfnamefont {S.~J.}\ \bibnamefont {Freeman}}, \bibinfo {author}
  {\bibfnamefont {T.}~\bibnamefont {Glasmacher}}, \bibinfo {author}
  {\bibfnamefont {G.~F.}\ \bibnamefont {Grinyer}}, \bibinfo {author}
  {\bibfnamefont {F.~G.}\ \bibnamefont {Kondev}}, \bibinfo {author}
  {\bibfnamefont {S.}~\bibnamefont {McDaniel}}, \bibinfo {author}
  {\bibfnamefont {A.}~\bibnamefont {Poves}}, \bibinfo {author} {\bibfnamefont
  {A.}~\bibnamefont {Ratkiewicz}}, \bibinfo {author} {\bibfnamefont {E.~A.}\
  \bibnamefont {McCutchan}}, \bibinfo {author} {\bibfnamefont {D.~K.}\
  \bibnamefont {Sharp}}, \bibinfo {author} {\bibfnamefont {I.}~\bibnamefont
  {Stefanescu}}, \bibinfo {author} {\bibfnamefont {K.~A.}\ \bibnamefont
  {Walsh}}, \bibinfo {author} {\bibfnamefont {D.}~\bibnamefont {Weisshaar}}, \
  and\ \bibinfo {author} {\bibfnamefont {S.}~\bibnamefont {Zhu}},\ }\href
  {\doibase 10.1103/PhysRevC.86.011305} {\bibfield  {journal} {\bibinfo
  {journal} {Phys. Rev. C}\ }\textbf {\bibinfo {volume} {86}},\ \bibinfo
  {pages} {011305} (\bibinfo {year} {2012})}\BibitemShut {NoStop}%
\bibitem [{\citenamefont {Adrich}\ \emph {et~al.}(2008)\citenamefont {Adrich},
  \citenamefont {Amthor}, \citenamefont {Bazin}, \citenamefont {Bowen},
  \citenamefont {Brown}, \citenamefont {Campbell}, \citenamefont {Cook},
  \citenamefont {Gade}, \citenamefont {Galaviz}, \citenamefont {Glasmacher},
  \citenamefont {McDaniel}, \citenamefont {Miller}, \citenamefont {Obertelli},
  \citenamefont {Shimbara}, \citenamefont {Siwek}, \citenamefont {Tostevin},\
  and\ \citenamefont {Weisshaar}}]{ADR08}%
  \BibitemOpen
  \bibfield  {author} {\bibinfo {author} {\bibfnamefont {P.}~\bibnamefont
  {Adrich}}, \bibinfo {author} {\bibfnamefont {A.~M.}\ \bibnamefont {Amthor}},
  \bibinfo {author} {\bibfnamefont {D.}~\bibnamefont {Bazin}}, \bibinfo
  {author} {\bibfnamefont {M.~D.}\ \bibnamefont {Bowen}}, \bibinfo {author}
  {\bibfnamefont {B.~A.}\ \bibnamefont {Brown}}, \bibinfo {author}
  {\bibfnamefont {C.~M.}\ \bibnamefont {Campbell}}, \bibinfo {author}
  {\bibfnamefont {J.~M.}\ \bibnamefont {Cook}}, \bibinfo {author}
  {\bibfnamefont {A.}~\bibnamefont {Gade}}, \bibinfo {author} {\bibfnamefont
  {D.}~\bibnamefont {Galaviz}}, \bibinfo {author} {\bibfnamefont
  {T.}~\bibnamefont {Glasmacher}}, \bibinfo {author} {\bibfnamefont
  {S.}~\bibnamefont {McDaniel}}, \bibinfo {author} {\bibfnamefont
  {D.}~\bibnamefont {Miller}}, \bibinfo {author} {\bibfnamefont
  {A.}~\bibnamefont {Obertelli}}, \bibinfo {author} {\bibfnamefont
  {Y.}~\bibnamefont {Shimbara}}, \bibinfo {author} {\bibfnamefont {K.~P.}\
  \bibnamefont {Siwek}}, \bibinfo {author} {\bibfnamefont {J.~A.}\ \bibnamefont
  {Tostevin}}, \ and\ \bibinfo {author} {\bibfnamefont {D.}~\bibnamefont
  {Weisshaar}},\ }\href {\doibase 10.1103/PhysRevC.77.054306} {\bibfield
  {journal} {\bibinfo  {journal} {Phys. Rev. C}\ }\textbf {\bibinfo {volume}
  {77}},\ \bibinfo {pages} {054306} (\bibinfo {year} {2008})}\BibitemShut
  {NoStop}%
\bibitem [{\citenamefont {Mach}\ \emph {et~al.}(1989)\citenamefont {Mach},
  \citenamefont {Gill},\ and\ \citenamefont {Moszy\'nski}}]{MAC89}%
  \BibitemOpen
  \bibfield  {author} {\bibinfo {author} {\bibfnamefont {H.}~\bibnamefont
  {Mach}}, \bibinfo {author} {\bibfnamefont {R.}~\bibnamefont {Gill}}, \ and\
  \bibinfo {author} {\bibfnamefont {M.}~\bibnamefont {Moszy\'nski}},\ }\href
  {\doibase 10.1016/0168-9002(89)91272-2} {\bibfield  {journal} {\bibinfo
  {journal} {Nuclear Instruments and Methods in Physics Research Section A:
  Accelerators, Spectrometers, Detectors and Associated Equipment}\ }\textbf
  {\bibinfo {volume} {280}},\ \bibinfo {pages} {49 } (\bibinfo {year}
  {1989})}\BibitemShut {NoStop}%
\bibitem [{\citenamefont {Moszy\'nski}\ and\ \citenamefont
  {Mach}(1989)}]{MOS89}%
  \BibitemOpen
  \bibfield  {author} {\bibinfo {author} {\bibfnamefont {M.}~\bibnamefont
  {Moszy\'nski}}\ and\ \bibinfo {author} {\bibfnamefont {H.}~\bibnamefont
  {Mach}},\ }\href {\doibase 10.1016/0168-9002(89)90770-5} {\bibfield
  {journal} {\bibinfo  {journal} {Nuclear Instruments and Methods in Physics
  Research Section A: Accelerators, Spectrometers, Detectors and Associated
  Equipment}\ }\textbf {\bibinfo {volume} {277}},\ \bibinfo {pages} {407 }
  (\bibinfo {year} {1989})}\BibitemShut {NoStop}%
\bibitem [{\citenamefont {Browne}\ and\ \citenamefont {Tuli}(2010)}]{BRO10}%
  \BibitemOpen
  \bibfield  {author} {\bibinfo {author} {\bibfnamefont {E.}~\bibnamefont
  {Browne}}\ and\ \bibinfo {author} {\bibfnamefont {J.}~\bibnamefont {Tuli}},\
  }\href {\doibase 10.1016/j.nds.2010.09.002} {\bibfield  {journal} {\bibinfo
  {journal} {Nuclear Data Sheets}\ }\textbf {\bibinfo {volume} {111}},\
  \bibinfo {pages} {2425 } (\bibinfo {year} {2010})}\BibitemShut {NoStop}%
\bibitem [{\citenamefont {{Hannawald}}(2000)}]{HAN00}%
  \BibitemOpen
  \bibfield  {author} {\bibinfo {author} {\bibfnamefont {M.~W.}\ \bibnamefont
  {{Hannawald}}},\ }\emph {\bibinfo {title} {Kernspektroskopie an N $\approx$
  40 und N $\approx$ 82 Nukliden}},\ \href@noop {} {Ph.D. thesis},\ \bibinfo
  {school} {Univ. Johannes Gutenberg, Mainz} (\bibinfo {year}
  {2000})\BibitemShut {NoStop}%
\bibitem [{\citenamefont {Grzywacz}\ \emph {et~al.}(1998)\citenamefont
  {Grzywacz}, \citenamefont {B\'eraud}, \citenamefont {Borcea}, \citenamefont
  {Emsallem}, \citenamefont {Glogowski}, \citenamefont {Grawe}, \citenamefont
  {Guillemaud-Mueller}, \citenamefont {Hjorth-Jensen}, \citenamefont {Houry},
  \citenamefont {Lewitowicz}, \citenamefont {Mueller}, \citenamefont {Nowak},
  \citenamefont {P\l{}ochocki}, \citenamefont {Pf\"utzner}, \citenamefont
  {Rykaczewski}, \citenamefont {Saint-Laurent}, \citenamefont {Sauvestre},
  \citenamefont {Schaefer}, \citenamefont {Sorlin}, \citenamefont {Szerypo},
  \citenamefont {Trinder}, \citenamefont {Viteritti},\ and\ \citenamefont
  {Winfield}}]{GRZ98}%
  \BibitemOpen
  \bibfield  {author} {\bibinfo {author} {\bibfnamefont {R.}~\bibnamefont
  {Grzywacz}}, \bibinfo {author} {\bibfnamefont {R.}~\bibnamefont {B\'eraud}},
  \bibinfo {author} {\bibfnamefont {C.}~\bibnamefont {Borcea}}, \bibinfo
  {author} {\bibfnamefont {A.}~\bibnamefont {Emsallem}}, \bibinfo {author}
  {\bibfnamefont {M.}~\bibnamefont {Glogowski}}, \bibinfo {author}
  {\bibfnamefont {H.}~\bibnamefont {Grawe}}, \bibinfo {author} {\bibfnamefont
  {D.}~\bibnamefont {Guillemaud-Mueller}}, \bibinfo {author} {\bibfnamefont
  {M.}~\bibnamefont {Hjorth-Jensen}}, \bibinfo {author} {\bibfnamefont
  {M.}~\bibnamefont {Houry}}, \bibinfo {author} {\bibfnamefont
  {M.}~\bibnamefont {Lewitowicz}}, \bibinfo {author} {\bibfnamefont {A.~C.}\
  \bibnamefont {Mueller}}, \bibinfo {author} {\bibfnamefont {A.}~\bibnamefont
  {Nowak}}, \bibinfo {author} {\bibfnamefont {A.}~\bibnamefont {P\l{}ochocki}},
  \bibinfo {author} {\bibfnamefont {M.}~\bibnamefont {Pf\"utzner}}, \bibinfo
  {author} {\bibfnamefont {K.}~\bibnamefont {Rykaczewski}}, \bibinfo {author}
  {\bibfnamefont {M.~G.}\ \bibnamefont {Saint-Laurent}}, \bibinfo {author}
  {\bibfnamefont {J.~E.}\ \bibnamefont {Sauvestre}}, \bibinfo {author}
  {\bibfnamefont {M.}~\bibnamefont {Schaefer}}, \bibinfo {author}
  {\bibfnamefont {O.}~\bibnamefont {Sorlin}}, \bibinfo {author} {\bibfnamefont
  {J.}~\bibnamefont {Szerypo}}, \bibinfo {author} {\bibfnamefont
  {W.}~\bibnamefont {Trinder}}, \bibinfo {author} {\bibfnamefont
  {S.}~\bibnamefont {Viteritti}}, \ and\ \bibinfo {author} {\bibfnamefont
  {J.}~\bibnamefont {Winfield}},\ }\href {\doibase 10.1103/PhysRevLett.81.766}
  {\bibfield  {journal} {\bibinfo  {journal} {Phys. Rev. Lett.}\ }\textbf
  {\bibinfo {volume} {81}},\ \bibinfo {pages} {766} (\bibinfo {year}
  {1998})}\BibitemShut {NoStop}%
\bibitem [{\citenamefont {Daugas}\ \emph {et~al.}(2010)\citenamefont {Daugas},
  \citenamefont {Faul}, \citenamefont {Grawe}, \citenamefont {Pf\"utzner},
  \citenamefont {Grzywacz}, \citenamefont {Lewitowicz}, \citenamefont
  {Achouri}, \citenamefont {Ang\'elique}, \citenamefont {Baiborodin},
  \citenamefont {Bentida}, \citenamefont {B\'eraud}, \citenamefont {Borcea},
  \citenamefont {Bingham}, \citenamefont {Catford}, \citenamefont {Emsallem},
  \citenamefont {de~France}, \citenamefont {Grzywacz}, \citenamefont {Lemmon},
  \citenamefont {Lopez~Jimenez}, \citenamefont {de~Oliveira~Santos},
  \citenamefont {Regan}, \citenamefont {Rykaczewski}, \citenamefont
  {Sauvestre}, \citenamefont {Sawicka}, \citenamefont {Stanoiu}, \citenamefont
  {Sieja},\ and\ \citenamefont {Nowacki}}]{DAU10}%
  \BibitemOpen
  \bibfield  {author} {\bibinfo {author} {\bibfnamefont {J.~M.}\ \bibnamefont
  {Daugas}}, \bibinfo {author} {\bibfnamefont {T.}~\bibnamefont {Faul}},
  \bibinfo {author} {\bibfnamefont {H.}~\bibnamefont {Grawe}}, \bibinfo
  {author} {\bibfnamefont {M.}~\bibnamefont {Pf\"utzner}}, \bibinfo {author}
  {\bibfnamefont {R.}~\bibnamefont {Grzywacz}}, \bibinfo {author}
  {\bibfnamefont {M.}~\bibnamefont {Lewitowicz}}, \bibinfo {author}
  {\bibfnamefont {N.~L.}\ \bibnamefont {Achouri}}, \bibinfo {author}
  {\bibfnamefont {J.~C.}\ \bibnamefont {Ang\'elique}}, \bibinfo {author}
  {\bibfnamefont {D.}~\bibnamefont {Baiborodin}}, \bibinfo {author}
  {\bibfnamefont {R.}~\bibnamefont {Bentida}}, \bibinfo {author} {\bibfnamefont
  {R.}~\bibnamefont {B\'eraud}}, \bibinfo {author} {\bibfnamefont
  {C.}~\bibnamefont {Borcea}}, \bibinfo {author} {\bibfnamefont {C.~R.}\
  \bibnamefont {Bingham}}, \bibinfo {author} {\bibfnamefont {W.~N.}\
  \bibnamefont {Catford}}, \bibinfo {author} {\bibfnamefont {A.}~\bibnamefont
  {Emsallem}}, \bibinfo {author} {\bibfnamefont {G.}~\bibnamefont {de~France}},
  \bibinfo {author} {\bibfnamefont {K.~L.}\ \bibnamefont {Grzywacz}}, \bibinfo
  {author} {\bibfnamefont {R.~C.}\ \bibnamefont {Lemmon}}, \bibinfo {author}
  {\bibfnamefont {M.~J.}\ \bibnamefont {Lopez~Jimenez}}, \bibinfo {author}
  {\bibfnamefont {F.}~\bibnamefont {de~Oliveira~Santos}}, \bibinfo {author}
  {\bibfnamefont {P.~H.}\ \bibnamefont {Regan}}, \bibinfo {author}
  {\bibfnamefont {K.}~\bibnamefont {Rykaczewski}}, \bibinfo {author}
  {\bibfnamefont {J.~E.}\ \bibnamefont {Sauvestre}}, \bibinfo {author}
  {\bibfnamefont {M.}~\bibnamefont {Sawicka}}, \bibinfo {author} {\bibfnamefont
  {M.}~\bibnamefont {Stanoiu}}, \bibinfo {author} {\bibfnamefont
  {K.}~\bibnamefont {Sieja}}, \ and\ \bibinfo {author} {\bibfnamefont
  {F.}~\bibnamefont {Nowacki}},\ }\href {\doibase 10.1103/PhysRevC.81.034304}
  {\bibfield  {journal} {\bibinfo  {journal} {Phys. Rev. C}\ }\textbf {\bibinfo
  {volume} {81}},\ \bibinfo {pages} {034304} (\bibinfo {year}
  {2010})}\BibitemShut {NoStop}%
\bibitem [{\citenamefont {Daugas}\ \emph {et~al.}(2011)\citenamefont {Daugas},
  \citenamefont {Matea}, \citenamefont {Delaroche}, \citenamefont {Pf\"utzner},
  \citenamefont {Sawicka}, \citenamefont {Becker}, \citenamefont {B\'elier},
  \citenamefont {Bingham}, \citenamefont {Borcea}, \citenamefont {Bouchez},
  \citenamefont {Buta}, \citenamefont {Dragulescu}, \citenamefont {Georgiev},
  \citenamefont {Giovinazzo}, \citenamefont {Girod}, \citenamefont {Grawe},
  \citenamefont {Grzywacz}, \citenamefont {Hammache}, \citenamefont {Ibrahim},
  \citenamefont {Lewitowicz}, \citenamefont {Libert}, \citenamefont {Mayet},
  \citenamefont {M\'eot}, \citenamefont {Negoita}, \citenamefont
  {de~Oliveira~Santos}, \citenamefont {Perru}, \citenamefont {Roig},
  \citenamefont {Rykaczewski}, \citenamefont {Saint-Laurent}, \citenamefont
  {Sauvestre}, \citenamefont {Sorlin}, \citenamefont {Stanoiu}, \citenamefont
  {Stefan}, \citenamefont {Stodel}, \citenamefont {Theisen}, \citenamefont
  {Verney},\ and\ \citenamefont {\ifmmode~\dot{Z}\else
  \.{Z}\fi{}ylicz}}]{DAU11}%
  \BibitemOpen
  \bibfield  {author} {\bibinfo {author} {\bibfnamefont {J.~M.}\ \bibnamefont
  {Daugas}}, \bibinfo {author} {\bibfnamefont {I.}~\bibnamefont {Matea}},
  \bibinfo {author} {\bibfnamefont {J.-P.}\ \bibnamefont {Delaroche}}, \bibinfo
  {author} {\bibfnamefont {M.}~\bibnamefont {Pf\"utzner}}, \bibinfo {author}
  {\bibfnamefont {M.}~\bibnamefont {Sawicka}}, \bibinfo {author} {\bibfnamefont
  {F.}~\bibnamefont {Becker}}, \bibinfo {author} {\bibfnamefont
  {G.}~\bibnamefont {B\'elier}}, \bibinfo {author} {\bibfnamefont {C.~R.}\
  \bibnamefont {Bingham}}, \bibinfo {author} {\bibfnamefont {R.}~\bibnamefont
  {Borcea}}, \bibinfo {author} {\bibfnamefont {E.}~\bibnamefont {Bouchez}},
  \bibinfo {author} {\bibfnamefont {A.}~\bibnamefont {Buta}}, \bibinfo {author}
  {\bibfnamefont {E.}~\bibnamefont {Dragulescu}}, \bibinfo {author}
  {\bibfnamefont {G.}~\bibnamefont {Georgiev}}, \bibinfo {author}
  {\bibfnamefont {J.}~\bibnamefont {Giovinazzo}}, \bibinfo {author}
  {\bibfnamefont {M.}~\bibnamefont {Girod}}, \bibinfo {author} {\bibfnamefont
  {H.}~\bibnamefont {Grawe}}, \bibinfo {author} {\bibfnamefont
  {R.}~\bibnamefont {Grzywacz}}, \bibinfo {author} {\bibfnamefont
  {F.}~\bibnamefont {Hammache}}, \bibinfo {author} {\bibfnamefont
  {F.}~\bibnamefont {Ibrahim}}, \bibinfo {author} {\bibfnamefont
  {M.}~\bibnamefont {Lewitowicz}}, \bibinfo {author} {\bibfnamefont
  {J.}~\bibnamefont {Libert}}, \bibinfo {author} {\bibfnamefont
  {P.}~\bibnamefont {Mayet}}, \bibinfo {author} {\bibfnamefont
  {V.}~\bibnamefont {M\'eot}}, \bibinfo {author} {\bibfnamefont
  {F.}~\bibnamefont {Negoita}}, \bibinfo {author} {\bibfnamefont
  {F.}~\bibnamefont {de~Oliveira~Santos}}, \bibinfo {author} {\bibfnamefont
  {O.}~\bibnamefont {Perru}}, \bibinfo {author} {\bibfnamefont
  {O.}~\bibnamefont {Roig}}, \bibinfo {author} {\bibfnamefont {K.}~\bibnamefont
  {Rykaczewski}}, \bibinfo {author} {\bibfnamefont {M.~G.}\ \bibnamefont
  {Saint-Laurent}}, \bibinfo {author} {\bibfnamefont {J.~E.}\ \bibnamefont
  {Sauvestre}}, \bibinfo {author} {\bibfnamefont {O.}~\bibnamefont {Sorlin}},
  \bibinfo {author} {\bibfnamefont {M.}~\bibnamefont {Stanoiu}}, \bibinfo
  {author} {\bibfnamefont {I.}~\bibnamefont {Stefan}}, \bibinfo {author}
  {\bibfnamefont {C.}~\bibnamefont {Stodel}}, \bibinfo {author} {\bibfnamefont
  {C.}~\bibnamefont {Theisen}}, \bibinfo {author} {\bibfnamefont
  {D.}~\bibnamefont {Verney}}, \ and\ \bibinfo {author} {\bibfnamefont
  {J.}~\bibnamefont {\ifmmode~\dot{Z}\else \.{Z}\fi{}ylicz}},\ }\href {\doibase
  10.1103/PhysRevC.83.054312} {\bibfield  {journal} {\bibinfo  {journal} {Phys.
  Rev. C}\ }\textbf {\bibinfo {volume} {83}},\ \bibinfo {pages} {054312}
  (\bibinfo {year} {2011})}\BibitemShut {NoStop}%
\bibitem [{\citenamefont {Daugas}\ \emph {et~al.}(2006)\citenamefont {Daugas},
  \citenamefont {Sawicka}, \citenamefont {Pf\"{u}tzner}, \citenamefont {Matea},
  \citenamefont {Grawe}, \citenamefont {Grzywacz}, \citenamefont {Achouri},
  \citenamefont {Ang\'{e}lique}, \citenamefont {Baiborodin}, \citenamefont
  {Becker}, \citenamefont {B\'{e}lier}, \citenamefont {Bentida}, \citenamefont
  {B\'{e}raud}, \citenamefont {Bingham}, \citenamefont {Borcea}, \citenamefont
  {Borcea}, \citenamefont {Bouchez}, \citenamefont {Buta}, \citenamefont
  {Catford}, \citenamefont {Dragulescu}, \citenamefont {Emsallem},
  \citenamefont {de~France}, \citenamefont {Giovinazzo}, \citenamefont {Girod},
  \citenamefont {Goutte}, \citenamefont {Gorgiev}, \citenamefont
  {Grzywacz-Jones}, \citenamefont {Hammache}, \citenamefont {Ibrahim},
  \citenamefont {Lemmon}, \citenamefont {Lewitowicz}, \citenamefont
  {Lopez-Jimenez}, \citenamefont {Mayet}, \citenamefont {M\'{e}ot},
  \citenamefont {Negoita}, \citenamefont {de~Oliveira-Santos}, \citenamefont
  {Perru}, \citenamefont {Regan}, \citenamefont {Roig}, \citenamefont
  {Rykaczewski}, \citenamefont {Saint-Laurent}, \citenamefont {Sauvestre},
  \citenamefont {Sletten}, \citenamefont {Sorlin}, \citenamefont {Stanoiu},
  \citenamefont {Stefan}, \citenamefont {Stodel}, \citenamefont {Theisen},
  \citenamefont {Verney},\ and\ \citenamefont {Zylicz}}]{DAU06}%
  \BibitemOpen
  \bibfield  {author} {\bibinfo {author} {\bibfnamefont {J.~M.}\ \bibnamefont
  {Daugas}}, \bibinfo {author} {\bibfnamefont {M.}~\bibnamefont {Sawicka}},
  \bibinfo {author} {\bibfnamefont {M.}~\bibnamefont {Pf\"{u}tzner}}, \bibinfo
  {author} {\bibfnamefont {I.}~\bibnamefont {Matea}}, \bibinfo {author}
  {\bibfnamefont {H.}~\bibnamefont {Grawe}}, \bibinfo {author} {\bibfnamefont
  {R.}~\bibnamefont {Grzywacz}}, \bibinfo {author} {\bibfnamefont {N.~L.}\
  \bibnamefont {Achouri}}, \bibinfo {author} {\bibfnamefont {J.~C.}\
  \bibnamefont {Ang\'{e}lique}}, \bibinfo {author} {\bibfnamefont
  {D.}~\bibnamefont {Baiborodin}}, \bibinfo {author} {\bibfnamefont
  {F.}~\bibnamefont {Becker}}, \bibinfo {author} {\bibfnamefont
  {G.}~\bibnamefont {B\'{e}lier}}, \bibinfo {author} {\bibfnamefont
  {R.}~\bibnamefont {Bentida}}, \bibinfo {author} {\bibfnamefont
  {R.}~\bibnamefont {B\'{e}raud}}, \bibinfo {author} {\bibfnamefont
  {C.}~\bibnamefont {Bingham}}, \bibinfo {author} {\bibfnamefont
  {C.}~\bibnamefont {Borcea}}, \bibinfo {author} {\bibfnamefont
  {R.}~\bibnamefont {Borcea}}, \bibinfo {author} {\bibfnamefont
  {E.}~\bibnamefont {Bouchez}}, \bibinfo {author} {\bibfnamefont
  {A.}~\bibnamefont {Buta}}, \bibinfo {author} {\bibfnamefont {W.~N.}\
  \bibnamefont {Catford}}, \bibinfo {author} {\bibfnamefont {E.}~\bibnamefont
  {Dragulescu}}, \bibinfo {author} {\bibfnamefont {A.}~\bibnamefont
  {Emsallem}}, \bibinfo {author} {\bibfnamefont {G.}~\bibnamefont {de~France}},
  \bibinfo {author} {\bibfnamefont {J.}~\bibnamefont {Giovinazzo}}, \bibinfo
  {author} {\bibfnamefont {M.}~\bibnamefont {Girod}}, \bibinfo {author}
  {\bibfnamefont {H.}~\bibnamefont {Goutte}}, \bibinfo {author} {\bibfnamefont
  {G.}~\bibnamefont {Gorgiev}}, \bibinfo {author} {\bibfnamefont {K.~L.}\
  \bibnamefont {Grzywacz-Jones}}, \bibinfo {author} {\bibfnamefont
  {F.}~\bibnamefont {Hammache}}, \bibinfo {author} {\bibfnamefont
  {F.}~\bibnamefont {Ibrahim}}, \bibinfo {author} {\bibfnamefont {R.~C.}\
  \bibnamefont {Lemmon}}, \bibinfo {author} {\bibfnamefont {M.}~\bibnamefont
  {Lewitowicz}}, \bibinfo {author} {\bibfnamefont {M.~J.}\ \bibnamefont
  {Lopez-Jimenez}}, \bibinfo {author} {\bibfnamefont {P.}~\bibnamefont
  {Mayet}}, \bibinfo {author} {\bibfnamefont {V.}~\bibnamefont {M\'{e}ot}},
  \bibinfo {author} {\bibfnamefont {F.}~\bibnamefont {Negoita}}, \bibinfo
  {author} {\bibfnamefont {F.}~\bibnamefont {de~Oliveira-Santos}}, \bibinfo
  {author} {\bibfnamefont {O.}~\bibnamefont {Perru}}, \bibinfo {author}
  {\bibfnamefont {P.~H.}\ \bibnamefont {Regan}}, \bibinfo {author}
  {\bibfnamefont {O.}~\bibnamefont {Roig}}, \bibinfo {author} {\bibfnamefont
  {K.}~\bibnamefont {Rykaczewski}}, \bibinfo {author} {\bibfnamefont {M.~G.}\
  \bibnamefont {Saint-Laurent}}, \bibinfo {author} {\bibfnamefont {J.~E.}\
  \bibnamefont {Sauvestre}}, \bibinfo {author} {\bibfnamefont {G.}~\bibnamefont
  {Sletten}}, \bibinfo {author} {\bibfnamefont {O.}~\bibnamefont {Sorlin}},
  \bibinfo {author} {\bibfnamefont {M.}~\bibnamefont {Stanoiu}}, \bibinfo
  {author} {\bibfnamefont {I.}~\bibnamefont {Stefan}}, \bibinfo {author}
  {\bibfnamefont {C.}~\bibnamefont {Stodel}}, \bibinfo {author} {\bibfnamefont
  {C.}~\bibnamefont {Theisen}}, \bibinfo {author} {\bibfnamefont
  {D.}~\bibnamefont {Verney}}, \ and\ \bibinfo {author} {\bibfnamefont
  {J.}~\bibnamefont {Zylicz}},\ }\href {\doibase 10.1063/1.2200971} {\bibfield
  {journal} {\bibinfo  {journal} {AIP Conference Proceedings}\ }\textbf
  {\bibinfo {volume} {831}},\ \bibinfo {pages} {427} (\bibinfo {year}
  {2006})}\BibitemShut {NoStop}%
\bibitem [{\citenamefont {Block}\ \emph {et~al.}(2008)\citenamefont {Block},
  \citenamefont {Bachelet}, \citenamefont {Bollen}, \citenamefont {Facina},
  \citenamefont {Folden}, \citenamefont {Gu\'enaut}, \citenamefont
  {Kwiatkowski}, \citenamefont {Morrissey}, \citenamefont {Pang}, \citenamefont
  {Prinke}, \citenamefont {Ringle}, \citenamefont {Savory}, \citenamefont
  {Schury},\ and\ \citenamefont {Schwarz}}]{BLO08}%
  \BibitemOpen
  \bibfield  {author} {\bibinfo {author} {\bibfnamefont {M.}~\bibnamefont
  {Block}}, \bibinfo {author} {\bibfnamefont {C.}~\bibnamefont {Bachelet}},
  \bibinfo {author} {\bibfnamefont {G.}~\bibnamefont {Bollen}}, \bibinfo
  {author} {\bibfnamefont {M.}~\bibnamefont {Facina}}, \bibinfo {author}
  {\bibfnamefont {C.~M.}\ \bibnamefont {Folden}}, \bibinfo {author}
  {\bibfnamefont {C.}~\bibnamefont {Gu\'enaut}}, \bibinfo {author}
  {\bibfnamefont {A.~A.}\ \bibnamefont {Kwiatkowski}}, \bibinfo {author}
  {\bibfnamefont {D.~J.}\ \bibnamefont {Morrissey}}, \bibinfo {author}
  {\bibfnamefont {G.~K.}\ \bibnamefont {Pang}}, \bibinfo {author}
  {\bibfnamefont {A.}~\bibnamefont {Prinke}}, \bibinfo {author} {\bibfnamefont
  {R.}~\bibnamefont {Ringle}}, \bibinfo {author} {\bibfnamefont
  {J.}~\bibnamefont {Savory}}, \bibinfo {author} {\bibfnamefont
  {P.}~\bibnamefont {Schury}}, \ and\ \bibinfo {author} {\bibfnamefont
  {S.}~\bibnamefont {Schwarz}},\ }\href {\doibase
  10.1103/PhysRevLett.100.132501} {\bibfield  {journal} {\bibinfo  {journal}
  {Phys. Rev. Lett.}\ }\textbf {\bibinfo {volume} {100}},\ \bibinfo {pages}
  {132501} (\bibinfo {year} {2008})}\BibitemShut {NoStop}%
\bibitem [{\citenamefont {Ferrer}\ \emph {et~al.}(2010)\citenamefont {Ferrer},
  \citenamefont {Block}, \citenamefont {Bachelet}, \citenamefont {Barquest},
  \citenamefont {Bollen}, \citenamefont {Campbell}, \citenamefont {Facina},
  \citenamefont {Folden}, \citenamefont {Gu\'enaut}, \citenamefont
  {Kwiatkowski}, \citenamefont {Lincoln}, \citenamefont {Morrissey},
  \citenamefont {Pang}, \citenamefont {Prinke}, \citenamefont {Ringle},
  \citenamefont {Savory}, \citenamefont {Schury},\ and\ \citenamefont
  {Schwarz}}]{FER10}%
  \BibitemOpen
  \bibfield  {author} {\bibinfo {author} {\bibfnamefont {R.}~\bibnamefont
  {Ferrer}}, \bibinfo {author} {\bibfnamefont {M.}~\bibnamefont {Block}},
  \bibinfo {author} {\bibfnamefont {C.}~\bibnamefont {Bachelet}}, \bibinfo
  {author} {\bibfnamefont {B.~R.}\ \bibnamefont {Barquest}}, \bibinfo {author}
  {\bibfnamefont {G.}~\bibnamefont {Bollen}}, \bibinfo {author} {\bibfnamefont
  {C.~M.}\ \bibnamefont {Campbell}}, \bibinfo {author} {\bibfnamefont
  {M.}~\bibnamefont {Facina}}, \bibinfo {author} {\bibfnamefont {C.~M.}\
  \bibnamefont {Folden}}, \bibinfo {author} {\bibfnamefont {C.}~\bibnamefont
  {Gu\'enaut}}, \bibinfo {author} {\bibfnamefont {A.~A.}\ \bibnamefont
  {Kwiatkowski}}, \bibinfo {author} {\bibfnamefont {D.~L.}\ \bibnamefont
  {Lincoln}}, \bibinfo {author} {\bibfnamefont {D.~J.}\ \bibnamefont
  {Morrissey}}, \bibinfo {author} {\bibfnamefont {G.~K.}\ \bibnamefont {Pang}},
  \bibinfo {author} {\bibfnamefont {A.~M.}\ \bibnamefont {Prinke}}, \bibinfo
  {author} {\bibfnamefont {R.}~\bibnamefont {Ringle}}, \bibinfo {author}
  {\bibfnamefont {J.}~\bibnamefont {Savory}}, \bibinfo {author} {\bibfnamefont
  {P.}~\bibnamefont {Schury}}, \ and\ \bibinfo {author} {\bibfnamefont
  {S.}~\bibnamefont {Schwarz}},\ }\href {\doibase 10.1103/PhysRevC.81.044318}
  {\bibfield  {journal} {\bibinfo  {journal} {Phys. Rev. C}\ }\textbf {\bibinfo
  {volume} {81}},\ \bibinfo {pages} {044318} (\bibinfo {year}
  {2010})}\BibitemShut {NoStop}%
\bibitem [{\citenamefont {Pauwels}\ \emph {et~al.}(2009)\citenamefont
  {Pauwels}, \citenamefont {Ivanov}, \citenamefont {Bree}, \citenamefont
  {B\"uscher}, \citenamefont {Cocolios}, \citenamefont {Huyse}, \citenamefont
  {Kudryavtsev}, \citenamefont {Raabe}, \citenamefont {Sawicka}, \citenamefont
  {de~Walle}, \citenamefont {Duppen}, \citenamefont {Korgul}, \citenamefont
  {Stefanescu}, \citenamefont {Hecht}, \citenamefont {Hoteling}, \citenamefont
  {W\"ohr}, \citenamefont {Walters}, \citenamefont {Broda}, \citenamefont
  {Fornal}, \citenamefont {Krolas}, \citenamefont {Pawlat}, \citenamefont
  {Wrzesinski}, \citenamefont {Carpenter}, \citenamefont {Janssens},
  \citenamefont {Lauritsen}, \citenamefont {Seweryniak}, \citenamefont {Zhu},
  \citenamefont {Stone},\ and\ \citenamefont {Wang}}]{PAU09}%
  \BibitemOpen
  \bibfield  {author} {\bibinfo {author} {\bibfnamefont {D.}~\bibnamefont
  {Pauwels}}, \bibinfo {author} {\bibfnamefont {O.}~\bibnamefont {Ivanov}},
  \bibinfo {author} {\bibfnamefont {N.}~\bibnamefont {Bree}}, \bibinfo {author}
  {\bibfnamefont {J.}~\bibnamefont {B\"uscher}}, \bibinfo {author}
  {\bibfnamefont {T.~E.}\ \bibnamefont {Cocolios}}, \bibinfo {author}
  {\bibfnamefont {M.}~\bibnamefont {Huyse}}, \bibinfo {author} {\bibfnamefont
  {Y.}~\bibnamefont {Kudryavtsev}}, \bibinfo {author} {\bibfnamefont
  {R.}~\bibnamefont {Raabe}}, \bibinfo {author} {\bibfnamefont
  {M.}~\bibnamefont {Sawicka}}, \bibinfo {author} {\bibfnamefont {J.~V.}\
  \bibnamefont {de~Walle}}, \bibinfo {author} {\bibfnamefont {P.~V.}\
  \bibnamefont {Duppen}}, \bibinfo {author} {\bibfnamefont {A.}~\bibnamefont
  {Korgul}}, \bibinfo {author} {\bibfnamefont {I.}~\bibnamefont {Stefanescu}},
  \bibinfo {author} {\bibfnamefont {A.~A.}\ \bibnamefont {Hecht}}, \bibinfo
  {author} {\bibfnamefont {N.}~\bibnamefont {Hoteling}}, \bibinfo {author}
  {\bibfnamefont {A.}~\bibnamefont {W\"ohr}}, \bibinfo {author} {\bibfnamefont
  {W.~B.}\ \bibnamefont {Walters}}, \bibinfo {author} {\bibfnamefont
  {R.}~\bibnamefont {Broda}}, \bibinfo {author} {\bibfnamefont
  {B.}~\bibnamefont {Fornal}}, \bibinfo {author} {\bibfnamefont
  {W.}~\bibnamefont {Krolas}}, \bibinfo {author} {\bibfnamefont
  {T.}~\bibnamefont {Pawlat}}, \bibinfo {author} {\bibfnamefont
  {J.}~\bibnamefont {Wrzesinski}}, \bibinfo {author} {\bibfnamefont {M.~P.}\
  \bibnamefont {Carpenter}}, \bibinfo {author} {\bibfnamefont {R.~V.~F.}\
  \bibnamefont {Janssens}}, \bibinfo {author} {\bibfnamefont {T.}~\bibnamefont
  {Lauritsen}}, \bibinfo {author} {\bibfnamefont {D.}~\bibnamefont
  {Seweryniak}}, \bibinfo {author} {\bibfnamefont {S.}~\bibnamefont {Zhu}},
  \bibinfo {author} {\bibfnamefont {J.~R.}\ \bibnamefont {Stone}}, \ and\
  \bibinfo {author} {\bibfnamefont {X.}~\bibnamefont {Wang}},\ }\href {\doibase
  10.1103/PhysRevC.79.044309} {\bibfield  {journal} {\bibinfo  {journal} {Phys.
  Rev. C}\ }\textbf {\bibinfo {volume} {79}},\ \bibinfo {pages} {044309}
  (\bibinfo {year} {2009})}\BibitemShut {NoStop}%
\bibitem [{\citenamefont {{Gaudefroy}}(2005)}]{GAU05PhD}%
  \BibitemOpen
  \bibfield  {author} {\bibinfo {author} {\bibfnamefont {L.}~\bibnamefont
  {{Gaudefroy}}},\ }\emph {\bibinfo {title} {Etude de la fermeture de couche
  N=28 autour du noyau Ar-28 par reaction de transfert d'un neutron :
  application a l'astrophysique et Spectroscopie beta de noyaux riches en
  neutrons autour de N=32/34 et N=40}},\ \href@noop {} {Ph.D. thesis},\
  \bibinfo  {school} {Univ. de Paris XI U.F.R. scientifique D'Orsay} (\bibinfo
  {year} {2005})\BibitemShut {NoStop}%
\bibitem [{\citenamefont {Fedosseev}\ \emph {et~al.}(2012)\citenamefont
  {Fedosseev}, \citenamefont {Berg}, \citenamefont {Fedorov}, \citenamefont
  {Fink}, \citenamefont {Launila}, \citenamefont {Losito}, \citenamefont
  {Marsh}, \citenamefont {Rossel}, \citenamefont {Rothe}, \citenamefont
  {Seliverstov}, \citenamefont {Sj\"{o}din},\ and\ \citenamefont
  {Wendt}}]{FED12}%
  \BibitemOpen
  \bibfield  {author} {\bibinfo {author} {\bibfnamefont {V.~N.}\ \bibnamefont
  {Fedosseev}}, \bibinfo {author} {\bibfnamefont {L.-E.}\ \bibnamefont {Berg}},
  \bibinfo {author} {\bibfnamefont {D.~V.}\ \bibnamefont {Fedorov}}, \bibinfo
  {author} {\bibfnamefont {D.}~\bibnamefont {Fink}}, \bibinfo {author}
  {\bibfnamefont {O.~J.}\ \bibnamefont {Launila}}, \bibinfo {author}
  {\bibfnamefont {R.}~\bibnamefont {Losito}}, \bibinfo {author} {\bibfnamefont
  {B.~A.}\ \bibnamefont {Marsh}}, \bibinfo {author} {\bibfnamefont {R.~E.}\
  \bibnamefont {Rossel}}, \bibinfo {author} {\bibfnamefont {S.}~\bibnamefont
  {Rothe}}, \bibinfo {author} {\bibfnamefont {M.~D.}\ \bibnamefont
  {Seliverstov}}, \bibinfo {author} {\bibfnamefont {A.~M.}\ \bibnamefont
  {Sj\"{o}din}}, \ and\ \bibinfo {author} {\bibfnamefont {K.~D.~A.}\
  \bibnamefont {Wendt}},\ }\href {\doibase 10.1063/1.3662206} {\bibfield
  {journal} {\bibinfo  {journal} {Review of Scientific Instruments}\ }\textbf
  {\bibinfo {volume} {83}},\ \bibinfo {eid} {02A903} (\bibinfo {year}
  {2012})}\BibitemShut {NoStop}%
\bibitem [{\citenamefont {Kugler}(2000)}]{KUG00}%
  \BibitemOpen
  \bibfield  {author} {\bibinfo {author} {\bibfnamefont {E.}~\bibnamefont
  {Kugler}},\ }\href {\doibase 10.1023/A:1012603025802} {\bibfield  {journal}
  {\bibinfo  {journal} {Hyperfine Interactions}\ }\textbf {\bibinfo {volume}
  {129}},\ \bibinfo {pages} {23} (\bibinfo {year} {2000})}\BibitemShut
  {NoStop}%
\bibitem [{\citenamefont {Köster}(2002)}]{KOS02}%
  \BibitemOpen
  \bibfield  {author} {\bibinfo {author} {\bibfnamefont {U.}~\bibnamefont
  {Köster}},\ }\href {\doibase 10.1140/epja/i2001-10264-2} {\bibfield
  {journal} {\bibinfo  {journal} {The European Physical Journal A - Hadrons and
  Nuclei}\ }\textbf {\bibinfo {volume} {15}},\ \bibinfo {pages} {255} (\bibinfo
  {year} {2002})}\BibitemShut {NoStop}%
\bibitem [{\citenamefont {Mach}\ \emph {et~al.}(1991)\citenamefont {Mach},
  \citenamefont {Wohn}, \citenamefont {Moln\'ar}, \citenamefont {Sistemich},
  \citenamefont {Hill}, \citenamefont {Moszy\'nski}, \citenamefont {Gill},
  \citenamefont {Krips},\ and\ \citenamefont {Brenner}}]{MAC91}%
  \BibitemOpen
  \bibfield  {author} {\bibinfo {author} {\bibfnamefont {H.}~\bibnamefont
  {Mach}}, \bibinfo {author} {\bibfnamefont {F.}~\bibnamefont {Wohn}}, \bibinfo
  {author} {\bibfnamefont {G.}~\bibnamefont {Moln\'ar}}, \bibinfo {author}
  {\bibfnamefont {K.}~\bibnamefont {Sistemich}}, \bibinfo {author}
  {\bibfnamefont {J.~C.}\ \bibnamefont {Hill}}, \bibinfo {author}
  {\bibfnamefont {M.}~\bibnamefont {Moszy\'nski}}, \bibinfo {author}
  {\bibfnamefont {R.}~\bibnamefont {Gill}}, \bibinfo {author} {\bibfnamefont
  {W.}~\bibnamefont {Krips}}, \ and\ \bibinfo {author} {\bibfnamefont
  {D.}~\bibnamefont {Brenner}},\ }\href {\doibase 10.1016/0375-9474(91)90001-M}
  {\bibfield  {journal} {\bibinfo  {journal} {Nuclear Physics A}\ }\textbf
  {\bibinfo {volume} {523}},\ \bibinfo {pages} {197 } (\bibinfo {year}
  {1991})}\BibitemShut {NoStop}%
\bibitem [{XIA(2011)}]{XIA}%
  \BibitemOpen
  \href {http://www.xia.com} {\emph {\bibinfo {title} {User's Manual Digital
  Gamma Finder (DGF) Pixie-4}}},\ \bibinfo {organization} {XIA},\ \bibinfo
  {edition} {2nd}\ ed. (\bibinfo {year} {2011})\BibitemShut {NoStop}%
\bibitem [{\citenamefont {Naimi}\ \emph {et~al.}(2012)\citenamefont {Naimi},
  \citenamefont {Audi}, \citenamefont {Beck}, \citenamefont {Blaum},
  \citenamefont {B\"ohm}, \citenamefont {Borgmann}, \citenamefont
  {Breitenfeldt}, \citenamefont {George}, \citenamefont {Herfurth},
  \citenamefont {Herlert}, \citenamefont {Kellerbauer}, \citenamefont
  {Kowalska}, \citenamefont {Lunney}, \citenamefont {Minaya~Ramirez},
  \citenamefont {Neidherr}, \citenamefont {Rosenbusch}, \citenamefont
  {Schweikhard}, \citenamefont {Wolf},\ and\ \citenamefont {Zuber}}]{NAI12}%
  \BibitemOpen
  \bibfield  {author} {\bibinfo {author} {\bibfnamefont {S.}~\bibnamefont
  {Naimi}}, \bibinfo {author} {\bibfnamefont {G.}~\bibnamefont {Audi}},
  \bibinfo {author} {\bibfnamefont {D.}~\bibnamefont {Beck}}, \bibinfo {author}
  {\bibfnamefont {K.}~\bibnamefont {Blaum}}, \bibinfo {author} {\bibfnamefont
  {C.}~\bibnamefont {B\"ohm}}, \bibinfo {author} {\bibfnamefont
  {C.}~\bibnamefont {Borgmann}}, \bibinfo {author} {\bibfnamefont
  {M.}~\bibnamefont {Breitenfeldt}}, \bibinfo {author} {\bibfnamefont
  {S.}~\bibnamefont {George}}, \bibinfo {author} {\bibfnamefont
  {F.}~\bibnamefont {Herfurth}}, \bibinfo {author} {\bibfnamefont
  {A.}~\bibnamefont {Herlert}}, \bibinfo {author} {\bibfnamefont
  {A.}~\bibnamefont {Kellerbauer}}, \bibinfo {author} {\bibfnamefont
  {M.}~\bibnamefont {Kowalska}}, \bibinfo {author} {\bibfnamefont
  {D.}~\bibnamefont {Lunney}}, \bibinfo {author} {\bibfnamefont
  {E.}~\bibnamefont {Minaya~Ramirez}}, \bibinfo {author} {\bibfnamefont
  {D.}~\bibnamefont {Neidherr}}, \bibinfo {author} {\bibfnamefont
  {M.}~\bibnamefont {Rosenbusch}}, \bibinfo {author} {\bibfnamefont
  {L.}~\bibnamefont {Schweikhard}}, \bibinfo {author} {\bibfnamefont {R.~N.}\
  \bibnamefont {Wolf}}, \ and\ \bibinfo {author} {\bibfnamefont
  {K.}~\bibnamefont {Zuber}},\ }\href {\doibase 10.1103/PhysRevC.86.014325}
  {\bibfield  {journal} {\bibinfo  {journal} {Phys. Rev. C}\ }\textbf {\bibinfo
  {volume} {86}},\ \bibinfo {pages} {014325} (\bibinfo {year}
  {2012})}\BibitemShut {NoStop}%
\bibitem [{\citenamefont {Pauwels}\ \emph {et~al.}(2012)\citenamefont
  {Pauwels}, \citenamefont {Radulov}, \citenamefont {Walters}, \citenamefont
  {Darby}, \citenamefont {De~Witte}, \citenamefont {Diriken}, \citenamefont
  {Fedorov}, \citenamefont {Fedosseev}, \citenamefont {Fraile}, \citenamefont
  {Huyse}, \citenamefont {K\"oster}, \citenamefont {Marsh}, \citenamefont
  {Popescu}, \citenamefont {Seliverstov}, \citenamefont {Sj\"odin},
  \citenamefont {Van~den Bergh}, \citenamefont {Van~de Walle}, \citenamefont
  {Van~Duppen}, \citenamefont {Venhart},\ and\ \citenamefont {Wimmer}}]{PAU12}%
  \BibitemOpen
  \bibfield  {author} {\bibinfo {author} {\bibfnamefont {D.}~\bibnamefont
  {Pauwels}}, \bibinfo {author} {\bibfnamefont {D.}~\bibnamefont {Radulov}},
  \bibinfo {author} {\bibfnamefont {W.~B.}\ \bibnamefont {Walters}}, \bibinfo
  {author} {\bibfnamefont {I.~G.}\ \bibnamefont {Darby}}, \bibinfo {author}
  {\bibfnamefont {H.}~\bibnamefont {De~Witte}}, \bibinfo {author}
  {\bibfnamefont {J.}~\bibnamefont {Diriken}}, \bibinfo {author} {\bibfnamefont
  {D.~V.}\ \bibnamefont {Fedorov}}, \bibinfo {author} {\bibfnamefont {V.~N.}\
  \bibnamefont {Fedosseev}}, \bibinfo {author} {\bibfnamefont {L.~M.}\
  \bibnamefont {Fraile}}, \bibinfo {author} {\bibfnamefont {M.}~\bibnamefont
  {Huyse}}, \bibinfo {author} {\bibfnamefont {U.}~\bibnamefont {K\"oster}},
  \bibinfo {author} {\bibfnamefont {B.~A.}\ \bibnamefont {Marsh}}, \bibinfo
  {author} {\bibfnamefont {L.}~\bibnamefont {Popescu}}, \bibinfo {author}
  {\bibfnamefont {M.~D.}\ \bibnamefont {Seliverstov}}, \bibinfo {author}
  {\bibfnamefont {A.~M.}\ \bibnamefont {Sj\"odin}}, \bibinfo {author}
  {\bibfnamefont {P.}~\bibnamefont {Van~den Bergh}}, \bibinfo {author}
  {\bibfnamefont {J.}~\bibnamefont {Van~de Walle}}, \bibinfo {author}
  {\bibfnamefont {P.}~\bibnamefont {Van~Duppen}}, \bibinfo {author}
  {\bibfnamefont {M.}~\bibnamefont {Venhart}}, \ and\ \bibinfo {author}
  {\bibfnamefont {K.}~\bibnamefont {Wimmer}},\ }\href {\doibase
  10.1103/PhysRevC.86.064318} {\bibfield  {journal} {\bibinfo  {journal} {Phys.
  Rev. C}\ }\textbf {\bibinfo {volume} {86}},\ \bibinfo {pages} {064318}
  (\bibinfo {year} {2012})}\BibitemShut {NoStop}%
\bibitem [{\citenamefont {Hoteling}\ \emph {et~al.}(2006)\citenamefont
  {Hoteling}, \citenamefont {Walters}, \citenamefont {Janssens}, \citenamefont
  {Broda}, \citenamefont {Carpenter}, \citenamefont {Fornal}, \citenamefont
  {Hecht}, \citenamefont {Hjorth-Jensen}, \citenamefont {Kr\'olas},
  \citenamefont {Lauritsen}, \citenamefont {Paw\l{}at}, \citenamefont
  {Seweryniak}, \citenamefont {Wang}, \citenamefont {W\"ohr}, \citenamefont
  {Wrzesi\ifmmode~\acute{n}\else \'{n}\fi{}ski},\ and\ \citenamefont
  {Zhu}}]{HOT06}%
  \BibitemOpen
  \bibfield  {author} {\bibinfo {author} {\bibfnamefont {N.}~\bibnamefont
  {Hoteling}}, \bibinfo {author} {\bibfnamefont {W.~B.}\ \bibnamefont
  {Walters}}, \bibinfo {author} {\bibfnamefont {R.~V.~F.}\ \bibnamefont
  {Janssens}}, \bibinfo {author} {\bibfnamefont {R.}~\bibnamefont {Broda}},
  \bibinfo {author} {\bibfnamefont {M.~P.}\ \bibnamefont {Carpenter}}, \bibinfo
  {author} {\bibfnamefont {B.}~\bibnamefont {Fornal}}, \bibinfo {author}
  {\bibfnamefont {A.~A.}\ \bibnamefont {Hecht}}, \bibinfo {author}
  {\bibfnamefont {M.}~\bibnamefont {Hjorth-Jensen}}, \bibinfo {author}
  {\bibfnamefont {W.}~\bibnamefont {Kr\'olas}}, \bibinfo {author}
  {\bibfnamefont {T.}~\bibnamefont {Lauritsen}}, \bibinfo {author}
  {\bibfnamefont {T.}~\bibnamefont {Paw\l{}at}}, \bibinfo {author}
  {\bibfnamefont {D.}~\bibnamefont {Seweryniak}}, \bibinfo {author}
  {\bibfnamefont {X.}~\bibnamefont {Wang}}, \bibinfo {author} {\bibfnamefont
  {A.}~\bibnamefont {W\"ohr}}, \bibinfo {author} {\bibfnamefont
  {J.}~\bibnamefont {Wrzesi\ifmmode~\acute{n}\else \'{n}\fi{}ski}}, \ and\
  \bibinfo {author} {\bibfnamefont {S.}~\bibnamefont {Zhu}},\ }\href {\doibase
  10.1103/PhysRevC.74.064313} {\bibfield  {journal} {\bibinfo  {journal} {Phys.
  Rev. C}\ }\textbf {\bibinfo {volume} {74}},\ \bibinfo {pages} {064313}
  (\bibinfo {year} {2006})}\BibitemShut {NoStop}%
\bibitem [{\citenamefont {Endt}(1979)}]{END79}%
  \BibitemOpen
  \bibfield  {author} {\bibinfo {author} {\bibfnamefont {P.}~\bibnamefont
  {Endt}},\ }\href {\doibase http://dx.doi.org/10.1016/0092-640X(79)90030-5}
  {\bibfield  {journal} {\bibinfo  {journal} {Atomic Data and Nuclear Data
  Tables}\ }\textbf {\bibinfo {volume} {23}},\ \bibinfo {pages} {547 }
  (\bibinfo {year} {1979})}\BibitemShut {NoStop}%
\bibitem [{\citenamefont {Endt}(1981)}]{END81}%
  \BibitemOpen
  \bibfield  {author} {\bibinfo {author} {\bibfnamefont {P.}~\bibnamefont
  {Endt}},\ }\href {\doibase http://dx.doi.org/10.1016/0092-640X(81)90011-5}
  {\bibfield  {journal} {\bibinfo  {journal} {Atomic Data and Nuclear Data
  Tables}\ }\textbf {\bibinfo {volume} {26}},\ \bibinfo {pages} {47 } (\bibinfo
  {year} {1981})}\BibitemShut {NoStop}%
\end{thebibliography}%

\end{document}